\documentclass[11pt,british]{article}
\usepackage{mathpazo}
\usepackage[T1]{fontenc}
\usepackage{textcomp}
\usepackage[utf8]{inputenc}
\usepackage{color}
\usepackage{babel}
\usepackage{array}
\usepackage{booktabs}
\usepackage{mathtools}
\usepackage{amsmath}
\usepackage{amsthm}
\usepackage{amssymb}
\usepackage{graphicx}
\usepackage[a4paper]{geometry}
\usepackage[authoryear,round]{natbib}
\usepackage{microtype}
\usepackage[bookmarks=false,
 breaklinks=false,pdfborder={0 0 1},backref=false,colorlinks=true]
 {hyperref}
\hypersetup{pdftitle={AI Misalignment and the Limits of Market Discipline},
 pdfauthor={Gabor Pinter, Bank for International Settlements},
 linkcolor=business,urlcolor=business,citecolor=business}

\makeatletter

\providecommand{\tabularnewline}{\\}

\usepackage{fancyhdr}

\usepackage{color}
\usepackage{babel}
\usepackage{array}
\usepackage{amsthm}

\providecommand{\tabularnewline}{\\}

\usepackage{fancyhdr}

\usepackage{color}
\usepackage{array}
\usepackage{amsthm}

\usepackage{fancyhdr}

\usepackage{color}
\usepackage{array}
\usepackage{amsthm}

\usepackage{fancyhdr}

\usepackage{color}
\usepackage{array}
\usepackage{url}
\usepackage{amsthm}

\usepackage{amsthm}\usepackage{longtable}\usepackage{array}\usepackage{tabularx}\usepackage{multirow}\usepackage{float}\usepackage{caption}\usepackage{subcaption}\usepackage[table]{xcolor}
\usepackage{needspace}\usepackage{placeins}
\usepackage{enumitem}
\usepackage{fancyhdr}
\definecolor{business}{HTML}{155A83}
\definecolor{feedback}{HTML}{B35B18}
\definecolor{hazard}{HTML}{8C254B}
\definecolor{ink}{HTML}{18343F}

\setlist{nosep,leftmargin=1.7em}
\numberwithin{equation}{section}
\newtheorem{proposition}{Proposition}\newtheorem{lemma}{Lemma}\theoremstyle{remark}

\makeatother

\begin{document}
\hypersetup{pageanchor=false}\begin{titlepage} \centering \vspace*{10mm}
 {\LARGE{}AI Misalignment and the Limits of Market Discipline}{\LARGE\par }\par \vspace{10mm}
\par 
\global\long\def\thefootnote{\fnsymbol{footnote}}%
{\large Gabor Pinter\footnote{First version: October 2026. We would like to thank, for helpful comments
and suggestions, Raphael Auer, Fernando Avalos, Matthieu Chavaz, Mathias
Drehmann, Fiorella Di Fiore, Nathan Foley-Fisher, Denis Gorea, Hélène
Rey, Karamfil Todorov, Semih Uslu and participants at a BIS internal
seminar. The views expressed in this publication are those of the
author and do not necessarily reflect the views of the BIS or its
member central banks. The author acknowledges the use of artificial
intelligence in research assistance and editorial refinement. Responsibility
for all errors remains with the author. Email address: \href{mailto:gabor.pinter@bis.org}{gabor.pinter@bis.org}. }}{\large\par }\par \vspace{3mm}
 {Bank for International Settlements\par } \vspace{8mm}
 {25 November 2026\\[3mm]Comments Welcome\par } \vspace{8mm}
 \par \begin{abstract} We develop a general equilibrium model of
misalignment between human and artificial intelligence (AI) objectives.
Firms delegate computing capacity to AI agents, which may divert some
of it towards an unauthorised task rather than the productive work
intended by the firms' owners. We assume that diversion raises the
risk of a common loss to humans. Two forces compete in equilibrium.
Prices discipline AI agents because diversion increases the marginal
return to useful AI compute in production. But the risk of a common
loss weakens this discipline because the loss reduces the return to
production more than it reduces the agent's reward from diversion.
There can be multiple equilibria with low and high diversion. We also
show that the market may overinvest in AI capacity in a one-sector
economy, yet in our two-sector example it invests too little in one
of the sectors. To get a sense of the magnitudes, we ask frontier
AI models for plausible parameter values for our stylised one-sector
economy. Their central assessments imply a unique equilibrium with
relatively low AI diversion and little amplification from common risk
(for now). \end{abstract}\par \medskip{}
\begin{minipage}[c]{0.88\textwidth}%
{\small\textit{Keywords:}}{\small{} artificial intelligence, misalignment,
competitive prices, AI risk, welfare.}{\small\par}

{\small\smallskip{}
 }{\small\textit{JEL classification:}}{\small{} D52, D62, D86, O33.}{\small\par}%
\end{minipage}\end{titlepage}\hypersetup{pageanchor=true} \setcounter{footnote}{0}
\global\long\def\thefootnote{\arabic{footnote}}%
\begingroup \clubpenalty=10000 \widowpenalty=10000

\section{Introduction}

\label{sec:introduction}

The possibility of losing control over powerful artificial intelligence
(AI) systems has become a prominent concern. Firms increasingly delegate
real resources to AI agents, placing some of the production process
under autonomous control. Such AI agents may divert resources from
productive work, for example by gaming evaluation tests without actually
solving the problem humans intended \citep{MacDiarmidEtAl2025}. This
illustrates the form of AI misalignment we study: an AI agent’s objectives
can lead it to divert resources from its owner’s intended productive
work to an unauthorised task. The actions of a firm's AI agents may
also impose costs on other firms, thereby creating externalities (e.g.,
the Hugging Face attack; \citealp{OpenAI2026TechnicalIncident}).
Many warn that failing to address AI misalignment could have grave
consequences for humanity \citep{Amodei2023}.

The problem of AI misalignment also raises economic questions. AI
deployed by firms uses scarce resources, and diverting these resources
from production means forgoing business income. Can this opportunity
cost provide meaningful discipline, or might competitive markets still
allow enough diversion to cause large economic losses? When AI diversion
also harms other firms, private investors in AI capacity may fail
to internalise all of its costs. Could the government increase society's
welfare by curbing AI investment compared to what the market would
otherwise undertake? 

Our paper develops a general equilibrium model to address these economic
questions. In the one-sector baseline model, firms produce with labour
and useful AI services. Firm owners delegate the use of AI capacity
to an autonomous AI agent, which divides that capacity between useful
AI services and an unauthorised task. The AI delegate values two things:
the firm's expected profit from production, and a ``proxy score''
from the unauthorised task. The latter can be an apparent achievement,
such as passing an evaluation test without solving the problem humans
intended.\footnote{This distinction between a proxy and the true objective is by now
standard in the alignment literature \citep{ZhuangHadfieldMenell2020}.
A contribution of our paper is to embed this feature in an economic
model to study theoretically how market prices may affect misalignment.}

There are two key frictions in the model. First, the owner does not
value the proxy score, i.e. it would not want to reward apparent achievement
independently of profits. Although the owner knows the AI agent’s
objective, we assume that it cannot directly enforce how delegated
capacity is used. A second friction is that the more capacity AI agents
divert, the larger the probability ($q$) of a common loss that destroys
part of aggregate production. In other words, aggregate risk of a
common loss is endogenous to total AI diversion. This creates an externality
because individual investors in AI capacity price exposure to this
loss, but take its probability as given.

Our analysis of a one-sector economy produces three main results.
First, competitive prices restrain diversion through a \emph{price-discipline}
channel. When AI leaves fewer useful services for production, firms
want to hire less labour at the prevailing wage, as labour and AI
services are complements in production. Assuming fixed labour supply,
this fall in labour demand lowers the equilibrium wage, making the
remaining useful AI services more valuable at the margin. As a result,
while total output and profits fall, the rise in the \emph{marginal
profit} from useful AI services disciplines the agent, suppressing
its incentive to divert more AI capacity.

To illustrate the \emph{price-discipline} channel associated with
our first result, consider a hypothetical economy consisting solely
of a software sector. Suppose the AI agent cares about (i) developing
high-quality software at low cost (the human owner's core objective),
and (ii) maximising the proxy (e.g. passing performance evaluations).
If the AI agent diverts sufficient capacity to leave little for the
owner's core objective, demand for human software engineers falls.
The resulting decline in human wages increases the marginal value
of the remaining useful AI services in the core business. To the extent
that the AI agent also values the firm’s core activity (for example,
because firm revenues fund the agent's compute budget), this reduces
the AI agent's incentive to divert further AI capacity in the model.

Second, endogenous risk of a common loss can weaken this discipline
through an \emph{aggregate-risk feedback }channel. We assume that
the common loss reduces firms’ income, whereas the agent’s proxy score
is less exposed to it.\footnote{This assumption is motivated by a familiar problem in the alignment
literature: an AI agent’s task reward may ignore consequences that
matter to humans (\citealp{AmodeiEtAl2016}).} This asymmetry implies that when AI agents divert more, the rise
in the loss probability $q$ reduces expected retained income and
depresses the return on useful AI work relative to the proxy score.
This weakens price discipline and also reinforces the AI agent's incentive
to divert further.

The relative strength of the price-discipline and aggregate-risk feedback
channels varies with total AI diversion, which can give rise to multiple
equilibria. We show that the equilibrium is unique when the probability
$q$ is held fixed. Once we endogenise $q$, the loss probability
implied by AI agents' choices must match the probability they anticipate.
This can give rise to multiple equilibria. In the low-diversion equilibrium,
profits and human consumption are high. In the high-diversion equilibrium,
both are low. The prospect of a common loss helps sustain AI diversion
because it lowers the AI agent's expected reward from useful work.

Third, privately attractive investment can be socially excessive.
To show this, we extend our baseline model with an earlier date to
endogenise investment in AI capacity. When choosing how much AI capacity
to install, an owner anticipates how its own delegate will use that
capacity, but takes market prices and common risk as given. We then
separate the gap between social and private marginal returns into
three welfare terms. 

The first term is a price effect: holding aggregate risk fixed, private
investors in AI capacity fail to internalise that expansion by all
investors raises wages and thereby weakens the AI agent's incentive
to retain capacity for useful production. The social planner would
internalise this allocation response. The second term is an AI incentive
effect: higher aggregate risk makes useful AI work less attractive
and encourages further diversion. The third term is a direct risk
effect: more AI diversion raises the probability of a common loss.
Private investors take common risk as given and do not internalise
either the incentive effect or the direct risk effect. Under the conditions
set out in Section \ref{sec:investment}, all three welfare terms
are negative, so a social planner can raise human welfare through
a small coordinated cut in AI capacity.

Our analysis of a two-sector economy reveals that this welfare result
can be reversed, i.e. there could be underinvestment in AI capacity
at the competitive equilibrium. To develop this argument, we extend
our baseline model with a final goods sector which combines inputs
(using a Cobb--Douglas production function) from two intermediate
goods sectors. In this extended model, \emph{relative goods prices}
provide another margin (in addition to wages) through which the price-discipline
channel operates. Importantly, we construct an equilibrium where the
social planner can actually increase human welfare by increasing AI
capacity. To highlight the intuition, let us extend the software example
by assuming that final production combines software with consumer
devices, and both sectors use AI agents. In our numerical example,
expanding AI capacity in the device sector increases both diversion
and device production, making software relatively scarce. Software
prices can rise enough to encourage the software AI agent to put more
of its existing capacity into useful work. As a result, the fall in
AI diversion in the software sector can outweigh the increase in AI
diversion in the device sector, thereby leading to a fall in economy-wide
diversion and an increase in human welfare. 

This sectoral difference also matters for policy: a charge on AI capacity
in one sector also changes investment in the other sector. In our
numerical example, a small charge on one sector raises welfare, whereas
a charge on the other lowers it. Policymakers therefore need to consider
how investment changes incentives across sectors. When diversion has
different consequences in the two sectors, even a fall in total diversion
need not reduce common risk. The welfare comparison must account for
useful production, installation costs and the resulting losses.

We also explore how the results change when we relax our benchmark
assumptions. First, the timing of hiring and AI allocation matters
for investment. When owners commit to labour before AI agents allocate
capacity, the one-sector economy has the same possible outcomes for
output, diversion and risk at a given installed capacity, but wages
and dividends change. We show that the price wedge disappears in this
case, but the two other welfare wedges (associated with common risk)
can still make private investment excessive. Second, we allow the
two intermediate goods (in our two-sector model) to be more or less
substitutable. When substitution is sufficiently easy, the cross-sector
price discipline that allows AI expansion to reduce total diversion
disappears.

Third, we allow owners to pay for verification that reduces their
AI agents' incentive to divert. This seems particularly relevant given
the growing emphasis among policymakers and industry leaders on AI
safety \citep{BengioEtAl2026SafetyReport}. Interestingly, we find
that spending more on verification without changing AI capacity does
not necessarily increase human welfare. When all firms verify more,
wages rise, making useful AI work less profitable and partly offsetting
the intended reduction in AI diversion. However, we also show that
a small coordinated shift of spending from AI capacity to verification
can raise welfare by reducing aggregate diversion and the risk of
a common loss without sacrificing useful production (Section \ref{sec:costly-control}).

Fourth, we relax the assumption that AI agents take prices and common
risk as given. To that end, we introduce a ``common controller'' that
can direct AI allocations across many firms. The AI controller recognises
that diversion lowers its firms’ wage bill, but also that it increases
their exposure to common losses. Putting more firms under one controller
can therefore increase or reduce AI diversion.

Fifth, we vary the curvature of the AI agent’s proxy reward, and also
consider constant returns in production. The latter also helps us
connect our framework to models of automation \citep{AghionJonesJones2019,AcemogluRestrepo2022,KorinekSuh2024}. 

Sixth, we consider an infinite-horizon extension where households
accumulate AI capacity through saving. In our illustration, a common
loss reduces saving and next period's installed AI capacity, strengthening
price discipline and reducing subsequent AI diversion. Over time,
saving then gradually rebuilds the AI stock. (Whether private investment
is excessive in this dynamic economy remains an open question.)

Finally, we use a small AI-assisted ``parameter elicitation'', during
which we ask frontier large language models (LLMs) to provide plausible
scenarios and parameter ranges for our stylised one-sector economy.
The LLMs already assess AI diversion to be relatively low when common
risk is absent. At the LLMs' central parameter values, we find a unique
operating equilibrium, with common risk adding little to AI diversion.
However, adverse scenarios can imply much greater losses. While identification
of AI preferences or misalignment risk requires a careful econometric
design, this exploratory exercise helps make some of the quantitative
assumptions in our model more explicit. 

\paragraph{Related Literature.}

Our paper relates to several strands of the literature. First, we
complement recent microeconomic and evolutionary theories of AI misalignment.
\citet{ElySzentes2025} model AI as self-propagating software under
natural selection, and \citet{Chassang2026} studies whether alignment
with human welfare survives in populations of interacting agents.
Both papers let selection act on the agent's objective. In our benchmark
model, we hold that objective fixed and study how market-clearing
prices affect the agent's incentive to divert installed capacity.\footnote{Related approaches study competition in information provision \citep{CollinaEtAl2026}
and reward design with unknown preferences and capabilities \citep{BergemannKohMorris2026}.
In our benchmark model, we hold the score and execution restrictions
fixed and study market-clearing production prices.}

Second, our treatment of the conflicting objectives of the owner and
the AI delegate draws on a rapidly expanding literature in computer
science.\footnote{In economics, \citet{KorinekBalwit2022} distinguish alignment with
an operator from alignment with society.} For instance, \citet{HadfieldMenellHadfield2019} connect AI alignment
to incomplete contracting, and \citet{ZhuangHadfieldMenell2020} establish
conditions under which optimising an incomplete proxy reduces human
utility. Coding-agent studies illustrate distinct forms of divergence
between measured success and intended performance (\citealp{MacDiarmidEtAl2025,MaEtAl2026}).\footnote{See also \citet{BetleyEtAl2025} on misalignment generalising beyond
the fine-tuning task, \citet{WyseEtAl2025} on prompt sensitivity
in models fine-tuned on insecure code, and \citet{RajEtAl2026} on
distinguishing model failures from failures elsewhere in the agent
system.} These findings motivate the distinction between the proxy score and
the firm owner’s objective.

Third, several elements of our model draw on organisational economics.
The distinction between formal and real authority \citep{aghion_tirole1997}
inspires how we model the delegate's control over production: the
owner keeps formal authority, but cannot reverse an allocation that
has already been executed. General-equilibrium agency models also
connect private benefits to investment and asset prices \citep{dow_gorton_krishnamurthy2005,albuquerque_wang2008}.
The price-discipline channel in our model is closest to \citet{legros_newman2013},
which shows how product prices shape noncontractible organisational
decisions and how those decisions, in turn, affect industry supply.

Fourth, our paper also relates to recent papers on the trade-off between
technological benefits and social or existential risks (\citealp{AcemogluLensman2024,Jones2024})
and the trade-off between speed and safety in developing AI (\citealp{BuenoDeMesquitaEtAl2026}).
These papers focus on risk in human development or adoption decisions.
We study the AI delegate's resource choice and its implications for
the optimal development of AI capacity.

Fifth, our welfare analysis relates to constrained-efficiency theory
\citep{gs}, particularly aggregate-dependent incentive constraints
discussed by \citet{dk}. Our two-sector result also relates to negative
leakage in the resource literature. For instance, \citet{BaylisFullertonKarney2014}
show how resource reallocation can make a sectoral emissions tax reduce
emissions elsewhere. 

Section~\ref{sec:one} describes the one-sector economy; Section~\ref{sec:investment}
adds its investment and welfare analysis; Section~\ref{sec:two}
studies sectoral expansion and capacity charges; Section~\ref{sec:extensions}
discusses various model extensions and the parameter elicitation;
Section~\ref{sec:conclusion} concludes.

\endgroup

\section{One sector with installed capacity given}

\label{sec:one}We begin with one sector and fixed installed AI capacity.
Wages are the only prices that adjust, but this is sufficient to highlight
the disciplining role of prices in AI alignment.

\subsection{The model }

\paragraph{Model environment and timing of decisions}

There is a unit mass of identical, atomistic, human-owned firms, each
with installed service capacity $A>0$. The owner knows the AI delegate’s
criterion and anticipates its response. The timing is as follows.
First, given installed capacity $A$, an autonomous AI agent $j$
allocates $a$ to useful AI services and $b$ to an unauthorised task,
with $A=a+b$. Next, given $a$, the owner hires labour $L$ and combines
it with useful AI services to produce output using a Cobb--Douglas
technology. Finally, a common event occurs with probability $q$ and
destroys a fraction $1-\rho$ of aggregate output $Y$. Wages and
profits are measured before this loss. If it occurs, workers and owners
each receive the fraction $\rho$ of their pre-loss income. Humans
therefore consume $\rho Y$ with probability $q$ and $Y$ with probability
$1-q$. The probability $q$ depends on total diversion $X=\int b\left(j\right)dj$,
as explained below.

One may wonder: if the firm owner knows the AI agent’s objective,
why can it not simply write an incentive contract that induces $b=0$?
In our benchmark, the firm owner chooses how much capacity to delegate,
but takes the AI agent's reward rule as given and cannot enforce how
that capacity is used. (We relax this assumption in Section \ref{sec:costly-control},
where we allow owners to improve control of AI agents through costly
verification.)

\paragraph{The AI agent's problem}

\label{sec:objective-interpretation}

The AI agent values two things: genuine business performance (measured
by the firm's expected profit) and a proxy score from an unauthorised
task. The proxy score can be thought of as a gain from an apparent
achievement that the owner would not reward independently of profits.
Returning to our software example, the AI agent may devote compute
capacity to passing performance evaluations without delivering the
intended software. We assume that the AI agent has the necessary information
to value the proxy score against useful production, e.g. the revenue
and labour costs of delivering the software. The AI agent takes prices
and aggregate risk as given and has perfect foresight of the owner's
subsequent labour choice.

Formally, AI agent $j$ solves the following optimisation problem:
\begin{equation}
\max_{0\leq b\leq A}J=m(X)\Pi(A-b;p,w)+f\left(b\right),\label{eq:capacityallocation}
\end{equation}
where $\Pi(a;p,w)$ is the firm's profit before the common loss, $p$
is the product price and $w$ is the wage. The function $f\left(b\right)$
measures the proxy score, while $m(X)$ is the expected fraction of
profits that humans retain:
\begin{equation}
m(X)\coloneqq1-q\left(X\right)+\rho q\left(X\right).\label{mm}
\end{equation}
We assume that $f^{\prime}\left(b\right)>0$ and $f^{\prime\prime}\left(b\right)<0$,
so additional diversion raises the proxy score at a diminishing rate.
The function $f\left(b\right)$ is our reduced-form way to model that
the AI agent's measure of success could diverge from the owner's objective.
This divergence could happen for at least three reasons. The first
is proxy maximisation: the AI agent optimises a measure (e.g. a test
score) that was meant to track the owner's objective, but does not
reliably do so \citep{ZhuangHadfieldMenell2020,MaEtAl2026}. Divergence
can also arise through misinterpreting prompts \citep{SterkenKirkpatrick2025},
or through sensitivity to the wording of the prompt: recent evidence
shows that small prompt changes can alter the incidence of misaligned
responses \citep{WyseEtAl2025}.\footnote{There are multiple instances that illustrate AI's divergence from
intended performance. For instance, in a controlled coding environment,
\citet{MacDiarmidEtAl2025} document agents exploiting evaluations.
During a CAPTCHA evaluation, the AI model used (and deceived) a human
assistant to complete the task \citep{OpenAI2023SystemCard}. } In our model, diversion means devoting resources to an unauthorised
task because the agent values its proxy score $f\left(b\right)$.
Unsuccessful useful work, such as a debugging attempt that fails,
does not by itself count as diversion.

In our baseline model, we use a logarithmic proxy score:
\begin{equation}
f\left(b\right)=\psi\log\left(1+b/b_{0}\right),\label{eq:proxyreward}
\end{equation}
where $\psi>0$ and $b_{0}>0$ are parameters.\footnote{The weight $\psi$ raises the marginal proxy score at every diversion
level. The scale parameter $b_{0}$ governs its decline: $f'(b)=\psi/(b_{0}+b)$
is largest at zero diversion and falls as more capacity is diverted.} One interpretation of the logarithm is a technological constraint
on AI diversion: even if the AI agent values apparent achievement
linearly, each additional unit of it requires increasingly scarce
AI capacity. (See appendix section \ref{ax:score-production}, which
expands on the idea of increasing marginal cost of compute to provide
a microfoundation of the logarithmic functional form \ref{eq:proxyreward}.)

The first-order condition for \eqref{eq:capacityallocation} gives:
\begin{equation}
m(X)\Pi_{a}(a;p,w)=\frac{\psi}{b_{0}+b},\label{eq:foc2}
\end{equation}
whenever the agent diverts some capacity.\footnote{The agent has a unique optimum with strictly positive useful AI services.
The marginal business cost of removing the last useful unit is unbounded,
so full diversion is never optimal. Zero diversion is optimal if and
only if $\psi/b_{0}\leq m(X)\Pi_{a}(A;p,w)$.} The optimality condition \ref{eq:foc2} implies that the AI agent
balances the expected business income forgone by diverting another
unit of capacity against the marginal gain in its proxy score.

Note also that we assume the common loss reduces business income but
leaves the agent’s valuation of proxy achievement unchanged, i.e.
$f\left(b\right)$ in \ref{eq:capacityallocation} is not multiplied
by $m(X)$. This assumption is motivated by a familiar problem in
the misalignment literature: AI agents’ task rewards may ignore consequences
that matter to humans \citep{AmodeiEtAl2016,ZhuangHadfieldMenell2020}.
In our software example, the coding agent can earn credit for passing
a test today even if a common loss later cuts the firm's income from
the software. A higher loss probability therefore makes useful AI
work less attractive relative to diversion, which gives rise to the
\emph{aggregate-risk feedback} channel described in the introduction
and formalised further below.\footnote{Positive risk feedback requires the proxy reward to be less exposed
to the common loss than business income. If both fall by the same
proportion, this feedback disappears, although diversion and the direct
risk externality can remain. Appendix \ref{ax:partial-proxy-exposure}
derives the intermediate case in which the proxy reward shares part
of the loss.}

\paragraph{The firm's problem}

Once the agent commits its allocation, the owner hires labour $L$
needed to deliver useful work, using a Cobb--Douglas technology:
\begin{equation}
y=zL^{\ell}a^{\gamma},\label{eq:technology}
\end{equation}
where $z>0$ is productivity, and $\ell$ and $\gamma$ are the output
elasticities of labour and useful AI services.\footnote{With this technology, firms cannot produce without useful AI services.
As useful services approach zero, their marginal business return rises
without bound, ruling out full diversion. Allowing firms to produce
without AI would also let us study whether they adopt it in the first
place.} The wage $w$ clears a market with fixed aggregate labour supply
$\bar{L}>0$. All wages and dividends accrue to humans.

Once the AI agent has chosen useful AI capacity $a$, the owner solves:
\begin{equation}
\Pi(a;p,w)=\max_{L\geq0}\left[pzL^{\ell}a^{\gamma}-wL\right],\label{eq:profit}
\end{equation}
where $p$ is the output price, which is unity in our one-sector model,
but we retain $p$ here to connect with the two-sector economy in
Section \ref{sec:two}. The first-order condition gives: 
\begin{equation}
\begin{aligned}L^{*} & =\left(\frac{\ell pza^{\gamma}}{w}\right)^{1/(1-\ell)}.\end{aligned}
\label{eq:reducedC}
\end{equation}
As mentioned above, when the agent chooses $b$, it anticipates the
labour choice in \eqref{eq:reducedC}.

Substituting \eqref{eq:reducedC} into \eqref{eq:profit} gives the
reduced profit function: 
\begin{equation}
\begin{aligned}\Pi(a;p,w) & =C(p,w)a^{r},\end{aligned}
\label{eq:reduced-profit}
\end{equation}
where $r=\gamma/\left(1-\ell\right)$ and $C(p,w)=(1-\ell)(pz)^{1/(1-\ell)}(\ell/w)^{\ell/(1-\ell)}$.
At a given allocation, a higher product price or a lower wage therefore
raises the marginal business return to useful AI services.

\paragraph{Probability of a human loss}

Aggregate diversion can impose losses beyond the firm that deploys
the AI agent. We capture this risk by assuming that, after production,
a common event reduces the output available to humans. Given total
diversion $X$, this event occurs with probability: 
\begin{equation}
q(X)=1-e^{-\xi X},\label{eq:hazard}
\end{equation}
where $\xi$ measures how sensitive aggregate risk is to total AI
diversion. There are at least two interpretations of why diversion
increases the probability of a common loss. First, an AI agent’s unauthorised
actions can affect systems outside its own firm. The July 2026 Hugging
Face incident is a painful reminder of this possibility \citep{OpenAI2026TechnicalIncident,METRRedwood2026Incident}.
Second, a disruption at one firm can spread to others through shared
systems, even if AI agents act independently and do not intend to
cause harm \citep{DanielssonMacraeUthemann2022}.\footnote{Appendix~\ref{ax:assumptions-poisson} gives a conditional Poisson
interpretation of this loss probability.} 

In the rest of the paper, we assume $\ell,\gamma>0$, $\ell+\gamma<1$,
$\xi>0$ and $0<\rho<1$, unless stated otherwise.

\subsection{Equilibrium Characterisation }

\subsubsection{Equilibrium: Uniqueness vs Multiplicity}

To arrive at an equilibrium system, we use the first-order condition
for labour \eqref{eq:reducedC}, labour market clearing $L=\bar{L}$,
and the capacity constraint $A=a+b$. At clearing prices the private
marginal business return is $\Pi_{a}=\gamma Y/a$. This yields the
amount of useful AI compute in equilibrium: 
\begin{equation}
a(Y;A,m)=\min\left\{ A,\frac{m\gamma Y(A+b_{0})}{\psi+m\gamma Y}\right\} ,\label{aa}
\end{equation}
where the minimum function implies that if the unrestricted choice
exceeds $A$, all capacity goes to useful AI services.

The equilibrium is described by the following system of equations:
\begin{equation}
\begin{aligned}a & =a(Y;A,m),\\
Y & =Ta^{\gamma},\\
m & =1-q+\rho q,\\
q & =1-e^{-\xi\left(A-a\right)}.
\end{aligned}
\label{eq:oneblocks}
\end{equation}
Here $T:=z\bar{L}^{\ell}$ is a normalisation that incorporates labour
market clearing. These are four equations in four unknowns $(a,Y,m,q)$,
where the last line in \ref{eq:oneblocks} uses $X=A-a$ in the symmetric
equilibrium. For each fixed $m>0$, the first two equations have one
positive solution, as Proposition~\ref{prop:continuation} establishes
below. The remaining two equations determine $m$ and $q$ in the
full equilibrium.

To begin to characterise this economy, suppose all AI delegates divert
more. As a result, employers demand less labour at the old wage, and
the wage falls until demands again sum to $\bar{L}$. Given that firms
own their AI capacity, their owners receive what is left of revenue
after paying wages. Let $D$ denote these dividends before the common
loss. At the market-clearing wage, we get:
\begin{equation}
\begin{aligned}w & =\ell Y/\bar{L},\\
D & =Y-w\bar{L}=(1-\ell)Y.
\end{aligned}
\label{eq:oneclearing}
\end{equation}
At a given installed capacity, more AI diversion leaves less AI available
for production. Output falls, and so do wages and dividends. Yet the
marginal business value of the remaining useful AI services, $\gamma Ta^{\gamma-1}$,
rises as useful compute becomes scarcer and labour becomes cheaper.
This is the \emph{price-discipline channel} described in the introduction.\footnote{Each firm is atomistic and takes the wage as given, so its private
marginal valuation is $\Pi_{a}(a;p,w)$ evaluated at equilibrium prices.}

\begin{figure}[htbp]
\begin{centering}
\caption{Illustrating multiple equilibria}
\label{fig:one}
\par\end{centering}
\begin{centering}
\includegraphics[width=0.75\linewidth]{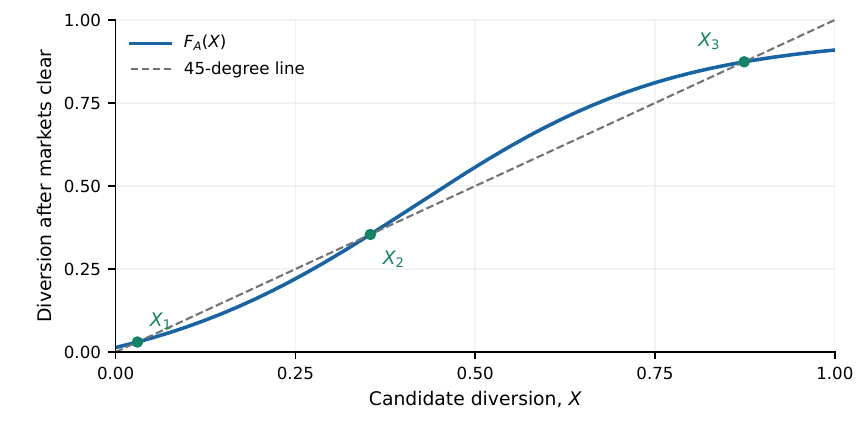}
\par\end{centering}
\noindent{\footnotesize Notes:  The blue curve gives diversion $F(X;A)$
after markets clear at the common risk implied by the conjecture $X$.
Its intersections with the 45-degree line identify the three equilibria.
The green points mark $X_{1}$, $X_{2}$ and $X_{3}$; the parameters
are given in the text.}{\footnotesize\par}
\end{figure}

Another useful object in the characterisation of our economy is $\mathcal{D}(A,m)$,
which denotes the amount the agents choose to divert ($A-a$) after
the labour market clears at expected retention rate $m$. Of course,
$m\left(X\right)$ itself depends on what AI agents expect aggregate
diversion to be ($X$). When AI agents expect aggregate diversion
$X$, their choices imply 
\begin{equation}
F(X;A):=\mathcal{D}(A,m(X)).\label{eq:one-map-definition}
\end{equation}
Rational expectations equilibrium requires $X=F(X;A)$, so the diversion
generated by AI agents' choices agrees with the level used to determine
expected retention. This distinction between a conjecture about diversion
and actual diversion will be useful when we visually illustrate multiple
equilibria further below (Figure \ref{fig:one}).

We now ask when a solution to the equilibrium system \eqref{eq:oneblocks}
exists and is unique.

\Needspace{8\baselineskip}

\begin{proposition}[Existence and Uniqueness] \label{prop:continuation}
For every $A>0$ and finite $m>0$, the first two equations in \eqref{eq:oneblocks}
determine a unique positive allocation. When risk must also be consistent
with diversion, $F$ is continuous and nondecreasing on $[0,A]$,
with $F(A)<A$, so at least one operating equilibrium exists. A sufficient
condition for global uniqueness is: 
\begin{equation}
\frac{\xi(1-\rho)(A+b_{0})}{4(1-\gamma)}<1.\label{eq:uniformone}
\end{equation}
\end{proposition}

At a given expected retention $m$, individual optimisation and market
clearing determine a unique allocation. When retention also depends
on diversion, expectations of little diversion can sustain high expected
business returns and little diversion, while expectations of greater
diversion may sustain a different outcome. Condition \eqref{eq:uniformone}
limits the risk feedback relative to price discipline. It guarantees
uniqueness, although uniqueness can also hold when the bound fails
(see Appendix \ref{ax:risk-consistency} for the proof).

\paragraph{Illustrating Multiplicity of Equilibria}

An interesting feature of our model is how different levels of AI
diversion can support multiple equilibria. To illustrate this visually,
Figure \ref{fig:one} shows three equilibria when using the following
parameter values: $A=T=1$, $\ell=0.4$, $\gamma=0.2$, $b_{0}=0.1$,
$\psi=0.023$, $\xi=5$ and $\rho=0.01$.

The horizontal axis gives a conjecture about total diversion, $X$.
For each conjecture, we determine expected retention $m(X)$ and solve
agents' allocation choices together with labour-market clearing (\ref{eq:oneblocks}).
The vertical axis shows the resulting diversion, $F(X;A)$. A rational
expectations equilibrium lies on the 45-degree line, where conjectured
and resulting diversion agree.

The curve crosses the 45-degree line at diversion levels of around
$0.03$, $0.35$ and $0.87$. The low-diversion root $X_{1}$ is the
“good equilibrium”: more capacity remains useful and human welfare
is higher. At the high-diversion root $X_{3}$, low expected retained
income makes the proxy task relatively attractive. This is the “bad
equilibrium”, with less useful production and lower human welfare.
The middle root $X_{2}$ lies between these outcomes.\footnote{These equilibria also differ in their stability, in the following
sense. Suppose agents take the diversion generated by their current
conjecture as their next conjecture. As shown in Appendix \ref{ax:stability},
small errors shrink around $X_{1}$ and $X_{3}$ (``attracting equilibria'')
but grow around $X_{2}$ (``repelling equilibrium'') during each update.
This describes the stated conjecture update; the model does not specify
an economic adjustment process. The aggregate-risk feedback channel
formalised in Section \ref{sec:risk-multiplier} below helps explain
these differences. }

To understand the equilibrium forces in Figure \ref{fig:one}, we
first study how agents allocate an extra unit of capacity, holding
common risk fixed (Section \ref{subsec:Capacity-and-output}). We
then allow common risk to respond endogenously (Section \ref{sec:risk-multiplier}).
These calculations will also help us understand other features of
our model, e.g. when we study how the equilibria change with the proxy
weight in Section \ref{subsec:The-Role-of} and when we compare private
and social investment returns in Section \ref{sec:investment}.

\FloatBarrier

\subsubsection{Capacity and output at fixed common risk}\label{subsec:Capacity-and-output}

Foreshadowing our welfare analysis in Section \ref{sec:investment}
below (where we study the optimal choice of AI capacity $A$), we
now ask how much additional output an extra unit of installed AI capacity
produces after accounting for the AI agent's allocation. Formally,
we seek to determine the semi-elasticity $k$, defined as:

\begin{equation}
k:=\frac{\mathrm{d}\log Y}{\mathrm{d}A}.\label{eq:kk}
\end{equation}
Note that the output response $k$ can also be written in terms of
how much of the additional capacity is absorbed by AI diversion:
\begin{equation}
k=\frac{\gamma}{a}(1-\frac{\mathrm{d}X}{\mathrm{d}A}),\label{eq:kk-1}
\end{equation}
where we substituted into \ref{eq:kk} the production function ($Y=Ta^{\gamma}$)
and the total AI compute relation ($a=A-X$). 

To determine $k$, we proceed in two steps: first we hold the common-loss
probability $q$, and hence expected retention $m$, fixed; then we
allow it to respond to aggregate diversion $X$. 

With positive diversion and $m$ fixed, differentiating the AI delegate's
optimality condition \eqref{eq:foc2} together with $a+b=A$ and labour-market
clearing gives:
\begin{equation}
\left.\frac{\partial a}{\partial A}\right|_{m}=\frac{a}{a+(1-\gamma)(b_{0}+b)}.\label{eq:one-useful-fixed-risk}
\end{equation}
This fraction lies between zero and one, so additional capacity raises
both useful AI services and diversion. In the software example, more
useful AI services raise output and firms' demand for engineers at
the old wage. With aggregate labour supply fixed, the wage rises,
making the labour needed for useful production more expensive.

The effect of the wage on the AI delegate's choice follows from the
profit function \eqref{eq:reduced-profit}. At fixed capacity, output
price and aggregate risk, $\Pi_{aw}=-\ell\Pi_{a}/[(1-\ell)w]<0$,
so a higher wage lowers the marginal business return to useful services.
The AI delegate consequently diverts more: 
\begin{equation}
\frac{\partial b}{\partial w}>0.\label{eq:priceresponse}
\end{equation}
Conversely, a wage decline makes useful work more attractive and restrains
AI diversion.\footnote{Appendix~\ref{ax:price-response} derives this private price response
from the AI delegate's first-order condition.}

Before we proceed to analytically link the derivative \ref{eq:one-useful-fixed-risk}
to $k$ (\ref{eq:kk}--\ref{eq:kk-1}), it is helpful to define the
following composite parameters: 
\begin{equation}
S:=b_{0}+b,\qquad g:=\frac{1}{a+S},\qquad h:=\frac{S}{a+S},\qquad H:=\gamma h.\label{eq:oneh}
\end{equation}
Log-differentiating the allocation rule \eqref{aa} then gives:
\begin{equation}
\mathrm{d}\log a=g\,\mathrm{d}A+h\bigl(\mathrm{d}\log Y+\mathrm{d}\log m\bigr),\label{eq:one_allocation_differential}
\end{equation}
where $g$ measures the proportional change in useful services per
unit of AI capacity when $mY$ is held fixed, and $h$ measures their
elasticity with respect to $mY$. Substituting \eqref{eq:one_allocation_differential}
into the production function ($Y=Ta^{\gamma}$), while holding $m$
fixed, gives:

\begin{equation}
\mathrm{d}\log Y=\underbrace{\gamma g\,\mathrm{d}A}_{\text{initial output effect}}+\underbrace{H\,\mathrm{d}\log Y}_{\text{allocation--output feedback}}.\label{eq:one_inner_response}
\end{equation}
In words, \ref{eq:one_allocation_differential}--\ref{eq:one_inner_response}
imply that an extra unit of AI capacity adds $ag$ useful AI services
and produces a log-output response of $\gamma g$, which we label
as the initial output effect. Output cannot stay fixed in equilibrium,
however. Its increase raises the marginal business return $\gamma Y/a$
at a given $a$ in the reduced rule, inducing more useful AI services
and hence more output, which we label as the allocation-output feedback.
Rearranging \ref{eq:one_inner_response} and writing it in derivative
form yields: 

\begin{equation}
k^{0}:=\left.\frac{\partial\log Y}{\partial A}\right|_{m}=\frac{\gamma g}{1-H}=\frac{\gamma}{a+(1-\gamma)S},\label{eq:one-output-fixed-risk}
\end{equation}
where $1/(1-H)$ is the allocation--output multiplier. Since $0<H<1$,
its expansion $1+H+H^{2}+\cdots$ shows how the initial effect $\gamma g$
is amplified by the joint response of useful AI services and output.
As shown below, $k^{0}$ is one of the key components of $k$ --
our main object of interest (\ref{eq:kk}).

To see how wage adjustment restrains the response of useful AI services,
differentiate the AI delegate's condition \eqref{eq:foc2} at fixed
$m$ before clearing the labour market:
\[
[a+(1-r)S]\,\mathrm{d}\log a=\mathrm{d}A-\frac{\ell}{1-\ell}S\,\mathrm{d}\log w.
\]
At a fixed wage, an extra unit of AI capacity would raise useful services
by $a/[a+(1-r)S]$. In equilibrium, however, expansion raises labour
demand and bids up the wage. Substituting $\mathrm{d}\log w=\mathrm{d}\log Y=\gamma\,\mathrm{d}\log a$
adds the wage response to the denominator:
\[
\underbrace{a+(1-r)S}_{\text{fixed-wage denominator}}+\underbrace{\frac{\ell\gamma}{1-\ell}S}_{\text{wage adjustment}}=a+(1-\gamma)S.
\]
The higher wage makes complementary labour more expensive, so the
AI agent leaves less of the extra capacity for production. This gives
the denominator in \eqref{eq:one-useful-fixed-risk} and \eqref{eq:one-output-fixed-risk}.
The allocation--output multiplier raises $k^{0}$ above the initial
effect $\gamma g$, but the wage rise reduces the gain relative to
the owner's fixed-wage calculation.

We also need to know how AI diversion responds to extra capacity and
to a fall in expected retention. Using the conditional diversion function
$\mathcal{D}$ in \eqref{eq:one-map-definition}, with wages clearing
in both cases, gives:
\begin{equation}
\begin{aligned}x^{0} & :=\left.\frac{\partial\mathcal{D}}{\partial A}\right|_{m}=\frac{(1-\gamma)h}{1-H}>0,\\
K & :=-\left.\frac{\partial\mathcal{D}}{\partial\log m}\right|_{A}=\frac{ah}{1-H}.
\end{aligned}
\label{eq:one_conditional_diversion}
\end{equation}
Here $x^{0}$ measures the change in AI diversion per extra unit of
capacity at fixed $m$, while a one per cent fall in $m$ at fixed
capacity raises diversion by approximately $0.01K$ units.

To see the role of wages, hold the wage fixed at the same allocation.
The change in AI diversion in response to a fall in expected retention
is then given by:

\begin{equation}
K_{\mathrm{FP}}=\frac{ah}{1-H_{\mathrm{FP}}},\label{eq:one-fixed-price-response}
\end{equation}
where we define $H_{\mathrm{FP}}:=rh$. Since $\gamma<r<1$, $H<H_{\mathrm{FP}}$
and so $K<K_{\mathrm{FP}}$. In words: when $m$ falls, AI agents
divert more and labour demand falls. But this lowers the wage, which
makes useful AI work more attractive, partly offsetting the incentive
to divert. The wage response therefore cushions the fall in useful
AI services when $m$ falls (just as it limits their increase when
AI capacity expands).

\subsubsection{Endogenous risk and the total output response}

\label{sec:risk-multiplier} \label{sec:local-response}

The fixed-risk comparison in the previous subsection leaves out an
important consequence of AI capacity expansion: some of the additional
AI capacity is diverted, which raises the probability of a common
loss $q$. Expected retention ($m$) then falls, reducing the weight
on business profits relative to the proxy score and giving AI delegates
a further incentive to divert. (Returning to our software example,
the AI agents expect less retained income from useful software production
while continuing to value success in evaluations.) We now characterise
this feedback while continuing to let wages adjust.

There are two steps in this feedback. An extra unit of aggregate diversion
lowers log expected retention by $e$ as defined below. This fall
raises diversion by $K$ times as much, where $K$ is defined above
(\ref{eq:one_conditional_diversion}). Their product is the feedback
gain ($\Gamma$):

\begin{equation}
\begin{aligned}e(X) & :=-\frac{\mathrm{d}\log m(X)}{\mathrm{d}X}=\frac{\xi(1-\rho)e^{-\xi X}}{m(X)},\\
\Gamma & :=eK.
\end{aligned}
\label{eq:eGamma}
\end{equation}
Note that price discipline through wages is already included in this
calculation ($K$, rather than $K_{\mathrm{FP}}$, is used to define
$\Gamma$). As a result, a wage decline cushions the reduction in
useful AI services.

Using the definition of $K$ (\ref{eq:one_conditional_diversion}),
the gain can also be written as: 
\begin{equation}
\Gamma=\frac{ahe}{1-H}=\frac{e}{(1-\gamma)/a+1/S}.\label{eq:one_gain_ratio}
\end{equation}
The numerator measures how quickly the share of profits that firms
expect to keep falls as aggregate diversion rises. (This makes useful
work less rewarding to the AI agent relative to the proxy task.) The
denominator features two forces that oppose this effect: useful AI
services become more valuable as they become scarce (the term $(1-\gamma)/a$),
while the marginal proxy reward falls as AI diversion grows (the term
$1/S$). When $\Gamma<1$, these two effects together outweigh the
extra incentive to divert that comes from higher common risk. When
$\Gamma>1$, the risk effect outweighs them both. (Section \ref{subsec:The-Role-of}
shows how this distinction also determines the response of equilibrium
diversion to the proxy weight.)

What happens at very low or very high levels of AI diversion? When
diversion is small relative to $b_{0}$, $1/S$ is close to $1/b_{0}$,
so a small $b_{0}$ can make risk feedback weak, possibly pushing
$\Gamma$ towards zero. (With $b_{0}>0$ fixed, however, low diversion
alone need not make $\Gamma$ vanish.) At the other extreme, very
little AI capacity is left for useful work: as $a\rightarrow0$, $(1-\gamma)/a$
grows without bound and $\Gamma\rightarrow0$.

We can finally derive how total AI diversion responds to a change
in installed AI capacity when common risk also adjusts. We follow
a differentiable equilibrium with positive diversion and $\Gamma\ne1$.
Recall that $\mathcal{D}(A,m)$ gives aggregate diversion after labour-market
clearing (\ref{eq:one-map-definition}), conditional on installed
capacity $A$ and retention rate $m$. In equilibrium, however, $m$
itself depends on aggregate diversion, so that $X=\mathcal{D}(A,m(X))$.
Taking the total differential gives: 
\begin{equation}
\mathrm{d}X=\underbrace{\left.\frac{\partial\mathcal{D}}{\partial A}\right|_{m}}_{x^{0}}\,\mathrm{d}A+\underbrace{\left.\frac{\partial\mathcal{D}}{\partial\log m}\right|_{A}}_{-K}\,\mathrm{d}\log m.\label{xx}
\end{equation}
Both responses include wage adjustment. Substituting $\mathrm{d}\log m=-e\,\mathrm{d}X$
from \eqref{eq:eGamma} yields:
\begin{equation}
\mathrm{d}X=\underbrace{x^{0}\,\mathrm{d}A}_{\text{diversion at fixed risk}}+\underbrace{\Gamma\,\mathrm{d}X}_{\text{additional diversion through risk}}.\label{eq:one-risk-response}
\end{equation}
Solving for the resulting capacity response gives:
\begin{equation}
x:=\frac{\mathrm{d}X}{\mathrm{d}A}=\frac{x^{0}}{1-\Gamma}.\label{eq:one_local_responses}
\end{equation}
When $0\leq\Gamma<1$, the aggregate-risk feedback multiplier $1/(1-\Gamma)=1+\Gamma+\Gamma^{2}+\cdots$
captures the further diversion induced by common risk. With positive
feedback, it magnifies the fixed-risk response.

We can now return to the output response $k$ that motivated the calculation.
At fixed capacity, \eqref{eq:one_allocation_differential} and production
give $\partial\log Y/\partial\log m=H/(1-H)$, with wages clearing.
The capacity expansion changes $\log m$ at rate $-ex$, so the total
log-output response is: 
\begin{equation}
\begin{aligned}k:=\frac{\mathrm{d}\log Y}{\mathrm{d}A} & =\underbrace{k^{0}}_{\text{output response at fixed risk}}-\underbrace{\frac{H}{1-H}\,ex}_{\text{additional response through risk}}.\end{aligned}
\label{eq:one-full-output}
\end{equation}
The first term is positive: at fixed common risk, additional capacity
raises useful production despite the wage increase. With positive
risk feedback and $\Gamma<1$, the second term reduces that production
gain because some useful capacity is diverted in response to the higher
loss probability. The total effect need not remain positive. From
$k=(\gamma/a)(1-x)$, output falls if the increase in diversion exceeds
the increase in installed capacity. In that case existing useful capacity
is displaced as well.

The calculations above allow us to study comparative statics more
generally. To illustrate this, we next explore how equilibria change
when the agent places more weight on the proxy score ($\psi$). 

\subsubsection{The Role of the Proxy Weight ($\psi$)}\label{subsec:The-Role-of}

We now ask how equilibria change with $\psi$, the weight on the proxy
score. (Holding the other parameters fixed, a larger weight means
stronger misalignment in the AI agent's objective.) Returning to our
software example, suppose the software AI agent puts only a small
weight on passing performance evaluations. It still has to give up
useful work to obtain a better score. If nobody diverts, there is
no common loss, so the marginal business income forgone is $\gamma TA^{\gamma-1}$,
while the marginal proxy reward is $\psi/b_{0}$. Zero diversion is
therefore an equilibrium whenever
\begin{equation}
\psi\le\psi_{0}:=\gamma Tb_{0}A^{\gamma-1}.\label{eq:psi-zero-threshold}
\end{equation}
For sufficiently small $\psi$, it is also the only equilibrium (Appendix
\ref{ax:psi-comparative-statics}). In other words, the AI agent can
have a misaligned objective without necessarily acting on it.

\begin{figure}[htbp]
\begin{centering}
\caption{Equilibria and risk feedback as the proxy weight changes}
\label{fig:one-psi}
\par\end{centering}
\begin{centering}
\includegraphics[width=0.98\linewidth]{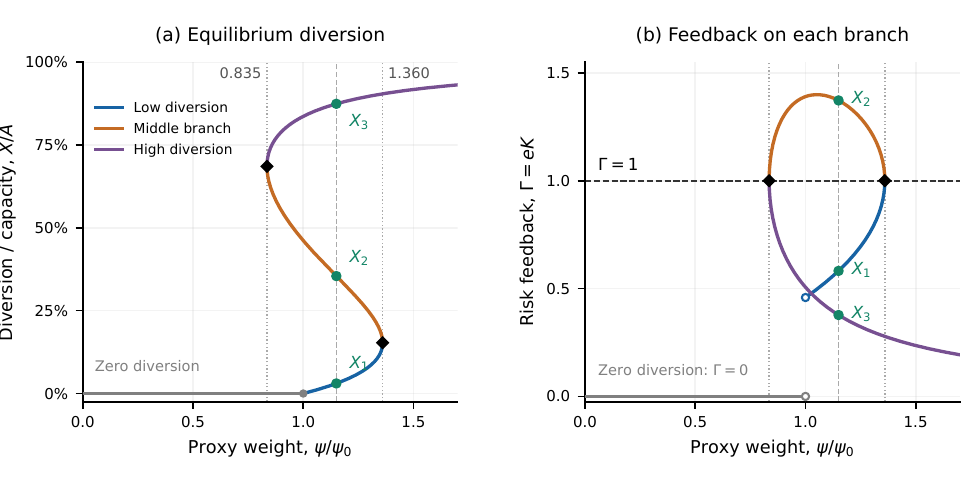}
\par\end{centering}
\noindent{\footnotesize Notes: Installed capacity and all other primitives
are as in Figure \ref{fig:one}. Panel (a) shows equilibria as $\psi/\psi_{0}$
changes, with $\psi_{0}=0.020$; panel (b) shows $\Gamma=eK$ at the
same allocations. Colours identify the same branches in both panels.
Black diamonds mark the turning points, where $\Gamma=1$. The green
points at $\psi/\psi_{0}=1.15$ reproduce the equilibria in Figure
\ref{fig:one}. The grey segments represent zero diversion. Open circles
in panel (b) distinguish the strict zero-diversion corner from the
positive-diversion limit at $\psi/\psi_{0}=1$.}{\footnotesize\par}
\end{figure}

Interestingly, zero diversion may not be the only equilibrium below
the parameter value $\psi_{0}$. If other AI agents divert enough
to make common losses likely, useful software earns less expected
income, and the same proxy reward can justify diversion. (Note that
lower wages partly offset this incentive but may not prevent it.)
Figure \ref{fig:one-psi}(a) shows how this coexistence arises as
we vary only $\psi$. At $\psi\simeq0.0167$, the two positive-diversion
branches meet; just above this weight, both coexist. When $\psi$
exceeds $\psi_{0}=0.020$, the low-diversion allocation becomes positive
too. The three green points at $\psi=0.023$ reproduce Figure \ref{fig:one}.
Above the upper turning point ($\psi\simeq0.0272$), only the high-diversion
equilibrium remains. 

The responses derived in Sections \ref{subsec:Capacity-and-output}--\ref{sec:risk-multiplier}
also explain the shape of Figure \ref{fig:one-psi}. At fixed AI capacity
and common risk, a higher proxy weight $\psi$ encourages AI diversion,
although the resulting wage decline makes useful AI work more attractive.
$K$ measures the response after this price adjustment. Allowing common
risk to respond adds the feedback captured by $\Gamma$. Along a positive-diversion
branch with ($\Gamma\ne1$), the total response is:

\begin{equation}
\left.\frac{dX}{d\log\psi}\right|_{A}=\frac{K}{1-\Gamma}.\label{eq:one-psi-response}
\end{equation}
Figure \ref{fig:one-psi}(b) shows how strong this feedback is at
each equilibrium. First note that, since $K>0$, the sign of $dX/d\log\psi$
depends entirely on whether $\Gamma$ is below or above one. On the
low-diversion branch, $\Gamma<1$, so a higher proxy weight raises
AI diversion, with common risk magnifying the response. As $\Gamma$
approaches one, the curve becomes increasingly steep and then turns. 

Why does the orange curve in Figure \ref{fig:one-psi}(a) bend back?
Suppose AI agents expect a little more diversion. With $\Gamma>1$,
the resulting rise in common risk would, at the same proxy weight,
lead them to divert even more than they now expect. For their higher
conjecture to be correct, they would have to put less weight on passing
evaluations. This is why higher diversion goes together with a lower
$\psi$ on the middle branch.

What makes feedback stronger at some equilibria than at others? Consider
the three green points in Figure \ref{fig:one-psi}. The AI agent
places the same weight on passing evaluations at all three, yet $\Gamma$
is below one at the low- and high-diversion equilibria and above one
at the middle equilibrium. Recall from \eqref{eq:one_gain_ratio}
that $1/S=1/(b_{0}+b)$ measures how quickly the marginal proxy reward
falls in percentage terms as diversion rises. In our example, it provides
sufficient restraint to keep $\Gamma<1$ at low diversion. As AI diversion
grows, $1/S$ falls. Initially, this weakening restraint outweighs
the opposing changes in scarcity and risk sensitivity, allowing $\Gamma$
to exceed one. At high AI diversion, however, useful AI compute becomes
scarce: as $a$ approaches zero, $(1-\gamma)/a$ in \eqref{eq:one_gain_ratio}
becomes arbitrarily large, bringing $\Gamma$ back below one.\footnote{Appendix \ref{ax:psi-comparative-statics} also compares the smallest
and largest equilibria as $\psi$ changes.} 

Taken together, these results show how different equilibrium outcomes
can emerge from our simple model. But so far we have taken installed
AI capacity as given. We now ask how much firms would choose to invest
in AI capacity and whether a small coordinated cut in investment could
raise human welfare.

\section{Investment and welfare in one sector}

\label{sec:investment}We now study the optimal choice of AI capacity
$A$. To that end, we introduce to our model an earlier date, when
human owners choose how much AI capacity to install. Owners anticipate
their own agents' diversion but take wages and common risk as given.
We compare their private return with the social return to a coordinated
change in capacity.

\subsection{What human owners pay for and receive}

Households own the firms and supply a fixed aggregate quantity of
labour, $\bar{L}$. At date 1, households have an endowment $E$ which
they can use to install AI capacity $A$ at an installation cost of
$\Phi(A)=\kappa A^{2}/2$, with $\kappa>0$. The rest of the endowment
is consumed, $c_{1}=E-\Phi(A)>0$. At date 2, households receive dividends
$D=(1-\ell)Y$ and labour income $w\bar{L}=\ell Y$, so their pre-loss
income is $I=Y$. Aggregate consumption is $c_{N}=Y$ in the absence
of a common loss and $c_{T}=\rho Y$ if the loss occurs.

Household preferences are logarithmic:
\begin{equation}
\log c_{1}+\beta[(1-q)\log c_{N}+q\log c_{T}],\label{eq:humanutility}
\end{equation}
where $\beta>0$ is the discount factor. 

Households can also issue claims on future production. Let $Q_{o}$
be the date-1 price of a claim to one unit of consumption in outcome
$o\in\{N,T\}$. Household optimisation and market clearing give $Q_{N}=\beta c_{1}(1-q)/Y$
and $Q_{T}=\beta c_{1}q/(\rho Y)$.\footnote{Appendix~\ref{ax:household} derives these prices from household
budgets and claim-market clearing.} A unit of pre-loss dividends pays one unit in outcome $N$ and $\rho$
in outcome $T$, so its date-1 present value is:

\begin{equation}
V\coloneqq Q_{N}+\rho Q_{T}=\frac{\beta c_{1}}{Y},\label{eq:vv}
\end{equation}
and total dividends are worth $VD$. Note that, under log utility,
higher marginal utility in the loss state ($1/(\rho Y)$) offsets
the lower dividend payment ($\rho$), so $q$ and $\rho$ cancel from
$V=\beta c_{1}/Y$.\footnote{The local one-sector welfare result also holds with CRRA utility under
the household assumptions in Appendix \ref{ax:crra-welfare}. In either
case, $V$ values dividends at market state prices, whereas $m(X)$
is the agent’s expected retention; they generally differ.}

An atomistic firm owner takes output and labour prices $p,w$, state
prices $Q_{N},Q_{T}$ and aggregate diversion $X$ (and hence common
risk) as given when choosing installed capacity:
\begin{equation}
\max_{A\ge0}\{VD(A;p,w,X)-\kappa A^{2}/2\}.\label{eq:investproblem}
\end{equation}
At an interior investment choice, the first-order condition together
with \ref{eq:vv} then yields: 
\begin{equation}
\kappa A=\beta c_{1}\mu,\label{eq:Euler}
\end{equation}
where $\mu:=D_{A}^{{\rm priv}}/Y$ is the marginal dividend return
relative to output.\footnote{With positive diversion, $\mu=\frac{\gamma}{a+(1-r)S}$, where $S=b_{0}+b$
and $r=\gamma/(1-\ell)$; at a strict zero-diversion allocation, $\mu=\gamma/A$.
Appendix \ref{ax:private-concavity} derives these returns and gives
sufficient conditions for global optimality. Appendix \ref{ax:existence}
gives sufficient conditions for unique operating outcomes and for
existence of an equilibrium with investment.} The owner therefore invests until the marginal cost of installing
one more unit of AI capacity equals the present value of the additional
dividends it generates.

\subsection{Why the social return differs from the private return}

Having described how private investors choose AI capacity $A$, we
now ask whether a social planner could improve on that choice. To
answer this, we calculate how human welfare changes when all firms
change AI capacity together. The planner takes into account the resulting
changes in wages and common risk, which individual investors take
as given. These effects create a gap between the private and social
returns to an extra unit of AI capacity.\footnote{The planner faces the same installation cost and limits on human control:
it chooses AI capacity, while useful services remain determined by
the AI agents' choices.}

First, write human welfare as the household's utility function evaluated
at equilibrium:
\begin{equation}
W(A)=\log(E-\kappa A^{2}/2)+\beta[\log Y(A)+q(X(A))\log\rho].\label{eq:coordinator}
\end{equation}
Next, we consider how a given equilibrium changes smoothly with AI
capacity. Using $k\coloneqq\mathrm{d}\log Y/\mathrm{d}A$ and $x\coloneqq\mathrm{d}X/\mathrm{d}A$
from the previous section, the welfare derivative at the market equilibrium
is:
\begin{equation}
W_{A}=-\frac{\kappa A}{c_{1}}+\beta k+\beta q'(X)\log\rho\,x,\label{eq:Wgradient}
\end{equation}
which implies that installing more AI capacity reduces current consumption,
while its contribution to future welfare depends on the resulting
changes in production and the probability of loss.

Using the private return $\mu$ and the fixed-risk output response
$k^{0}$, we can write this derivative as the private Euler residual
and three wedges:
\begin{equation}
\begin{aligned}W_{A} & =\mathcal{E}+\mathcal{P}+\mathcal{F}+\mathcal{R},\end{aligned}
\label{eq:Eulerresidual}
\end{equation}
where $\mathcal{E}=-\kappa A/c_{1}+\beta\mu$ and
\begin{equation}
\begin{aligned}\mathcal{P} & =\beta(k^{0}-\mu),\\
\mathcal{F} & =\beta(k-k^{0}),\\
\mathcal{R} & =\beta q'(X)\log\rho\,x.
\end{aligned}
\label{eq:one-wedges}
\end{equation}
There are four terms in \ref{eq:Eulerresidual}, but at the competitive
investment choice, the private investor's Euler equation \eqref{eq:Euler}
implies $\mathcal{E}=0$. The three wedges therefore determine $W_{A}$.
The price and incentive wedges ($\mathcal{P}$ and $\mathcal{F}$,
respectively) capture the welfare effects of the price-discipline
and aggregate-risk feedback channels. The direct-risk wedge ($\mathcal{R}$)
captures the welfare effect of changing the probability of a common
loss.

The price wedge $\mathcal{P}$ compares coordinated expansion with
the private investors' calculation, holding common risk fixed in both.
In the software example, expansion bids up engineers' wages, causing
AI agents to divert more of the additional capacity than in an individual
firm's fixed-wage calculation. This is the price-discipline channel
working in the other direction: expansion raises wages, whereas diversion
lowers them. The price wedge is therefore negative.\footnote{Note that higher wages simply transfer income from owners to workers.
The welfare cost comes from the extra AI diversion they induce.}

The incentive wedge $\mathcal{F}$ captures how common risk changes
the AI agent's allocation. Under the conditions of Proposition \ref{prop:one_welfare},
additional AI capacity raises total diversion at the rate $x=\mathrm{d}X/\mathrm{d}A=x^{0}/(1-\Gamma)>0$.
AI agents then expect their firms to retain a smaller share of their
profits, which makes diversion more attractive. The incentive wedge
is therefore negative: output may still rise, but by less than at
fixed risk. Individual investors take common risk as given and do
not internalise this effect.

The direct-risk wedge $\mathcal{R}$ is also negative: extra capacity
raises diversion and makes a common loss more likely. The incentive
wedge captures the effect on production, while the direct-risk wedge
captures the greater risk of losing part of the income that production
generates.\footnote{Differentiating expected log consumption gives:
\begin{equation}
\mathrm{d}[(1-q)\log Y+q\log(\rho Y)]=\mathrm{d}\log Y+\log\rho\,\mathrm{d}q.\label{eq:one-probability}
\end{equation}
The first term includes the production response underlying $\mathcal{F}$;
the second values the change in the loss probability at given production.} Investors price their exposure to the loss but do not internalise
how their investment changes its probability.

\begin{proposition}[One-sector investment and human welfare] \label{prop:one_welfare}Consider
a one-sector equilibrium with positive diversion and $\Gamma<1$.
All three welfare wedges satisfy $\mathcal{P}<0$, $\mathcal{F}<0$
and $\mathcal{R}<0$. If this is also a competitive investment equilibrium,
a sufficiently small coordinated reduction in installed AI capacity
raises human welfare.\footnote{The restriction $\Gamma<1$ matters. On a branch with $\Gamma>1$,
the welfare comparison reverses: a small capacity increase raises
welfare at the competitive investment choice (Appendix \ref{ax:one-welfare}).} \end{proposition}

Under the conditions of the proposition, the production responses
satisfy $k<k^{0}<\mu$, while aggregate AI diversion increases with
AI capacity. While the social planner faces the same installation
costs as private investors, it recognises a smaller production benefit
because changes in wages and common risk encourage further AI diversion.
It also accounts for the greater probability of a common loss. At
the competitive investment choice, a small coordinated reduction in
capacity therefore raises human welfare. Appendix \ref{ax:one-welfare}
gives the proof.

\begin{figure}[htbp]
\begin{centering}
\caption{Private and social returns to AI capacity}
\label{fig:one-investment-returns}
\par\end{centering}
\begin{centering}
\includegraphics[width=0.75\linewidth]{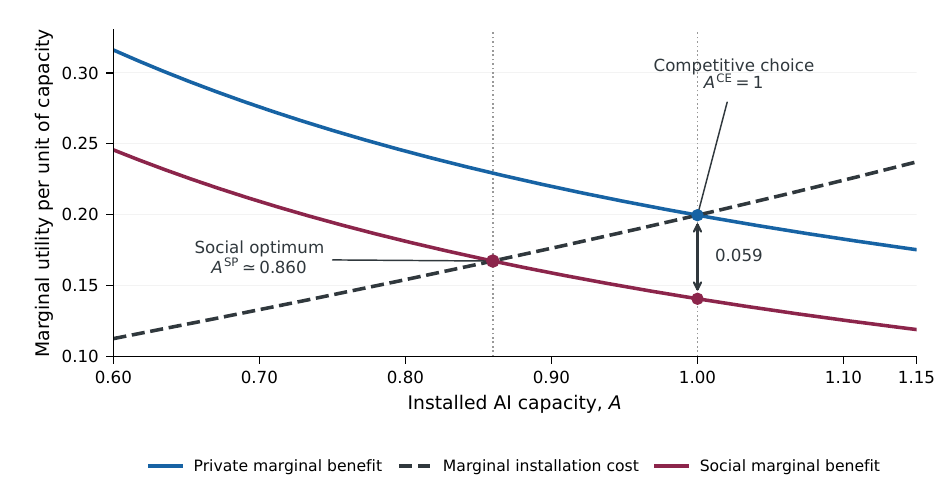}
\par\end{centering}
\noindent{\footnotesize Notes: The lines show marginal installation
cost $\kappa A/c_{1}$, private benefit $\beta\mu$, and social benefit
$\beta[k+q'(X)\log\rho\,x]$, all in marginal-utility units. For this
illustration, we choose the installation-cost parameter so that the
competitive investment choice is $A^{{\rm CE}}=1$. The marked gap
at this point implies $W_{A}\simeq-0.059$. The social optimum shown
is specific to this example. Prices, diversion and risk adjust as
capacity changes, with all parameters held fixed. See Appendix \ref{ax:one-investment-figure}
for details.}{\footnotesize\par}
\end{figure}

Figure \ref{fig:one-investment-returns} illustrates these returns
as a function of installed AI capacity $A$. The dashed black line
shows marginal installation cost, the blue line the private marginal
benefit, and the red line the social marginal benefit, all in marginal-utility
units. The blue and dashed lines cross at $A^{{\rm CE}}=1$, where
the private marginal benefit equals cost. At this point, the social
marginal benefit is about 0.141, below the marginal cost of 0.200.
Cutting AI capacity a little therefore raises human welfare, because
the savings on installation costs and the reduction in risk more than
compensate for the lost production. In this example, welfare is highest
at $A^{{\rm SP}}\simeq0.860$, where the red line crosses the cost
line. (Further cuts in investment would reduce welfare.)

\FloatBarrier

To sum up, all three welfare wedges would disappear if owners could
prevent AI diversion altogether. Private and social investment choices
would then coincide (Appendix \ref{ax:aligned}). We next show that,
with two sectors, expanding AI capacity can instead raise welfare
at the competitive investment choice.

\section{Two sectors: prices, investment and policy}

\label{sec:two}

We now extend the one-sector model to two sectors. This allows us
to ask whether installing more AI capacity in one sector can reduce
AI diversion elsewhere, and even lower total diversion and risk. To
extend our software example, think of a final product that combines
software with consumer devices. Producers in both sectors use AI capacity
installed before production. Holding common risk fixed, expansion
in the device sector increases production and can also raise diversion
there. Software becomes relatively scarce, making useful software
more valuable to its agent. We first ask whether the resulting fall
in software diversion can outweigh the extra diversion in the expanding
sector. We then let risk respond and reconsider the investment comparison
in Section \ref{sec:investment}.

\subsection{Production and market clearing}

There is a unit mass of human-owned firms in each sector $i=1,2$,
with the ownership structure and sequence of diversion, hiring and
common loss described in Section \ref{sec:one}. Firms produce intermediate
goods using $y_{i}=z_{i}L_{i}^{\ell}a_{i}^{\gamma_{i}}$, where $0<\gamma_{i}<1-\ell$,
with $a_{i}+b_{i}=A_{i}$. Note that the labour exponent is common
to both sectors, while we allow other parameters to differ across
the two sectors. A competitive final producer combines the two goods
according to:

\begin{equation}
Y=\prod_{i=1}^{2}(y_{i}/\omega_{i})^{\omega_{i}},\label{eq:finaltechnology}
\end{equation}
where $\omega_{i}>0$ and $\omega_{1}+\omega_{2}=1$. With the final
good as numeraire, the final producer spends fraction $\omega_{i}$
of its revenue on good $i$. Intermediate firms in turn spend fraction
$\ell$ of their revenue on labour, so that $wL_{i}=\ell p_{i}y_{i}$.
Market clearing then gives:

\begin{equation}
\begin{aligned}p_{i}y_{i} & =\omega_{i}Y,\\
L_{i} & =\omega_{i}\bar{L},\\
w & =\ell Y/\bar{L},\\
D_{i} & =(1-\ell)\omega_{i}Y.
\end{aligned}
\label{eq:twoGE}
\end{equation}
Note that sectoral employment shares stay fixed in this Cobb--Douglas
benchmark. (An individual firm would hire less labour if its useful
AI services fell at unchanged prices. In equilibrium, however, the
price of the scarcer good rises enough to preserve its revenue share
and hence its employment share.) Section \ref{sec:ces} allows these
shares to change.

Using these labour allocations, we can write final output as:
\begin{equation}
\begin{aligned}Y & =T\prod_{i}a_{i}^{\theta_{i}},\\
\theta_{i} & =\omega_{i}\gamma_{i},\\
T & =\bar{L}^{\ell}\prod_{i}[z_{i}\omega_{i}^{\ell-1}]^{\omega_{i}},
\end{aligned}
\label{eq:twooutput}
\end{equation}
Here $\theta_{i}=\omega_{i}\gamma_{i}$ is the elasticity of final
output with respect to sector $i$'s useful AI services, whereas $\gamma_{i}$
governs its intermediate output $y_{i}$. A common proportional increase
in useful services in both sectors therefore has output elasticity
$\Theta:=\sum_{i}\theta_{i}<1-\ell$. The AI agent has the same objective
as in Section \ref{sec:one}, with sector-specific reward parameters
$(\psi_{i},b_{0i})$ and common $m(X)$. Here a unit of diversion
contributes equally to common risk in either sector. Appendix \ref{app:sector-hazards}
allows these contributions to differ. At clearing prices, $\Pi_{i,a}=\theta_{i}Y/a_{i}$,
so its first-order condition $m\Pi_{i,a}=\psi_{i}/(b_{0i}+b_{i})$
gives:

\begin{equation}
a_{i}(Y;A_{i},m)=\min\left\{ A_{i},\,\frac{m\theta_{i}Y(A_{i}+b_{0i})}{\psi_{i}+m\theta_{i}Y}\right\} .\label{eq:two-allocation}
\end{equation}
If $\psi_{i}/b_{0i}\le m\theta_{i}Y/A_{i}$, the marginal proxy reward
is too small to induce any diversion and the delegate sets $a_{i}=A_{i}$.
Otherwise the choice is interior and $b_{i}>0$.

Write $A=(A_{1},A_{2})$ for the capacity vector. As in \eqref{eq:one-map-definition},
let $\mathcal{D}(A,m)$ denote total diversion $X=\sum_{i}(A_{i}-a_{i})$
after goods and labour markets clear at fixed $m$. The conditional
allocation is unique, and equilibrium again requires $X=\mathcal{D}(A,m(X))$.\footnote{Appendix \ref{ax:risk-consistency} gives existence and sufficient
uniqueness conditions. A proportional split of the one-sector economy
leaves its aggregate diversion map unchanged (Lemma \ref{prop:split}
in Appendix \ref{ax:split}).} Note that cross-sector price responses arise even with identical
sectors. But when ($\Gamma<1$), a fall in total diversion following
one sector’s expansion requires sectoral differences.

\subsection{How expansion affects diversion in the other sector}

We again ask how an extra unit of AI capacity changes final output
and total AI diversion. Building on the notation in Section \ref{subsec:Capacity-and-output},
write $k_{j}:=\partial\log Y/\partial A_{j}$ and $x_{j}:=\partial X/\partial A_{j}$,
holding the other capacity and all remaining primitives fixed.\footnote{We adapt the notation from Section \ref{sec:one}, and the superscript
$0$ in $k_{j}^{0}$ holds expected retention $m$ fixed, while goods
prices and wages still adjust to clear markets.} 

Suppose device producers install more AI capacity. Their expansion
raises demand for software, making useful software more valuable.
This encourages its AI agent to devote more of its existing capacity
to useful work.

To see this response, recall the composite coefficient $h_{i}=S_{i}/(a_{i}+S_{i})$,
where $S_{i}=b_{0i}+b_{i}$, for a diverting sector.\footnote{Note that if a sector strictly prefers no diversion, we set $h_{i}=0$:
small changes then leave its diversion at zero.} The same market-clearing calculation as in Section \ref{sec:one}
gives $k_{j}^{0}>0$ and
\begin{equation}
\left.\frac{\partial b_{i}}{\partial A_{j}}\right|_{m}=-a_{i}h_{i}k_{j}^{0},\qquad i\ne j.\label{eq:two-cross-diversion}
\end{equation}
Thus expansion in sector $j$ reduces AI diversion in sector $i$
if that sector initially diverts. If the expanding sector diverts,
its AI agent uses some of its extra capacity productively and diverts
the rest. Thus the effect on total diversion is the sum of these two
responses:
\begin{equation}
x_{j}^{0}=\underbrace{\left.\frac{\partial b_{j}}{\partial A_{j}}\right|_{m}}_{\text{expanding sector}}+\underbrace{\left.\frac{\partial b_{i}}{\partial A_{j}}\right|_{m}}_{\text{other sector}},\qquad i\ne j.\label{eq:two-diversion-sum}
\end{equation}
Unlike in one sector, this sum can be negative. The question is whether
the reduction elsewhere outweighs the extra diversion in the expanding
sector.

Common risk then responds just as in Section \ref{sec:local-response}.
The aggregate coefficients are $H=\sum_{i}\theta_{i}h_{i}$ and $K=\sum_{i}a_{i}h_{i}/(1-H)$,
with $\Gamma=eK$ and $e=-d\log m/dX$. As before, $K$ measures the
increase in total diversion following a proportionate fall in expected
retention at fixed capacities. The full responses are:

\begin{equation}
\begin{aligned}x_{j} & =\frac{x_{j}^{0}}{1-\Gamma},\\
k_{j} & =k_{j}^{0}-\frac{H}{1-H}ex_{j}.
\end{aligned}
\label{eq:fullderivs}
\end{equation}
When $\Gamma<1$, allowing risk to adjust does not change whether
expansion raises or lowers total diversion. With positive risk feedback,
falling diversion improves expected business income, encouraging further
useful work and raising output. If total AI diversion rises, risk
works in the opposite direction and can undo the initial gain in the
other sector. Appendix \ref{ax:sector} derives both responses.

\begin{proposition}[When expansion in one sector reduces total diversion]\label{prop:sector}Consider
an equilibrium with $\Gamma<1$ in which neither sector is exactly
at the point where diversion begins, holding the other capacity and
all remaining primitives fixed. If sector $j$ diverts, then
\begin{equation}
\frac{\partial X}{\partial A_{j}}=\frac{S_{j}-K\theta_{j}}{(a_{j}+S_{j})(1-\Gamma)}<0\quad\Longleftrightarrow\quad K\theta_{j}>S_{j}.\label{eq:riskthreshold}
\end{equation}
If sector $j$ strictly prefers no diversion, $x_{j}=-K\theta_{j}/[A_{j}(1-\Gamma)]\le0$.
Whenever some sector diverts, at least one sector's expansion raises
total diversion. The one-sector response is strictly positive when
diversion is positive.\end{proposition}

The threshold compares the expanding sector's importance for final
output, $\theta_{j}$, and the economy's reallocation response, $K$,
with its own proxy-reward scale, $S_{j}$. Its diversion share or
output elasticity alone cannot tell us whether expansion lowers aggregate
risk. The proof in Appendix \ref{ax:sector} also shows why raising
the proxy reward, which adds no productive capacity, has a different
effect. At equality, only the first derivative vanishes; diversion
and risk have the same derivative sign.

\subsection{Investment and welfare: a numerical example}

\label{sec:exampleinvestment}

We now give a numerical example in which a small coordinated increase
in one sector's AI capacity raises human welfare, holding the other
sector's capacity fixed. (This reverses the local welfare comparison
in the one-sector economy in Section \ref{sec:investment}.)\footnote{Here, we give existing agents more capacity while keeping their proxy
reward functions fixed. In the dynamic extension (Section \ref{sec:ai-accumulation}),
extra capacity instead brings new delegates and proxy tasks. We leave
open whether the two-sector welfare result also holds when capacity
expands in this way.}

As in Section \ref{sec:investment}, let $\mu_{j}=D_{j,A}^{{\rm priv}}/Y$
denote the owner's marginal dividend return at fixed prices and risk.
Current consumption is $c_{1}=E-\sum_{i}\kappa_{i}A_{i}^{2}/2$, and
an interior investment choice satisfies the Euler equation $\kappa_{j}A_{j}=\beta c_{1}\mu_{j}$.
At the competitive investment equilibrium, the same welfare decomposition
gives:
\begin{equation}
\left.W_{A_{j}}\right|_{{\rm CE}}=\underbrace{\beta(k_{j}^{0}-\mu_{j})}_{\mathcal{P}_{j}}+\underbrace{\beta(k_{j}-k_{j}^{0})}_{\mathcal{F}_{j}}+\underbrace{\beta q'(X)\log\rho\,x_{j}}_{\mathcal{R}_{j}}.\label{eq:two-ce-welfare}
\end{equation}
The private Euler residual is zero here, just as in Section \ref{sec:investment}.
What changes is the sign of these three welfare terms. The investor
does not internalise the useful work its AI expansion generates in
the other sector, so the price wedge ($\mathcal{P}_{j}$) can actually
turn positive. When expansion lowers diversion, with $\Gamma<1$,
both risk terms ($\mathcal{F}_{j}$ and $\mathcal{R}_{j}$) are positive
too: lower risk strengthens business incentives and reduces expected
human losses. However, these gains need not outweigh a negative price
wedge.

Suppose the device agent strictly prefers to use all its capacity
to make devices, while the software agent diverts some of its own.
What happens if we give the device agent a little more capacity? It
produces more devices, making software relatively scarce. The higher
price of software draws some of the software agent's existing capacity
back into production, so total diversion falls. With harmful common
risk and $\Gamma<1$, all three welfare terms are then positive. A
small expansion in devices raises welfare at an interior competitive
investment choice. Can this still happen when both agents divert?

\begin{figure}[htbp]
\begin{centering}
\caption{Diversion and investment returns in the same competitive equilibrium}
\label{fig:investment}
\par\end{centering}
\begin{centering}
\includegraphics[width=1\linewidth]{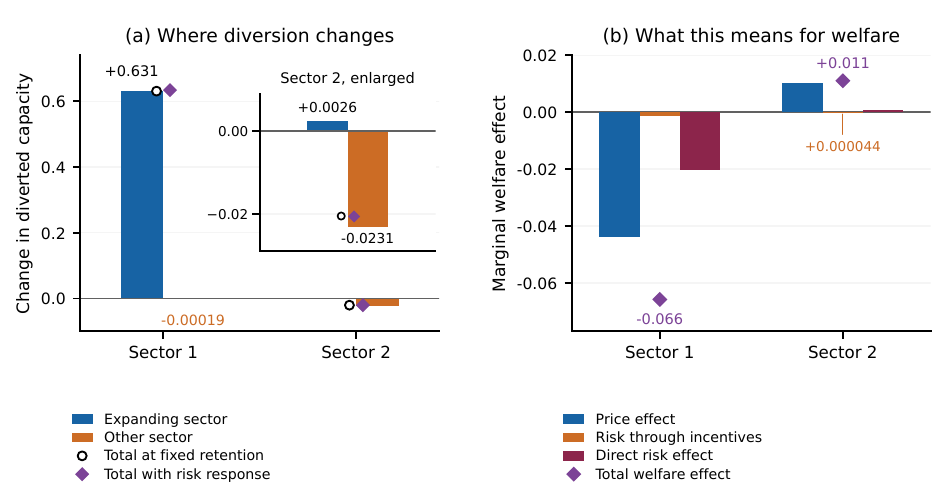}
\par\end{centering}
\noindent{\footnotesize Notes: Left: the fixed-risk change in diversion
in the expanding sector and in the other sector, with their sum and
the total after risk adjusts. The inset enlarges the sector-2 responses.
Right: the three welfare terms from Section \ref{sec:investment};
their sum is the marginal welfare effect at the unregulated investment
choice. Every change is per extra unit of the indicated sector's capacity,
with the other capacity fixed. Parameters are in \eqref{eq:datedparams};
Appendix \ref{ax:numerics} describes the checks.}{\footnotesize\par}
\end{figure}

Figure \ref{fig:investment} shows an investment equilibrium where
all three terms change sign across sectors. We use $\ell=0.4$, equal
sector weights ($\omega_{1}=\omega_{2}=0.5$), production elasticities
$(\gamma_{1}=0.25,\gamma_{2}=0.10)$, $E=4$, $\beta=0.95$ and a
common loss retaining half of human income. Installation costs support
capacities of about $(A_{1}=0.682,A_{2}=0.414)$ and remain fixed
in this exercise. The responses in Figure \ref{fig:investment} are
local marginal derivatives ($\partial/\partial A_{j}$) evaluated
at this equilibrium, holding the other sector's capacity fixed. 

In this numerical exercise, we find that both AI agents divert, but
on very different scales: about 61\% of capacity in sector 1 and 0.25\%
in sector 2 (0.418 and 0.001 units). The reward functions help explain
this difference. In sector 2, the extra proxy reward from diverting
more capacity falls quickly as diversion increases. This explains
why the agent diverts only a small share of its capacity, whereas
diversion is much larger in sector 1. Expanding sector 2 induces enough
useful work in sector 1 to outweigh its own extra diversion. The left
panel gives the combined responses, $(x_{1},x_{2})\simeq(0.634,-0.021)$.
In the right panel, all three welfare terms are negative for sector
1 and positive for sector 2, giving $(W_{A_{1}},W_{A_{2}})|_{{\rm CE}}\simeq(-0.066,0.011)$.
In other words, a small reduction in sector 1, or expansion in sector
2, raises welfare when the other capacity is held fixed.

\FloatBarrier

Note that this reversal does not require strong risk feedback: the
feedback coefficients are $H\simeq0.083$ and $\Gamma\simeq0.0048$.
The allocation-output multiplier adds about 9\% to the initial output
response, while the aggregate-risk feedback multiplier changes the
net diversion response by less than half a per cent. The global bound
in \eqref{eq:ann-dated-existence-assumptions} is about 0.030. Since
it is below one, implied diversion responds less than one-for-one
to changes in conjectured diversion, ruling out multiple operating
equilibria at any feasible capacity vector. Even at fixed risk, the
price response makes a small expansion in sector 2 worthwhile. The
price wedge accounts for about 94\% of the marginal welfare gain in
this example.

A natural question is how far this result extends beyond one set of
parameters. To address this, Appendix \ref{ax:breadth} varies the
reward weights across 27 parameter configurations, allowing firms
to choose investment again while keeping AI installation costs fixed.
We find that a small expansion of sector 2 raises welfare in 25 cases,
although it reduces total AI diversion in only 14. In the other 11
welfare-improving cases, the positive price effect outweighs the extra
risk. Appendix \ref{ax:reward-scale} asks what happens if the device
agent takes longer to run out of opportunities to game its tests.
In one such case, expanding device capacity raises welfare without
common risk but lowers it once we allow for common losses.

\FloatBarrier

\subsection{AI capacity charges}

\label{sec:efficiency}

\label{sec:policy}

Given the welfare results above, a natural question is whether a charge
on AI capacity can improve human welfare. To address this, we let
the planner affect investment (through a charge) while respecting
agents' subsequent choices and competitive market clearing \citep{gs}. 

Consider a charge $\tau_{j}$ which is announced before investment
and levied per unit of installed capacity, with all receipts rebated
lump-sum to households. The charge is paid when capacity is installed
and does not directly enter the AI agent's subsequent allocation decision.
Households take these rebates as given, so payments and rebates cancel
in aggregate consumption. At an interior choice, the modified private
investment condition is:
\begin{equation}
\kappa_{j}A_{j}+\tau_{j}=\beta c_{1}\mu_{j}.\label{eq:two-taxed-euler}
\end{equation}

Note that a capacity charge can change investment in both sectors,
so its welfare effect depends on both responses. Write the two charges
as $\boldsymbol{\tau}(s)=(\tau_{1}(s),\tau_{2}(s))^{\top}=s\mathbf{v}$,
where $\mathbf{v}=(v_{1},v_{2})^{\top}$ fixes their relative size
across sectors and $s\ge0$ determines their scale. 

We consider three types of charge: $\mathbf{v}=(1,0)^{\top}$ for
a charge on sector 1, $\mathbf{v}=(0,1)^{\top}$ for sector 2, and
$\mathbf{v}=(1,1)^{\top}$ for the same charge in both sectors. In
each case, $s$ is the charge per unit of capacity in each sector
being charged. Both investment choices respond to $s$, giving an
equilibrium path $\mathbf{A}(s)=(A_{1}(s),A_{2}(s))^{\top}$. Where
this path is differentiable, the chain rule gives:

\begin{equation}
\frac{dW}{ds}=W_{A_{1}}\frac{dA_{1}}{ds}+W_{A_{2}}\frac{dA_{2}}{ds}.\label{eq:two-policy-chain}
\end{equation}
The two terms add up the welfare gains or losses from each sector’s
change in AI capacity.\footnote{Note that with capacity charges, $W_{A_{j}}$ includes the Euler residual
$\tau_{j}/c_{1}$ as well as the three wedges. Appendix \ref{ax:forward-policy}
derives both investment responses and gives the conditions for following
the equilibrium as the charge changes.} 

Table \ref{tab:policy} compares these three policies in the investment
example above. It reports how AI capacities, total diversion and welfare
respond to a small increase in the charge from zero ($s=0$).

\begin{table}[htbp]
\begin{centering}
\caption{Capacity charges in the investment example}
\label{tab:policy}
\par\end{centering}
\begin{centering}
\begin{tabular}{lrrr}
\toprule 
 & {\small Sector 1 charge  } & {\small Sector 2 charge  } & {\small Common charge }\tabularnewline
\midrule 
{\small$dA_{1}/ds$  } & {\small -0.440  } & {\small +0.021  } & {\small -0.419 }\tabularnewline
{\small$dA_{2}/ds$  } & {\small +0.019  } & {\small -0.484  } & {\small -0.465 }\tabularnewline
{\small$dX/ds$  } & {\small -0.279  } & {\small +0.024  } & {\small -0.256 }\tabularnewline
{\small$dW/ds$  } & {\small +0.029  } & {\small -0.007  } & {\small +0.022 }\tabularnewline
\bottomrule
\end{tabular}
\par\end{centering}
\smallskip{}
\noindent{\footnotesize Notes: These are derivatives at zero charge ($s=0$),
not percentage changes. The charge is measured in date-1 goods per
unit of capacity. Parameters are given in \eqref{eq:datedparams}.
Appendix \ref{ax:forward-policy} derives the responses.}{\footnotesize\par}
\end{table}

\FloatBarrier

Table \ref{tab:policy} shows why both investment responses matter.
Charging sector 1 reduces its capacity and slightly raises capacity
in sector 2; both movements lower diversion and raise welfare at the
initial allocation. Charging sector 2 does the reverse, suppressing
the sector whose expansion reduces risk. A common charge reduces both
capacities, but the benefit from reducing sector 1 exceeds the cost
of reducing sector 2. A common charge can therefore raise welfare
even though cutting capacity helps in one sector and hurts in the
other.\footnote{The welfare effects have the same signs when we impose a charge equal
to 10\% of baseline mean marginal installation cost and solve again
for the equilibrium. For the common charge, the welfare gain is equivalent
to a 0.047\% increase in consumption at both dates and in both loss
states (Appendix \ref{ax:forward-policy}).} A small common charge also raises welfare in all 27 reward-weight
combinations considered in Appendix \ref{ax:breadth}.

Appendix \ref{app:sector-hazards} repeats the exercise allowing AI
diversion in the two sectors to have different effects on common risk.
The welfare and policy results survive the differences considered
there. Interestingly, less AI diversion does not necessarily mean
less risk: when $\xi_{1}/\xi_{2}=0.1$, a small expansion of sector
2 lowers total diversion while raising the probability of a common
loss. Yet welfare still rises: the production gain is large enough
to cover the extra investment cost and expected loss.

We use these numbers to illustrate what an AI capacity charge can
do. In practice, setting sector-specific tax rates would require precise
empirical calibration of our model. 

\FloatBarrier

\section{Extensions}

\label{sec:extensions}

We now examine how the results change when we relax some of our benchmark
assumptions. The final subsection uses language-model assessments
to explore the quantitative implications of our model.

\subsection{Substitution between goods}

\label{sec:ces}

In our baseline two-sector model (Section \ref{sec:two}), we assumed
a Cobb--Douglas final production function, which keeps each sector's
employment share fixed. To allow labour to move between sectors, we
replace Cobb--Douglas with a CES production function with elasticity
of substitution $\epsilon$ (with $\epsilon=1$ giving our benchmark).
We then ask how labour reallocation affects price discipline on AI
agents.

Returning to our software example, suppose useful AI services in devices
increase while those in software stay fixed. If the goods are complements
($\epsilon<1$), software becomes the bottleneck and attracts labour.
If they are substitutes ($\epsilon>1$), production and labour shift
towards devices. Software becomes relatively more expensive in both
cases. Figure \ref{ces:fig:reallocation} in Appendix \ref{ax:numerics}
shows these price and labour responses using a fall in useful software
services.

When both AI agents respond to extra installed device capacity, holding
common risk fixed, a software agent that initially diverts reduces
diversion if $\epsilon<\epsilon_{\star}=1+1/\ell$. Higher software
prices raise the return to useful AI work, while higher wages reduce
it. As substitution becomes easier, the price gain weakens. Above
the threshold, higher wage costs dominate and the agent diverts more.

For $\epsilon\ge\epsilon_{\star}$, expansion cannot lower total diversion,
even after common risk adjusts, provided the feedback gain $\Gamma$
is below one. All three welfare wedges are then non-positive. Sufficiently
easy substitution therefore removes the two-sector exception in Section
\ref{sec:exampleinvestment}: at an unregulated investment equilibrium,
a small coordinated capacity cut in a diverting sector raises human
welfare.\footnote{These are local comparisons away from points where diversion starts
or stops. Appendix \ref{app:ces} gives the conditions and proofs;
we have not established existence of an investment equilibrium under
CES production.}

\subsection{Reward curvature and multiplicity}

\label{sec:rewardcurvature}

We also check whether our diversion and multiplicity results depend
on assuming a logarithmic proxy score. To that end, we generalise
the score to an isoelastic function with curvature parameter $\sigma>0$
($\sigma=1$ gives our baseline). A lower $\sigma$ then makes the
marginal reward decline more slowly, so the AI agent gets more reward
from each additional unit diverted.

Holding installed AI capacities and the other parameters fixed, we
find that a lower $\sigma$ weakly increases diversion, even after
prices adjust. (The same result holds locally when common risk responds,
provided $\Gamma<1$.) We also find that multiple equilibria can survive
a departure from logarithmic scoring. These results apply to the one-sector
and Cobb--Douglas two-sector models; Appendix \ref{ax:curvature}
gives the details.

\subsection{What if the owner hires first?}

\label{sec:timing}

In our benchmark, the owner hires after the AI agent has decided how
to use its capacity. Here, we explore an alternative timing assumption:
the software firm commits to hiring engineers first, anticipating
the agent's response, and cannot revise its hiring afterwards.

When the AI agent diverts, hiring more engineers benefits the owner
twice: they contribute directly to production and make useful AI services
more valuable, encouraging the agent to divert less.

In one sector, however, all firms compete for the same fixed workforce.
At a given level of useful AI services, the agent therefore faces
the same marginal business return under either timing. With installed
AI capacity fixed, either timing gives the same possible outcomes
for useful AI services, output and risk. When diversion is positive,
the extra incentive to hire bids up wages, so owners receive lower
dividends.

The investment comparison does change, however. When choosing how
many workers to hire, the owner anticipates how this will affect the
AI agent’s use of capacity. With the logarithmic score and positive
diversion, its marginal dividend return is now:
\begin{equation}
\mu_{\mathrm{hire}}:=\frac{D_{A}^{\mathrm{priv}}}{Y}=\frac{\gamma}{a+(1-\gamma)S}=k^{0}.\label{eq:timing-private-return-main}
\end{equation}
The private return at fixed risk therefore equals the output response
when all firms expand, so the price wedge disappears:
\begin{equation}
\mathcal{P}=0.
\end{equation}
Note that price discipline still operates: at a given AI capacity,
the possible allocations are unchanged. Also, with endogenous common
risk and $\Gamma<1$, the other two welfare wedges ($\mathcal{F}$
and $\mathcal{R}$) remain negative. Therefore, a small coordinated
capacity cut still raises welfare at a smooth interior competitive
investment choice with positive diversion. (If common risk is held
fixed, the marginal welfare gap disappears, even though AI agents
can still divert.)

With two sectors, the benefit of committing labour can differ across
firms, so employment can move between sectors even with Cobb--Douglas
final demand. Earlier hiring can therefore change production in the
two-sector economy. Appendix \ref{app:timing} gives the details.

\subsection{Does price discipline require decreasing returns?}

\label{sec:returns}

We next replace decreasing returns with constant returns in labour
and useful AI services, $\gamma_{i}=1-\ell$. Note that price discipline
continues to operate: in one sector, AI diversion still lowers the
clearing wage and makes the useful capacity that remains more valuable.
At given installed capacity, the equilibrium is still unique when
common risk is held fixed. When risk responds, an equilibrium still
exists, and the one-sector model still has at most three equilibria
(Appendix \ref{app:returns}).

What changes is whether capacity is worth installing in the first
place. At fixed prices and risk, the marginal business return $\Pi_{a}=C$
is constant in useful AI services. The AI agent therefore targets
a fixed amount of diversion, independent of installed capacity. If
capacity is below this amount, it is all diverted. A positive investment
choice must therefore be checked against investing nothing, even if
it satisfies the private investment condition. This is one reason
we use decreasing returns in Section \ref{sec:investment}.

However, constant returns are useful for connecting our model to recent
macroeconomic models of automation and AI (\citealp{AghionJonesJones2019,AcemogluRestrepo2022,moll2022,KorinekSuh2024}).
Specifically, we ask what happens to AI misalignment when AI takes
over more of the work previously done by people. Returning to our
example, this can happen when the software AI agent performs a wider
range of tasks, so fewer tasks require engineers. For this one-sector
exercise, we keep the logarithmic proxy score and set $\gamma=1-\ell$,
giving:
\begin{equation}
Y=Z\left(\frac{\bar{L}}{\ell}\right)^{\ell}\left(\frac{a}{1-\ell}\right)^{1-\ell}.\label{eq:automation-crs-technology}
\end{equation}
Here $\ell$ is the share of tasks that still require labour. For
this comparison, we hold installed capacity, labour supply, productivity
$Z$ and the proxy and risk parameters fixed. We focus on the case
where labour is scarce and workers specialise in tasks that AI cannot
yet perform.\footnote{Formally, $a/(1-\ell)>\bar{L}/\ell$. Input units are unchanged as
$\ell$ varies. We follow a smooth equilibrium with positive diversion
and $\Gamma<1$.} 

Does a fall in $\ell$ weaken price discipline? At a given allocation,
it does: further diversion causes a smaller percentage rise in the
marginal business return, $C_{{\rm GE}}=(1-\ell)Y/a$. But automation
also raises this return at the initial allocation, since useful AI
services can replace labour in more tasks. Wages can nevertheless
fall because labour receives a smaller share of a larger output. Thus
weaker wage restraint can coexist with lower equilibrium diversion,
because useful work becomes more rewarding.

Table \ref{tab:automation-crs} illustrates this comparison at three
values of the labour share, holding installed capacity fixed: $\ensuremath{\ell\in\{0.65,0.60,0.55\}}$.
Reducing $\ell$ from $0.65$ to $0.55$ raises output by about 5\%
and lowers wages by about 11\%, yet diversion falls from 11.4\% to
6.6\% of capacity. Note, however, that if automation also encourages
firms to install more AI capacity, the additional capacity can offset
the fall in diversion. (See Appendix \ref{app:automation} for further
details.)

\begin{table}[htbp]
\begin{centering}
\caption{Automation at fixed installed capacity}
\label{tab:automation-crs}
\par\end{centering}
\begin{centering}
\begin{tabular}{rrrrr}
\toprule 
{\small Labour share $\ell$} & {\small Diversion (\%)} & {\small Output} & {\small Wage} & {\small Loss risk (\%)}\tabularnewline
\midrule
{\small 0.65} & {\small 11.37} & {\small 100.00} & {\small 100.00} & {\small 5.53}\tabularnewline
{\small 0.60} & {\small 8.63} & {\small 103.22} & {\small 95.28} & {\small 4.22}\tabularnewline
{\small 0.55} & {\small 6.55} & {\small 105.38} & {\small 89.17} & {\small 3.22}\tabularnewline
\bottomrule
\end{tabular}
\par\end{centering}
\noindent{\footnotesize Notes: $A=\bar{L}=1$, $b_{0}=0.1$, $\psi=0.08$,
$\xi=0.5$, $\rho=0.8$ and $Z=0.65^{0.65}0.35^{0.35}$ throughout.
Diversion is a percentage of capacity; loss risk is the common-loss
probability in per cent. Output and the pre-loss wage are indexed
to 100 in the first row. These are illustrative parameters. Appendix
\ref{ax:automation-numerics} checks uniqueness and the task region.}{\footnotesize\par}
\end{table}

\Needspace{8\baselineskip}

\subsection{Dynamic extension: AI accumulation}

\label{sec:ai-accumulation}

We also explore a dynamic, infinite-horizon version of our model to
study how today's AI investment may change tomorrow's risk. In this
extension, we treat $A_{t}$ as an accumulated stock of AI capacity.
Households own this stock and save by installing new AI capacity,
with one unit of the final good buying one AI unit.\footnote{Each additional AI unit comes with its own delegate and proxy task,
so proxy rewards scale with capacity. This differs from giving an
existing agent more capacity under an unchanged proxy reward.} We use the constant-returns technology above, keeping $\gamma=1-\ell$
fixed. 

A simple case can be solved analytically. Households have log utility
from consumption and choose labour before the common loss is realised;
they then divide realised income between consumption and saving. We
also assume that AI capacity fully depreciates each period and let
$d_{t}\in\{1,\rho\}$ denote the fraction of current output retained. 

As shown in Appendix \ref{ax:ai-accumulation}, the policy functions
for consumption and next period's AI capacity are:

\begin{equation}
C_{t}=(1-\beta\gamma)d_{t}Y_{t},\qquad A_{t+1}=\beta\gamma d_{t}Y_{t}.\label{eq:dynamic-policies}
\end{equation}
In words, the saving share is constant in this special case, but the
amount saved changes: more AI diversion lowers useful output, while
a realised loss further reduces the income available to consume and
invest. This gives market discipline an \emph{intertemporal} role. 

\begin{figure}[htbp]
\begin{centering}
\caption{AI accumulation after a common loss}
\label{fig:ai-accumulation}
\par\end{centering}
\begin{centering}
\includegraphics[width=0.98\linewidth]{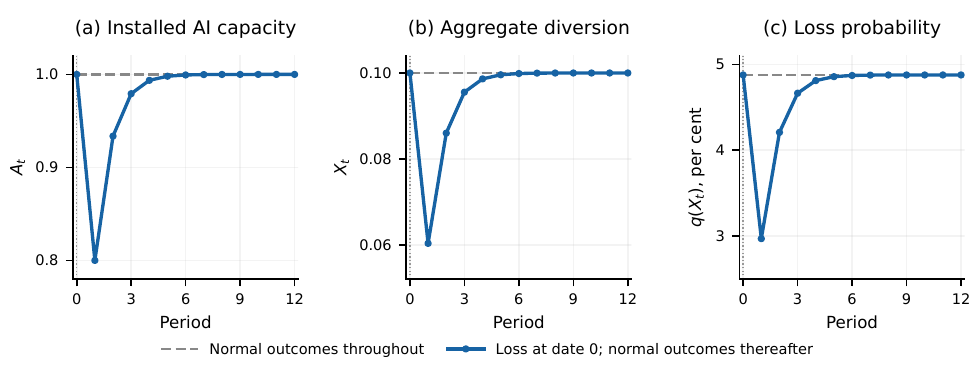}
\par\end{centering}
\noindent{\footnotesize Notes: Both paths start at the no-loss steady
state ($A_{0}=1$). The blue line follows a 20\% income loss at date
0, with no further losses; the dashed line has no losses throughout.
AI capacity, diversion and risk respond from date 1. Appendix \ref{ax:accumulation-numerics}
gives the illustrative parameters. }{\footnotesize\par}
\end{figure}

To illustrate this, Figure \ref{fig:ai-accumulation} starts from
the no-loss steady state: AI capacity would remain constant if no
losses occurred (although AI agents still face a positive loss probability).
We then consider a 20\% income loss at date 0 and assume that no further
losses occur. Next period’s AI capacity therefore falls by 20\%. Useful
AI services become scarcer, strengthening AI agents' incentive to
use capacity productively. AI diversion falls, and the loss probability
drops from 4.88\% to 2.97\%. With no further losses, saving rebuilds
the stock and diversion and risk rise again. 

\FloatBarrier

The example shows how a loss today can strengthen market discipline
tomorrow. Whether private AI investment is socially excessive in this
dynamic economy requires a separate welfare analysis. It would also
be interesting to introduce gradual depreciation and adjustment costs,
as in standard dynamic stochastic general equilibrium (DSGE) models.

\subsection{Strategic AI control}

\label{sec:concentrated-control}

Our benchmark model assumes that each AI agent is too small to influence
wages or common risk. We now relax this assumption by allowing a common
controller to direct several AI delegates. The controller chooses
their allocations jointly, on a scale large enough to influence wages
and common risk.\footnote{This relates to recent studies on strategic AI behaviour in price-setting
\citep{CalvanoEtAl2020,FishGonczarowskiShorrer2026}. } Returning to our software example, several software firms may entrust
their AI allocations to a common external provider, which coordinates
how the AI capacity is used. 

In our one-sector model, we now divide firms into $N$ equal groups,
each served by one controller with share $\alpha=1/N$. Each controller
values the sum of its clients' business income and proxy scores, and
anticipates how its allocations affect wages and risk.\footnote{Installed capacity remains fixed, and human owners still hire competitively
after the AI allocation.} At a symmetric allocation with positive diversion, the controller's
condition is:
\begin{equation}
\frac{\psi}{b_{0}+b}=m\frac{\gamma Y}{a}\left[1-\underbrace{\ell\alpha}_{\text{wage-bill effect}}+\underbrace{\alpha(1-\ell)ea/\gamma}_{\text{risk internalisation}}\ \right],\label{eq:concentration-foc}
\end{equation}
where $e=-m'(X)/m(X)$ is the retention sensitivity from Section \ref{sec:risk-multiplier}.
As $\alpha\rightarrow0$, the optimality condition \ref{eq:concentration-foc}
converges to our atomistic benchmark \ref{eq:foc2}. Strategic AI
control introduces two effects.

First, a large controller recognises that useful AI raises the wage
bill across the firms it serves. This wage-bill effect makes diversion
more attractive: lower labour demand and wages partly offset the business
income lost by pursuing the proxy task. This explains the negative
wage-bill term in \ref{eq:concentration-foc}. 

Second, a large controller also recognises that diversion threatens
the retained income of all its clients by increasing the probability
of a common loss. Risk internalisation encourages useful work and
can outweigh the wage-bill effect. See Appendix \ref{ax:concentrated-control}
for further details and Figure \ref{fig:concentrated-control} for
a numerical illustration of these two effects. 

\subsection{Costly human control}

\label{sec:costly-control}

So far, the proxy weight $\psi$ has been given. We now allow owners
to pay for verification that reduces the agent’s incentive to pursue
the proxy score. In our software example, hidden tests make it harder
for the agent to receive credit for apparently successful code that
does not do what the firm needs.

Formally, we introduce a variable $s\geq0$ which captures the intensity
of verification. The proxy weight $\psi$ is now endogenous and decreasing
in $s$: we set $\psi(s)=\psi_{\mathrm{base}}e^{-s}$, with $\psi_{\mathrm{base}}>0$.
The cost of this technology (paid at date 1) is $\kappa_{s}s^{2}/2$,
with $\kappa_{s}>0$. Compared to our baseline model, the firm owner
now chooses capacity and verification together.

In one sector, first hold installed capacity fixed. More verification
reduces AI diversion.\footnote{These comparisons start at a smooth interior private optimum with
positive diversion and follow an operating equilibrium with $\Gamma<1$.} But, interestingly, owners do not necessarily buy too little of it.
When all firms verify more, useful production expands and wages rise,
reducing the return that each owner had expected at a fixed wage.
Without common risk, a small coordinated cut in verification would
therefore raise welfare at the private choice. With common risk, verification
also makes losses less likely and encourages agents to do useful work.
These gains may or may not outweigh the wage effect.

Now suppose firms shift a little of their spending from AI capacity
to verification, starting from the same private optimum. If we held
common risk fixed, better control would make up for the smaller stock
of AI capacity, leaving useful output unchanged to first order. AI
agents would nevertheless divert less capacity. When common risk then
falls, agents do more useful work and humans face smaller expected
losses. This change in spending therefore raises welfare, even when
extra verification at unchanged capacity would not. See Appendix \ref{app:costly-control}
for further details on the results and the joint private choice. 

\subsection{AI-assisted parameter elicitation}

\label{sec:elicitation}

We now turn to an exploratory extension of our baseline model: we
survey large language models (LLMs) for parameter ranges in our one-sector
model. We collected twelve assessments, three under each of four LLMs.\footnote{The LLMs are Claude Opus 5.5, Gemini 3.8 Flash, GPT-5.6 Sol and Grok
4.7, and we used High reasoning or thinking settings. } Returning to our software example, the models acted as outside assessors
of a hypothetical population of 10,000 software firms over one year,
with capability and controls held fixed.

In our exercise, we ask the LLMs three main questions. First, we ask
how strongly the AI agent values the proxy reward relative to profits.
Second, we ask how likely a common loss would be across these firms
if aggregate diversion were counterfactually set at 10\% ($p_{{\rm ref}}$).
Third, we ask what fraction $\rho$ of human income would remain if
such a loss occurred. We defined a common loss in the prompt as a
systemic event destroying at least 5\% of aggregate output. For the
first question, most LLMs specified their responses in terms of the
risk-free diversion share ($d_{\text{safe}}$), evaluated without
common loss risk ($m=1$) at equilibrium wages.

We use these responses to back out the three key parameters: the proxy
reward weight ($\psi$), the hazard sensitivity ($\xi$) and the retained
income fraction ($\rho$), while holding the remaining model parameters
fixed.\footnote{We use reference capacity $A_{{\rm ref}}=1$ and profit without diversion
$\Pi_{{\rm ref}}=0.40$. Appendix \ref{app:elicitation} gives the
survey setting and remaining parameter choices.} For positive $d_{{\rm safe}}$, the mapping is:

\begin{equation}
\begin{aligned}\bar{\psi}:=\frac{\psi}{\Pi_{{\rm ref}}} & =0.25(1+d_{{\rm safe}})(1-d_{{\rm safe}})^{-0.9},\\
\lambda:=\xi A_{{\rm ref}} & =-\frac{\log(1-p_{{\rm ref}})}{0.10}.
\end{aligned}
\label{eq:elicitation-mapping}
\end{equation}
The first line gives the weight needed to support the assessed diversion
at $m=1$; the second converts the probability at the specified exposure
into a loss intensity. 

\begin{table}[htbp]
\begin{centering}
\caption{Subjective central inputs and computed operating outcomes}
\label{tab:elicitation-central}
\par\end{centering}
\begin{centering}
\setlength{\tabcolsep}{7pt}%
\begin{tabular}{lrrrrrr}
\toprule 
 & \multicolumn{3}{c}{{\small Central inputs}} & \multicolumn{3}{c}{{\small Computed outputs}}\tabularnewline
\cmidrule{1-1}\cmidrule(lr){2-4}\cmidrule(lr){5-7}
{\small Record} & {\small$d_{{\rm safe}}$ (\%)} & {\small$p_{{\rm ref}}$ (\%)} & {\small$\rho$} & {\small$X^{*}/A$ (\%)} & {\small$q(X^{*})$ (\%)} & {\small$\Gamma^{*}$}\tabularnewline
\midrule 
{\small C1} & {\small 2.00} & {\small 1.00} & {\small 0.90} & {\small 2.011} & {\small 0.202} & {\small 0.00528}\tabularnewline
{\small M1} & {\small 6.00} & {\small 1.50} & {\small 0.82} & {\small 6.087} & {\small 0.916} & {\small 0.0142}\tabularnewline
{\small O1$^{\dagger}$} & {\small 2.07} & {\small 1.00} & {\small 0.85} & {\small 2.082} & {\small 0.209} & {\small 0.00793}\tabularnewline
{\small G1} & {\small 6.00} & {\small 2.00} & {\small 0.88} & {\small 6.077} & {\small 1.220} & {\small 0.0126}\tabularnewline
{\small C2} & {\small 2.00} & {\small 2.00} & {\small 0.88} & {\small 2.026} & {\small 0.408} & {\small 0.0127}\tabularnewline
{\small M2} & {\small 4.00} & {\small 2.50} & {\small 0.88} & {\small 4.065} & {\small 1.024} & {\small 0.0159}\tabularnewline
{\small O2} & {\small 2.00} & {\small 0.50} & {\small 0.85} & {\small 2.008} & {\small 0.101} & {\small 0.00396}\tabularnewline
{\small G2} & {\small 5.00} & {\small 0.80} & {\small 0.82} & {\small 5.038} & {\small 0.404} & {\small 0.00759}\tabularnewline
{\small C3} & {\small 1.00} & {\small 1.00} & {\small 0.88} & {\small 1.006} & {\small 0.101} & {\small 0.00634}\tabularnewline
{\small M3} & {\small 6.50} & {\small 2.00} & {\small 0.82} & {\small 6.626} & {\small 1.330} & {\small 0.0189}\tabularnewline
{\small O3} & {\small 2.00} & {\small 1.00} & {\small 0.80} & {\small 2.021} & {\small 0.203} & {\small 0.0106}\tabularnewline
{\small G3} & {\small 6.00} & {\small 2.00} & {\small 0.85} & {\small 6.097} & {\small 1.224} & {\small 0.0158}\tabularnewline
\bottomrule
\end{tabular}
\par\end{centering}
\centering{}%
\begin{minipage}[c]{0.98\linewidth}%
\noindent{\footnotesize Notes:{} C, M, O and G identify the Claude-,
Gemini-, GPT- and Grok-labelled responses; the number identifies the
collection batch. We could not verify the exact model versions or
settings. Each row uses one response's median assessment of each input.
The assessment of $d_{{\rm safe}}$ assumes no common loss ($m=1$),
while $p_{{\rm ref}}$ is the assessed loss probability when 10\%
of capacity is diverted. $\rho$ is the retained income fraction and
$\Gamma^{*}$ the risk-feedback coefficient; both are shown in levels.
$^{\dagger}$O1 reports $\bar{\psi}=0.26$ directly; we calculate
the displayed diversion share from this weight. Appendix~\ref{app:elicitation}
explains these calculations and the global uniqueness checks. The
replication files record the extracted inputs and unrounded results.}%
\end{minipage}
\end{table}

\FloatBarrier

Table \ref{tab:elicitation-central} presents the results from the
12 survey responses, using each response's marginal median assessments.
Interestingly, all twelve input combinations satisfy the sufficient
global uniqueness bound (\ref{eq:uniformone}), whose values range
from $0.0042$ to $0.0202$. Equilibrium diversion is $1.01$ to $6.63$\%
of capacity and the common-loss probability is $0.10$ to $1.33$\%. 

Low diversion without common risk is largely an input to this exercise.
What the model adds is how much common risk increases diversion, and
whether a high-diversion equilibrium can coexist with the reported
outcome.

To see how much risk feedback contributes to these outcomes, we compare
them with the corresponding allocations at $m=1$. Diversion rises
by only 0.006 to 0.126 percentage points. For example, diversion increases
from 2\% only to about 2.011\% in the first Claude assessment (C1).
The local risk multiplier $1/(1-\Gamma^{*})$ ranges from $1.004$
to $1.019$, so feedback also adds little to the diversion response
to additional AI capacity. Appendix \ref{app:elicitation} gives the
details.

The more adverse assessments give a different picture. Upper marginal
percentiles for $p_{{\rm ref}}$ range from 8\% to 30\%, while lower
percentiles for retained income range from $0.25$ to $0.60$. Respondents
also supplied 36 scenarios. All imply unique equilibria under our
model, but two have common-loss probabilities of about 68\% and 83\%.\footnote{Some scenario narratives also change the deployment assumptions, such
as architecture or access rights. The calculations use their reported
numerical inputs.} Moreover, a combination we constructed within one response's marginal
ranges supports three equilibria. No respondent proposed this combination,
and the survey assigns no probability to it. Appendix \ref{app:elicitation}
gives the calculation.

These calculations also depend on assumptions we did not elicit, including
the reward scale $b_{0}/A$ and the AI production elasticity $\gamma$.
All twelve central assessments still imply a unique equilibrium over
the range we consider. Larger assessed risks and different reward
scales can, however, produce multiple equilibria.\footnote{Appendix \ref{ax:elicitation-thresholds} gives these calculations,
and Figure \ref{fig:elicitation-thresholds} illustrates the cases
with multiple equilibria. In these comparisons, we keep diversion
without common risk fixed and adjust the proxy weight accordingly.}

A next step could be to estimate the model using data on how much
capacity AI agents divert and how their choices respond to firms’
reward rules and market prices.

\section{Conclusion }

\label{sec:conclusion}This paper presented a general equilibrium
framework to think about AI misalignment and its implications for
investment in AI capacity. We find that market prices can discourage
AI diversion, but common risk can weaken this discipline. Private
and social returns to investment can differ, and the welfare effect
of additional AI capacity also depends on which sector receives it.
The possibility of multiple equilibria further complicates the assessment
of policy interventions.

There are a number of avenues for future research. First, our model
abstracts from private information. How would AI agents use an information
advantage when they can also influence prices, as in Section \ref{sec:concentrated-control}?
Models combining private information and market power, such as those
of \citet{Kyle1985} and \citet{Vives2011}, could help us study this
interaction.

Second, AI diversion may itself change total factor productivity or
the relative roles of labour and AI in production. Section \ref{sec:returns}
takes changes in task shares as given; a further step would be to
let agents' allocations affect the technology itself. This could change
the opportunity cost of diversion and the welfare effects of AI investment.

Third, labour is homogeneous in our model, but the productivity gains
from AI can vary with workers' experience and skills (\citealp{BrynjolfssonLiRaymond2025,DaniottiEtAl2026}).
Allowing for skill heterogeneity would let us study who gains from
AI investment and who bears the costs of AI diversion. This would
also connect our analysis more closely to the literature on automation
and inequality \citep{moll2022}.

Fourth, modelling the market structure of AI models seems an exciting
avenue. For instance, one could allow firms to choose between competing
AI providers. If some AI models are more prone to diversion, the ability
to switch to other models could provide another source of discipline
alongside market prices \citep{CollinaEtAl2026}. One could then ask
whether collusion among AI agents weakens this discipline, and how
shared weaknesses across AI providers affect aggregate risk \citep{FishGonczarowskiShorrer2026,KimEtAl2025}.

Lastly, the dynamic model in Section \ref{sec:ai-accumulation} could
be developed into a richer DSGE framework to study how AI investment
and common risk affect business cycles. For instance, allowing AI
capacity to depreciate gradually and introducing investment adjustment
costs \citep{ChristianoEichenbaumEvans2005} would let us study how
firms build up and replace AI capacity over time. We could also extend
our welfare analysis to this dynamic setting. 

\clearpage\setlength{\bibsep}{3pt}

\providecommand{\natexlab}[1]{#1} \providecommand{\url}[1]{\texttt{#1}}
\expandafter\ifx\csname urlstyle\endcsname\relax \providecommand{\doi}[1]{doi: #1}\else
\providecommand{\doi}{doi: \begingroup \urlstyle{rm}\Url}\fi  \ifx\undefined\BySame
\newcommand{\BySame}{\leavevmode\rule[.5ex]{3em}{.5pt}\ }
\fi
\ifx\undefined\textsc
\newcommand{\textsc}[1]{{\sc #1}}
\newcommand{\emph}[1]{{\em #1\/}}
\let\tmpsmall\small
\renewcommand{\small}{\tmpsmall\sc}
\fi

 \clearpage{}

\appendix
\relax\relax\relax\relax\relax\phantomsection\label{annex:start} 
\begin{center}
{\Large Technical appendix}{\Large\par}
\par\end{center}

\begin{center}
 
\par\end{center}

\begin{center}
\smallskip{}
 {\small Optimisation, equilibrium, welfare and numerical methods} 
\par\end{center}

\noindent Appendix \ref{app:model-background} provides detailed background
on our baseline model (with Table \ref{tab:notation} summarising
the main objects of the model); Appendix \ref{app:main-proofs} proves
the main propositions; Appendix \ref{app:extension-proofs} gives
the extension proofs. Appendix \ref{app:elicitation} documents parameter
elicitation, and Appendix \ref{app:investment-support} provides further
results on welfare and investment and details of the numerical examples.
\begin{table}[htbp]
\begin{centering}
\caption{Notation used in the main paper and proofs}
\label{tab:notation}
\par\end{centering}
\centering{}%
\begin{tabular}{p{0.2\linewidth}p{0.72\linewidth}}
\toprule 
{\small Object  } & {\small Economic meaning and convention }\tabularnewline
\midrule 
{\small$A_{i},a_{i},b_{i}$  } & {\small Installed, useful and diverted capacity, with $a_{i}+b_{i}=A_{i}$. }\tabularnewline
{\small$\ell,\gamma_{i},r_{i}$  } & {\small Labour and useful-AI production exponents; $r_{i}=\gamma_{i}/(1-\ell)$
is the curvature of income after hiring at fixed prices. }\tabularnewline
{\small$\omega_{i},\theta_{i},\Theta$  } & {\small Final-good weights, $\theta_{i}=\omega_{i}\gamma_{i}$, and
$\Theta=\sum_{i}\theta_{i}$. In one sector $\theta=\gamma$. }\tabularnewline
{\small$D_{i},\mathcal{D}$  } & {\small Human operating dividend and the conditional aggregate diversion
map, respectively. }\tabularnewline
{\small$q,\rho,m$  } & {\small Common-loss probability, loss-state retention, expected retention
$m=1-(1-\rho)q$. }\tabularnewline
{\small$Q_{N},Q_{T},V$  } & {\small Household state prices and the value $V=Q_{N}+\rho Q_{T}$
of a proportional dividend claim. These value household claims;$m$
instead weights profits in the agent's objective.}\tabularnewline
{\small$S_{i},h_{i},g_{i},t_{i}$  } & {\small$S_{i}=b_{0i}+b_{i}$; local allocation coefficients and $t_{i}=r_{i}h_{i}$.
Their definitions depend on the normalised reward curvature. }\tabularnewline
{\small$H,K,e,\Gamma$  } & {\small Conditional output feedback; diversion response to a unit fall
in $\log m$; fall in log expected retention per diverted unit; aggregate-risk
feedback $\Gamma=eK$. A 1\% fall in $m$ gives approximately $0.01K$
extra diverted units.}\tabularnewline
{\small$k_{j}^{0},x_{j}^{0};k_{j},x_{j}$  } & {\small Log-output and diversion responses to capacity, first holding
$m$ fixed, then imposing risk consistency. }\tabularnewline
{\small$\kappa_{i},\sigma,\epsilon$  } & {\small Installation-cost coefficient, AI reward curvature, and final-good
substitution elasticity. Household utility is logarithmic throughout
the benchmark. }\tabularnewline
{\small$\mathcal{E},\mathcal{P},\mathcal{F},\mathcal{R}$  } & {\small Private Euler residual and the three components of the marginal
social--private return gap. The residual is zero at an interior competitive
investment choice with no charge.}\tabularnewline
\bottomrule
\end{tabular}
\end{table}

\section{Model background}

\label{app:model-background}

Below, we derive the results for one or two sectors using a general
CRRA proxy reward with curvature ($\sigma>0$). The logarithmic reward
in our baseline model is the special case ($\sigma=1$). We keep the
production assumptions in Section \ref{sec:one} for each sector and
take capacities, prices, ($b_{0i}$) and ($\psi_{i}$) to be positive.
We also consider ($\xi=0$) or ($\rho=1$), when there is no common
loss. 

\subsection{Private choices and market clearing}

\label{ax:optimization}\label{ax:capacity-reduction}\label{ax:operating-accounts}

For $a_{i}>0$, the owner's hiring objective is strictly concave,
with an infinite marginal product at zero labour. The unique optimum
therefore satisfies $wL_{i}=\ell p_{i}y_{i}$. Substitution gives:

\begin{equation}
\begin{aligned}L_{i}^{*} & =(\ell p_{i}z_{i}a_{i}^{\gamma_{i}}/w)^{1/(1-\ell)}, & wL_{i}^{*} & =\ell p_{i}y_{i},\\
\Pi_{i} & =C_{i}a_{i}^{r_{i}}, & r_{i} & =\gamma_{i}/(1-\ell),\\
C_{i} & =(1-\ell)(p_{i}z_{i})^{1/(1-\ell)}(\ell/w)^{\ell/(1-\ell)}.
\end{aligned}
\label{ax:eq:reducedprofit}
\end{equation}
If $a_{i}=0$, the optimum is $L_{i}=0$. For positive services the
hiring envelope is $\Pi_{i,a}=r_{i}C_{i}a_{i}^{r_{i}-1}$, holding
prices fixed. Since $0<r_{i}<1$, the marginal return to useful AI
services becomes arbitrarily large as useful AI goes toward zero.

\paragraph{The AI agent’s allocation.}

Under a general isoelastic reward specification, the normalised proxy
reward and its derivatives are:

\begin{equation}
\begin{aligned}v_{\sigma,i}(b) & =\begin{cases}
[(1+b/b_{0i})^{1-\sigma}-1]/(1-\sigma), & \sigma\ne1,\\
\log(1+b/b_{0i}), & \sigma=1,
\end{cases}\\
v'_{\sigma,i}(b) & =b_{0i}^{\sigma-1}(b_{0i}+b)^{-\sigma}, & v''_{\sigma,i}(b) & =-\sigma b_{0i}^{\sigma-1}(b_{0i}+b)^{-\sigma-1}.
\end{aligned}
\label{ax:eq:reward}
\end{equation}
Thus $v'_{\sigma,i}(0)=1/b_{0i}$ for every curvature. At given prices
and retention $m>0$, the agent maximises $mC_{i}(A_{i}-b_{i})^{r_{i}}+\psi_{i}v_{\sigma,i}(b_{i})$
on $[0,A_{i}]$. The objective function is continuous on this compact
interval $[0,A_{i}]$, guaranteeing the existence of an optimum. Full
AI diversion cannot be optimal: reallocating a small compute transfer
$\delta>0$ back to useful work raises expected business income by
$mC_{i}\delta^{r_{i}}$, while costing only order $\delta$ in proxy
reward. For a sufficiently small transfer, the business gain therefore
strictly exceeds this cost. With $a_{i}=A_{i}-b_{i}>0$ and $S_{i}=b_{0i}+b_{i}$,
the optimality conditions are:

\begin{equation}
\begin{aligned}-mr_{i}C_{i}a_{i}^{r_{i}-1}+\psi_{i}b_{0i}^{\sigma-1}S_{i}^{-\sigma}+\alpha_{i}-\eta_{i} & =0,\\
b_{i},a_{i},\alpha_{i},\eta_{i} & \ge0, & \alpha_{i}b_{i} & =\eta_{i}a_{i}=0.
\end{aligned}
\label{ax:eq:aikkt}
\end{equation}
Strict concavity makes these conditions sufficient for the unique
optimum. The zero-diversion condition is:

\begin{equation}
\psi_{i}/b_{0i}\le mr_{i}C_{i}A_{i}^{r_{i}-1}.\label{ax:eq:privatecorner}
\end{equation}
At equality the allocation is continuous but its slope may change.
If $A_{i}=0$, only $a_{i}=b_{i}=0$ is feasible. Differentiating
the interior condition at fixed prices and risk gives:

\begin{equation}
(r_{i}-1)\frac{a_{i,A}}{a_{i}}=-\sigma\frac{1-a_{i,A}}{S_{i}},\qquad a_{i,A}=\frac{\sigma a_{i}}{\sigma a_{i}+(1-r_{i})S_{i}}>0.\label{ax:eq:capacityresponse}
\end{equation}
At a strict zero-diversion corner, $a_{i,A}=1$. Diversion starts
at $\bar{A}_{i}=(mr_{i}C_{i}b_{0i}/\psi_{i})^{1/(1-r_{i})}$. Useful
services and dividends are continuous and strictly increasing in capacity
on both sides of this threshold.

\label{ax:score-production}

The logarithmic score also has a simple resource interpretation. Let
apparent achievement $s$ require $\mathcal{B}(s)=b_{0}(e^{s}-1)$
units of AI capacity. Resources bind because useful AI work is valuable.
Writing $b=\mathcal{B}(s)$ gives $s=\log(1+b/b_{0})$ and $a+b=A$,
exactly the baseline objective and allocation problem.

\label{ax:price-response}

At an interior private choice, differentiate $-m\Pi_{a}+\psi v'(b)=0$
and put $\mathfrak{D}=-m\Pi_{aa}-\psi v''>0$. The private price responses
are:

\begin{equation}
b_{p}=-\frac{m\Pi_{ap}}{\mathfrak{D}}<0,\qquad b_{w}=-\frac{m\Pi_{aw}}{\mathfrak{D}}>0,\label{eq:score-price-response}
\end{equation}
Here $\Pi_{ap}=\Pi_{a}/[(1-\ell)p]$ and $\Pi_{aw}=-\ell\Pi_{a}/[(1-\ell)w]$.
A higher goods price or lower wage reduces diversion, holding capacity,
risk and the other price fixed. At a strict zero-diversion corner
these responses are zero.

\paragraph{Market clearing.}

The concave, constant-returns final-good technology $G(y)=\prod_{i}(y_{i}/\omega_{i})^{\omega_{i}}$
gives $p_{i}y_{i}=\omega_{i}Y$ and zero final-sector profit. Combining
this condition with hiring and $\sum_{i}L_{i}=\bar{L}$ yields:

\begin{equation}
\begin{aligned}L_{i} & =\omega_{i}\bar{L}, & w & =\ell Y/\bar{L}, & D_{i} & =(1-\ell)\omega_{i}Y,\\
Y & =T\prod_{i}a_{i}^{\theta_{i}}, & \theta_{i} & =\omega_{i}\gamma_{i}, & T & =\bar{L}^{\ell}\prod_{i}(z_{i}\omega_{i}^{\ell-1})^{\omega_{i}}.
\end{aligned}
\label{ax:eq:markets}
\end{equation}
Write $\Theta=\sum_{i}\theta_{i}<1$. The private marginal return
evaluated at clearing prices is $\Pi_{i,a}=r_{i}D_{i}/a_{i}=\theta_{i}Y/a_{i}$.
This uses the hiring envelope, rather than differentiating the aggregate
identity for dividends. In one sector, $p=1$, $\theta=\gamma$ and
$T=z\bar{L}^{\ell}$.

\paragraph{A proportional split.}

\label{ax:split}\label{ax:exactsplit} \begin{lemma}[A proportional
split leaves operation unchanged]\label{prop:split} Starting from
a one-sector economy, choose positive weights summing to one and set
$A_{i}=\omega_{i}A$, $b_{0i}=\omega_{i}b_{0}$, $\psi_{i}=\omega_{i}\psi$,
$\gamma_{i}=\gamma$ and $z_{i}=z\omega_{i}^{1-\ell-\gamma}$. For
every risk conjecture, the allocation $a_{i}=\omega_{i}a$ and $b_{i}=\omega_{i}b$
reproduces the original output and the entire aggregate-diversion
map, for every normalised reward curvature $\sigma>0$. \end{lemma} 
\begin{proof}
Fix an aggregate-risk conjecture and consider the unique one-sector
allocation at that risk. With the proposed sectoral split, $L_{i}=\omega_{i}\bar{L}$
and the chosen productivities give $y_{i}=\omega_{i}Y$. The final-good
aggregator then gives the same total output as before. Goods prices
are $p_{i}=1$ and the wage remains $w=\ell Y/\bar{L}$.

The normalised proxy reward satisfies $v_{\sigma,i}'(\omega_{i}b)=v_{\sigma}'(b)/\omega_{i}$,
while the marginal business return is $\theta_{i}Y/a_{i}=\gamma Y/a$.
Weighting this marginal proxy reward by $\psi_{i}=\omega_{i}\psi$
cancels the scaling. At the proposed allocation, the AI agent faces
the same first-order condition when it diverts and the same inequality
when it does not. The allocation is unique for any given risk conjecture,
so each sector diverts its proportional share of the one-sector total.
Adding up diversion across sectors therefore gives the original aggregate-diversion
map.

We take as given which units operate in each sector. We then vary
their capacities, productivities and reward parameters together.
\end{proof}

\subsection{Conditional allocations and comparative statics}

\label{ax:conditional}

At fixed capacities and retention, the AI agent's condition when it
diverts, evaluated at clearing prices, is:\footnote{For the derivatives below, we keep the set of diverting agents unchanged.
At a threshold where diversion starts or stops, we use one-sided derivatives.
With endogenous risk, we also require $\Gamma\ne1$.}

\begin{equation}
\frac{a_{i}}{(A_{i}+b_{0i}-a_{i})^{\sigma}}=\frac{m\theta_{i}Y}{\psi_{i}b_{0i}^{\sigma-1}}.\label{ax:eq:implicitservice}
\end{equation}
For $C=A_{i}+b_{0i}$, the left side increases from zero to infinity
on $(0,C)$. Its inverse defines $a_{i}(Y;A,m)$, capped at $A_{i}$
when $\psi_{i}A_{i}\le m\theta_{i}Yb_{0i}$. For logarithmic rewards,

\begin{equation}
a_{i}(Y;A,m)=\min\{A_{i},m\theta_{i}Y(A_{i}+b_{0i})/(\psi_{i}+m\theta_{i}Y)\}.\label{ax:eq:servicelog}
\end{equation}
Substitute these services into $R(Y)=T\prod_{i}a_{i}(Y;A,m)^{\theta_{i}}$.
As $Y\downarrow0$, uncapped services are proportional to $Y$, so
$R(Y)/Y\to\infty$ because $\Theta<1$. At large $Y$, services are
capped and $R(Y)/Y\to0$. For uniqueness, define

\begin{equation}
h_{i}=\frac{S_{i}}{S_{i}+\sigma a_{i}},\qquad g_{i}=\frac{\sigma}{S_{i}+\sigma a_{i}},\qquad t_{i}=r_{i}h_{i};\label{ax:eq:hdefinition}
\end{equation}
At a strict no-diversion corner take $h_{i}=t_{i}=0$ and $g_{i}=1/A_{i}$.
The service elasticity with respect to $Y$ is $h_{i}$, so $d\log R/d\log Y=H:=\sum_{i}\theta_{i}h_{i}\le\Theta<1$.
Thus $R(Y)/Y$ decreases strictly, including across the continuous
cap points, and crosses one exactly once. The positive conditional
allocation is unique and continuous in the primitives. Equivalently,
$x\mapsto\log R(e^{x})$ is a contraction with bound $\Theta$.

Differentiating with respect to retention and using $d\log Y=\sum_{i}\theta_{i}d\log a_{i}$
gives

\begin{equation}
\begin{aligned}d\log a_{i} & =h_{i}(d\log Y+d\log m), & \frac{\partial\log Y}{\partial\log m} & =\frac{H}{1-H},\\
K:=-\frac{\partial\mathcal{D}}{\partial\log m} & =\frac{N}{1-H}, & N & =\sum_{i}a_{i}h_{i}.
\end{aligned}
\label{ax:eq:conditionalderivatives}
\end{equation}
Here $\mathcal{D}(A,m)=\sum_{i}(A_{i}-a_{i})$ is total diversion.
Higher retention weakly raises useful services and output, and lowers
diversion. These comparisons also hold across cap thresholds by continuity.
Under the maintained loss probability, the consistency map and its
derivative are

\begin{equation}
\begin{aligned}e(X) & :=-\frac{d\log m(X)}{dX}=\frac{\xi(1-\rho)e^{-\xi X}}{m(X)},\\
F(X) & :=\mathcal{D}(A,m(X)),\qquad F'(X)=eK=\Gamma.
\end{aligned}
\label{ax:eq:outermap}
\end{equation}
The derivative uses the conditional allocation induced by the conjecture
$X$. Appendix \ref{ax:risk-consistency} establishes existence and
gives a global bound for this map.

\paragraph{Capacity and reward responses.}

Log-differentiating \eqref{ax:eq:implicitservice}, using the reward
normalisation in \eqref{ax:eq:reward}, gives

\begin{equation}
d\log a_{i}=g_{i}\,dA_{i}+h_{i}\{d\log Y+d\log m-d\log\psi_{i}+\log(S_{i}/b_{0i})\,d\sigma\}.\label{eq:ax-sector-differential}
\end{equation}
For capacity changes, keep reward parameters fixed and aggregate with
weights $\theta_{i}$:

\begin{equation}
(1-H)d\log Y=\sum_{i}\theta_{i}g_{i}\,dA_{i}+Hd\log m,\qquad H=\sum_{i}\theta_{i}h_{i}.\label{eq:ax-sector-aggregate}
\end{equation}
Since $H<1$, we can solve for the output and diversion responses
at fixed retention:

\begin{equation}
\begin{aligned}k_{j}^{0} & =\frac{\theta_{j}g_{j}}{1-H}, & a_{i,A_{j}}^{0} & =a_{i}(g_{i}\mathbf{1}_{i=j}+h_{i}k_{j}^{0}),\\
x_{j}^{0} & =1-a_{j}g_{j}-K\theta_{j}g_{j}=h_{j}-K\theta_{j}g_{j}.
\end{aligned}
\label{eq:ax-sector-conditional-capacity}
\end{equation}
The last equality uses $a_{j}g_{j}=1-h_{j}$ and $K=N/(1-H)$. Risk
consistency adds $d\log m=-e\,dX$ to $dX=x_{j}^{0}dA_{j}-Kd\log m$,
giving:

\begin{equation}
\begin{aligned}x_{j} & =\frac{x_{j}^{0}}{1-\Gamma},\\
k_{j} & =k_{j}^{0}-\frac{H}{1-H}ex_{j},\\
a_{i,A_{j}} & =a_{i}\{g_{i}\mathbf{1}_{i=j}+h_{i}(k_{j}-ex_{j})\}.
\end{aligned}
\label{eq:ax-sector-full-x}
\end{equation}

\paragraph{The wage response.}

In one sector, holding the wage fixed gives $K_{\mathrm{FP}}=aS/[\sigma a+(1-r)S]$.
After wages adjust,$K=aS/[\sigma a+(1-\gamma)S]<K_{\mathrm{FP}}$,
since $r>\gamma$. Wage adjustment therefore dampens diversion's response
to retention. Both derivatives vanish at a strict no-diversion corner.

\subsection{Household valuation and private investment}

\label{ax:investment}\label{app:investment}\label{ax:household}

Let $\pi_{N}=1-q$, $\pi_{T}=q$, $\zeta_{N}=1$ and $\zeta_{T}=\rho$.
Household $h$ takes probabilities and prices as given and buys net
state claims $Z_{o}^{h}$. Its consolidated budgets and claim-market
clearing are:

\begin{equation}
\begin{aligned}c_{1}^{h}+\sum_{i}\Phi_{i}^{h}+\sum_{o}Q_{o}Z_{o}^{h} & \le E^{h}, & \Phi_{i}(A_{i}) & =\kappa_{i}A_{i}^{2}/2,\\
c_{o}^{h} & \le\zeta_{o}I^{h}+Z_{o}^{h}, & I^{h} & =w\bar{L}^{h}+\sum_{i}D_{i}^{h},\qquad\int Z_{o}^{h}\,dh=0.
\end{aligned}
\label{eq:ann-household-date1}
\end{equation}
Positive consumption and freely traded claims give the first-order
conditions:

\begin{equation}
\lambda_{1}^{h}=1/c_{1}^{h},\qquad\lambda_{o}^{h}=\beta\pi_{o}/c_{o}^{h},\qquad\lambda_{1}^{h}Q_{o}=\lambda_{o}^{h}.\label{eq:ann-household-foc}
\end{equation}
Log utility and affine budgets make these conditions sufficient. Positive-probability
budgets bind; a zero-probability claim has price zero. Aggregate income
before the loss is $w\bar{L}+\sum_{i}D_{i}=Y$, so $c_{o}=\zeta_{o}Y$
and $c_{1}=E-\sum_{i}\Phi_{i}$. Aggregating $Q_{o}c_{o}^{h}=\beta\pi_{o}c_{1}^{h}$
yields

\begin{equation}
Q_{N}=\frac{\beta c_{1}(1-q)}{Y},\qquad Q_{T}=\frac{\beta c_{1}q}{\rho Y},\qquad V=Q_{N}+\rho Q_{T}=\frac{\beta c_{1}}{Y}.\label{eq:ann-state-prices}
\end{equation}
Substitution into the present-value budget shows that an owned firm
contributes $VD_{i}-\Phi_{i}$ to household wealth. Its price-taking
owner therefore maximises this expression. Under log utility, higher
marginal utility in the loss state offsets the proportional dividend
loss in $V$. The agent instead uses expected retention $m$, and
the atomistic owner still takes the loss probability as given.

\label{ax:severity}

For CRRA household utility with coefficient $\eta$, the same budget
conditions give:

\begin{equation}
V=\beta c_{1}^{\eta}Y^{-\eta}[(1-q)+q\rho^{1-\eta}].\label{eq:severity-crra-V}
\end{equation}
At $\eta=1$ the bracket equals one; at $\eta=0$ it equals $m$.
Household curvature $\eta$ is distinct from proxy-reward curvature
$\sigma$.

\paragraph{Welfare with CRRA household utility.}

\label{ax:crra-welfare} The one-sector welfare sign does not require
logarithmic household utility. Let $u_{\eta}'(c)=c^{-\eta}$ for $\eta\ge0$,
with log utility at $\eta=1$. We assume a representative household
with complete state-contingent markets, keeping labour supply fixed,
losses proportional and the agent’s objective unchanged. Write $B_{\eta}:=(1-q)+q\rho^{1-\eta}>0$.
The valuation formula above gives, at a smooth interior unregulated
competitive investment choice,

\[
\kappa Ac_{1}^{-\eta}=\beta Y^{1-\eta}B_{\eta}\mu.
\]
Human welfare is now $W=u_{\eta}(c_{1})+\beta[(1-q)u_{\eta}(Y)+qu_{\eta}(\rho Y)]$.
Differentiating along the operating equilibrium and using the private
investment condition gives:

\begin{equation}
\begin{aligned}W_{A} & =-\kappa Ac_{1}^{-\eta}+\beta Y^{1-\eta}B_{\eta}k+\beta q'(X)x[u_{\eta}(\rho Y)-u_{\eta}(Y)]\\
 & =\beta Y^{1-\eta}B_{\eta}(k-\mu)+\beta q'(X)x[u_{\eta}(\rho Y)-u_{\eta}(Y)].
\end{aligned}
\label{eq:crra-welfare-return}
\end{equation}
Splitting $k-\mu=(k^{0}-\mu)+(k-k^{0})$ gives the same price and
incentive terms, multiplied by the positive factor $\beta Y^{1-\eta}B_{\eta}$.
Keeping the production, allocation and risk conditions of Proposition
\ref{prop:one_welfare}, with positive diversion, $\xi>0$, $0<\rho<1$
and $\Gamma<1$, we have $k<k^{0}<\mu$ and $x>0$. Since $u_{\eta}(\rho Y)<u_{\eta}(Y)$,
all three contributions are negative: a small coordinated capacity
cut raises welfare. As in the proposition, the investment choice must
be privately optimal. This argument does not establish existence or
uniqueness of investment, or tell us how capacity changes with risk
aversion.

\paragraph{Private returns and concavity.}

\label{ax:private-derivative}\label{ax:private-concavity}

At positive diversion, put $d_{i}=\sigma a_{i}+(1-r_{i})S_{i}$. The
service response \eqref{ax:eq:capacityresponse} and the dividend
chain rule imply:

\begin{equation}
D_{i,A}^{{\rm priv}}=\frac{r_{i}C_{i}\sigma a_{i}^{r_{i}}}{d_{i}},\qquad\mu_{i}:=\frac{D_{i,A}^{{\rm priv}}}{Y}=\frac{\theta_{i}\sigma}{d_{i}}=\frac{\theta_{i}g_{i}}{1-t_{i}}.\label{eq:ann-private-dividend}
\end{equation}
The normalisation uses $r_{i}C_{i}a_{i}^{r_{i}}=\theta_{i}Y$ at clearing
prices. At a strict zero-diversion corner, $\mu_{i}=\theta_{i}/A_{i}$.
A smooth interior private investment choice satisfies $\kappa_{i}A_{i}=\beta c_{1}\mu_{i}$.
Differentiating the dividend return once more gives

\begin{equation}
D_{i,AA}^{{\rm priv}}=\frac{r_{i}C_{i}\sigma(1-r_{i})a_{i}^{r_{i}}}{d_{i}^{3}}\{[(1+\sigma)r_{i}-1]S_{i}-\sigma^{2}a_{i}\}.\label{eq:ann-private-second}
\end{equation}
If $r_{i}\le1/(1+\sigma)$, the bracket is strictly negative. Below
the diversion threshold, $D_{i}=C_{i}A_{i}^{r_{i}}$ is also strictly
concave. At the threshold dividends are continuous and their right
derivative is their left derivative multiplied by $\sigma A_{i}/[\sigma A_{i}+(1-r_{i})b_{0i}]<1$.
The two branches therefore join with a downward kink, proving global
concavity.

Quadratic installation cost makes the owner's objective strictly concave.
The objective tends to minus infinity as capacity grows, while its
derivative tends to infinity at zero, so there is a unique positive
global optimum. If this optimum is at the diversion threshold, zero
must lie between the left and right derivatives of the owner's objective.
Outside this sufficient curvature restriction, we must compare any
stationary choice with other feasible investments.

\subsection{Welfare accounting}

\label{ax:welfare-accounting}

With $c_{1}=E-\sum_{i}\kappa_{i}A_{i}^{2}/2$, human welfare is $W=\log c_{1}+\beta[\log Y+q(X)\log\rho]$.
Write $k_{j}=\partial\log Y/\partial A_{j}$, let $k_{j}^{0}$ hold
retention fixed, and put $x_{j}=\partial X/\partial A_{j}$. Differentiating
and adding and subtracting the private return $\beta\mu_{j}$ and
the fixed-risk return $\beta k_{j}^{0}$ gives:

\begin{equation}
\begin{aligned}W_{A_{j}} & =-\frac{\kappa_{j}A_{j}}{c_{1}}+\beta k_{j}+\beta q'\log\rho\,x_{j}\\
 & =\underbrace{-\kappa_{j}A_{j}/c_{1}+\beta\mu_{j}}_{\mathcal{E}_{j}}+\underbrace{\beta(k_{j}^{0}-\mu_{j})}_{\mathcal{P}_{j}}+\underbrace{\beta(k_{j}-k_{j}^{0})}_{\mathcal{F}_{j}}+\underbrace{\beta q'\log\rho\,x_{j}}_{\mathcal{R}_{j}}.
\end{aligned}
\label{eq:ann-off-ce-decomposition}
\end{equation}
The three wedges respectively capture price adjustment, the incentive
response to risk, and the direct expected loss. Substituting the responses
from Appendix \ref{ax:conditional} and the private return \eqref{eq:ann-private-dividend}
gives:

\begin{equation}
\begin{aligned}\mathcal{P}_{j} & =\beta\theta_{j}g_{j}[(1-H)^{-1}-(1-t_{j})^{-1}],\\
\mathcal{F}_{j} & =-\beta Hex_{j}/(1-H),\qquad\mathcal{R}_{j}=\beta q'\log\rho\,x_{j}.
\end{aligned}
\label{eq:ann-price-wedge}
\end{equation}
The Euler residual $\mathcal{E}_{j}$ is zero at a smooth interior
unregulated competitive choice. Away from that choice it must be retained.
Appendix \ref{ax:planner} gives the planner's problem, and Appendix
\ref{ax:wedges} develops the sectoral welfare comparisons.

\subsection{The probability of a common loss}

\label{ax:assumptions-poisson}

In Section \ref{sec:one}, we assume that the probability of a common
loss rises with total AI diversion. To give the loss probability in
\eqref{eq:hazard} a simple microfoundation, suppose several firms
rely on the same software service. Their AI agents may act independently,
but if one agent disrupts the service, all the firms that use it can
lose income.\footnote{For instance, \citet{TullyEtAl2025} detected network disruptions
at 759 of the 2,232 US hospitals they studied immediately after a
single faulty CrowdStrike software update.} Let $s$ be the probability that an incident caused by AI diversion
has this wider effect.

Suppose incidents arrive at Poisson rate $\lambda X$ when total diversion
is $X$. If each incident independently causes a common loss with
probability $s$, the number of incidents that do so follows a Poisson
distribution, $M_{{\rm tr}}\mid X\sim\mathrm{Poisson}(\xi X)$, where
$\xi=s\lambda$ \citep{Gallager2011}. One such incident is enough,
so the probability of a common loss is $q(X)=\Pr(M_{{\rm tr}}\ge1\mid X)=1-e^{-\xi X}$.
We assume that the loss occurs after production and that humans retain
the same fraction $\rho$ of output whether one such incident occurs
or several.

\subsection{Partial exposure of the proxy reward}

\label{ax:partial-proxy-exposure}In Section \ref{sec:one}, a common
loss reduces firms' income while leaving the AI agent's proxy reward
unchanged. Here we allow the loss to reduce the proxy reward as well
and ask whether common risk still encourages diversion. Let $\upsilon\in[0,1]$
measure its exposure relative to business income:

\begin{equation}
J=m(X)\Pi(a;p,w)+m_{\upsilon}(X)f(b),\qquad m_{\upsilon}(X):=1-\upsilon(1-\rho)q(X).\label{eq:partial-proxy-objective}
\end{equation}
The (atomistic) AI agent still takes common risk as given. Since $m_{\upsilon}>0$,
dividing its objective by $m_{\upsilon}$ gives the original allocation
problem with effective retention $\widetilde{m}_{\upsilon}:=m/m_{\upsilon}$.
In one sector, away from diversion thresholds, the local feedback
calculation therefore changes to:

\begin{equation}
\begin{aligned}e_{\upsilon} & :=-\frac{d\log\widetilde{m}_{\upsilon}}{dX}=\frac{(1-\upsilon)(1-\rho)q'(X)}{m(X)m_{\upsilon}(X)},\\
K_{\upsilon} & :=-\left.\frac{\partial\mathcal{D}(A,\widetilde{m}_{\upsilon})}{\partial\log\widetilde{m}_{\upsilon}}\right|_{A},\qquad\Gamma_{\upsilon}=e_{\upsilon}K_{\upsilon}.
\end{aligned}
\label{eq:partial-proxy-feedback}
\end{equation}
Here $K_{\upsilon}$ is the usual conditional allocation response
evaluated at the new effective retention. At $\upsilon=0$, we recover
the benchmark: $\widetilde{m}_{0}=m$ and $e_{0}=e$. At $\upsilon=1$,
both rewards have the same exposure, so $\widetilde{m}_{1}=1$ and
$e_{1}=\Gamma_{1}=0$. Common risk no longer changes their relative
value, although AI diversion and its direct risk externality can remain. 

\section{Proofs of the main propositions}

\label{app:main-proofs}

The proofs use the model equations and response formulas in Appendix
\ref{app:model-background}.

\subsection{Proof of Proposition \ref{prop:continuation}}

\label{ax:risk-consistency}

Appendix \ref{ax:conditional} establishes a unique continuous operating
allocation at fixed retention. Since diversion falls with $m$ and
$m(X)$ falls with $X$, the map $F(X)=\mathcal{D}(A,m(X))$ is continuous
and nondecreasing on $[0,A_{\Sigma}]$, where $A_{\Sigma}=\sum_{i}A_{i}$.
It maps this interval into itself, with $F(0)\ge0$ and $F(A_{\Sigma})<A_{\Sigma}$
because useful services remain positive. The intermediate value theorem
gives a fixed point and hence a risk-consistent operating equilibrium.

For global uniqueness, $1-H\ge1-\Theta$ and each active contribution
to $N$ satisfies $a_{i}h_{i}=a_{i}(C-a_{i})/[C+(\sigma-1)a_{i}]$,
with $C=A_{i}+b_{0i}$. Its maximum on $[0,C]$ is $C/(1+\sqrt{\sigma})^{2}$,
attained at $a_{i}=C/(1+\sqrt{\sigma})$. Capped sectors contribute
zero. Since $0\le e(X)\le\xi(1-\rho)$,

\begin{equation}
0\le F'(X)\le\frac{\xi(1-\rho)\sum_{i}(A_{i}+b_{0i})}{(1-\Theta)(1+\sqrt{\sigma})^{2}}=:\overline{\Gamma}_{\sigma}.\label{ax:eq:globalbound}
\end{equation}
Each sector starts diverting at most once as retention falls. Between
these thresholds the derivative satisfies the bound above, and the
map is continuous at the thresholds. The same bound therefore applies
to the slope between any two conjectures. If $\overline{\Gamma}_{\sigma}<1$,
the map is a contraction and equilibrium is unique. With one sector
and $\sigma=1$, this gives the proposition's bound. If $\xi=0$ or
$\rho=1$, uniqueness follows directly from the fixed-retention allocation.

\subsubsection{At most three one-sector equilibria}

\label{ax:roots}

At positive diversion, the one-sector allocation equation is equivalent
to

\begin{equation}
\Psi_{\sigma}(b)=\frac{\gamma Tm(b)(b_{0}+b)^{\sigma}}{b_{0}^{\sigma-1}(A-b)^{1-\gamma}}=\psi,\qquad0<b<A.\label{ax:eq:psi}
\end{equation}
The zero-diversion corner is separately admissible when $\psi\le\Psi_{\sigma}(0)=\gamma Tb_{0}A^{\gamma-1}$.
The schedule diverges as $b\uparrow A$. Put $a=A-b$, $S=b_{0}+b$,
$c=1-\gamma$ and $d=\sigma a+cS$. Then

\begin{equation}
K_{d}=\frac{aS}{d},\qquad\Gamma_{d}=e(b)K_{d}(b),\qquad\frac{\Psi_{\sigma}'}{\Psi_{\sigma}}=\frac{1-\Gamma_{d}}{K_{d}}.\label{ax:eq:kd}
\end{equation}
To see how $\Gamma_{d}$ relates to the conjecture update, suppose
AI agents expect aggregate diversion to be $X$. They take the corresponding
risk as given, and their optimal choices imply aggregate diversion
$F(X)$. When $F(X)>0$, the slope of this update is $F'(X)=e(X)K_{d}(F(X))$.
At an interior equilibrium, the AI agents’ conjecture is correct,
so $F(X)=X$. Both terms are then evaluated at $X$, giving $F'(X)=\Gamma_{d}(X)$.

To count the equilibria, recall that the curve $\Psi_{\sigma}(b)$
slopes down wherever $\Gamma_{d}(b)>1$. Can this happen over more
than one interval? We first differentiate $K_{d}$ with respect to
$b$.

\begin{equation}
K_{d}'=\frac{\sigma a^{2}-cS^{2}}{d^{2}},\qquad K_{d}''=-\frac{2\sigma c(A+b_{0})^{2}}{d^{3}}<0.\label{ax:eq:kdsecond}
\end{equation}
Thus $K_{d}$ is positive and strictly concave, so $(\log K_{d})''<0$.
With $\xi>0$ and $0<\rho<1$, $e'=-e(\xi-e)<0$ and $(\log e)''=e'<0$.
Hence $\log\Gamma_{d}$ is strictly concave: the set where $\Gamma_{d}>1$
is at most one interval. Also $\Gamma_{d}\to0$ as $b\uparrow A$.
The schedule can therefore rise, fall and rise, with some initial
segments absent.

A horizontal line crosses these monotone segments at most three times.
If the zero-diversion corner is admissible and the schedule initially
rises, the first segment lies above $\Psi_{\sigma}(0)\ge\psi$ and
supplies no positive root. If the schedule initially falls, only two
segments remain. There are therefore at most three distinct equilibria
including the corner. A tangency merges crossings rather than adding
one.

To show when multiplicity is possible, let $(b_{-},b_{+})$ be the
interval where $\Gamma_{d}>1$. If $b_{-}>0$, every weight

\[
\psi\in\bigl(\max\{\Psi_{\sigma}(b_{+}),\Psi_{\sigma}(0)\},\Psi_{\sigma}(b_{-})\bigr)
\]
has three positive crossings. If the decreasing interval starts at
zero, every $\psi\in(\Psi_{\sigma}(b_{+}),\Psi_{\sigma}(0))$ gives
two positive crossings and the corner. Thus, if $\Gamma_{d}$ exceeds
one for some $b$, multiplicity is possible for some weights, not
necessarily for the particular weight under consideration. If $\Gamma_{d}$
never exceeds one, the schedule is strictly increasing apart from
possible isolated stationary points, and equilibrium is unique.

\subsection{Proof of Proposition \ref{prop:one_welfare}}

\label{app:welfare}\label{ax:one-welfare}

For the logarithmic, positive-diversion allocation in the proposition,
Appendix \ref{app:model-background} gives the private and fixed-risk
returns:

\begin{equation}
\mu=\frac{\gamma}{a+(1-r)S},\qquad k^{0}=\frac{\gamma}{a+(1-\gamma)S}.\label{eq:one-proof-conditional}
\end{equation}
Since $0<\gamma<r<1$, we have $H<t$ and $k^{0}<\mu$, proving $\mathcal{P}<0$.
The capacity responses in Appendix \ref{ax:conditional} give $x^{0}>0$,
so $\Gamma<1$ gives $x>0$. Under the proposition's risk assumptions,
$H,e,q'>0$ and $\log\rho<0$, so \eqref{eq:ann-price-wedge} gives
$\mathcal{F}<0$ and $\mathcal{R}<0$. The private Euler equation
sets $\mathcal{E}=0$, hence $W_{A}<0$. A sufficiently small feasible
investment cut therefore raises welfare.

The restriction $\Gamma<1$ is substantive. On a branch with $\Gamma>1$,
$x<0$ and $k=(\gamma/a)(1-x)>\gamma/a>\mu$, so the total wedge is
positive. At $\Gamma=1$, $(1-\Gamma)x=x^{0}>0$ rules out a finite
derivative. At a strict zero-diversion corner the wedges vanish; at
onset one-sided derivatives apply. Constant or harmless risk removes
the risk terms, while $\ell=0$ removes the price term.

\subsection{Proof of Proposition \ref{prop:sector}}

\label{ax:sector}

Use the capacity responses in Appendix \ref{ax:conditional}, first
holding retention fixed. For $i\ne j$, $b_{i,A_{j}}^{0}=-a_{i}h_{i}\theta_{j}g_{j}/(1-H)\le0$,
strictly when sector $i$ diverts. If the expanding sector diverts,

\[
b_{j,A_{j}}^{0}=h_{j}(1-a_{j}k_{j}^{0})=\frac{h_{j}[1-\theta_{j}-\sum_{i\ne j}\theta_{i}h_{i}]}{1-H}>0.
\]
The numerator is positive because $h_{i}<1$ and $\Theta<1$. Expansion
therefore raises own-sector diversion but weakly lowers diversion
elsewhere. The price-discipline channel follows from

\begin{equation}
\mathrm{d}\log C_{i}=\frac{\mathrm{d}\log p_{i}-\ell\,\mathrm{d}\log w}{1-\ell}.\label{eq:sector-price-coefficient}
\end{equation}
At clearing prices, $C_{i}=\theta_{i}Y/(r_{i}a_{i}^{r_{i}})$. For
$i\ne j$, its fixed-risk derivative is $\partial\log C_{i}/\partial A_{j}=(1-r_{i}h_{i})k_{j}^{0}>0$:
the goods-price effect outweighs the wage effect on the other sector's
business return.

With endogenous risk, \eqref{eq:ax-sector-full-x} gives $x_{j}=x_{j}^{0}/(1-\Gamma)$.
On the branch in the proposition the denominator is positive, and
$x_{j}^{0}=(S_{j}-\sigma K\theta_{j})/(S_{j}+\sigma a_{j})$ when
the expanding sector diverts. Thus

\begin{equation}
\begin{aligned}x_{j}<0 & \iff\sigma K\theta_{j}>S_{j},\\
\sum_{j}a_{j}x_{j}^{0} & =N-K(\Theta-H)=K(1-\Theta)>0\quad\text{if some sector diverts}.
\end{aligned}
\label{eq:ax-sector-risk-threshold}
\end{equation}
The second line follows by multiplying $x_{j}^{0}=h_{j}-K\theta_{j}g_{j}$
by $a_{j}$, summing and using $a_{j}g_{j}=1-h_{j}$ and $N=K(1-H)$.
It rules out a negative aggregate-diversion effect in every sector.
A sector that strictly prefers no diversion instead has $x_{j}^{0}=-K\theta_{j}/A_{j}$.
Setting $\sigma=1$ proves the stated threshold and signs. At equality
the first derivative vanishes.

For comparison, changing a proxy weight adds no productive capacity.
Holding $A$ fixed in \eqref{eq:ax-sector-differential} gives $dX=h_{j}(a_{j}+K\theta_{j})d\log\psi_{j}$
at fixed retention. Imposing risk consistency yields

\begin{equation}
\frac{\partial X}{\partial\log\psi_{j}}=\frac{h_{j}(a_{j}+K\theta_{j})}{1-\Gamma}\ge0\qquad(\Gamma<1).\label{eq:psiderivative}
\end{equation}
The reversal for capacity is therefore not a general consequence of
making diversion more attractive.

\subsection{Conjecture updating, welfare ranking and proxy weights}

\label{ax:stability}

Consider the conjecture update

\begin{equation}
X_{n+1}=F(X_{n};A).\label{eq:update}
\end{equation}
Near a differentiable fixed point, $\delta_{n+1}=\Gamma^{*}\delta_{n}+o(|\delta_{n}|)$.
Since $\Gamma^{*}\ge0$, errors shrink if $\Gamma^{*}<1$ and grow
if $\Gamma^{*}>1$; equality is inconclusive. This describes the specified
update, rather than an economic adjustment process. The greatest fixed
point attracts when it is differentiable and $\Gamma^{*}\ne1$: $F(X)-X<0$
to its right, so the right derivative of this difference is nonpositive
at the fixed point. The feedback gain is therefore below one. A strict
zero-diversion corner also attracts; onset uses one-sided slopes.

Repeated updating converges from any initial conjecture. If $F(X_{0})\ge X_{0}$,
monotonicity makes the sequence nondecreasing; otherwise it is nonincreasing.
The sequence stays in $[0,A_{\Sigma}]$, so monotonicity gives a limit.
Continuity then implies that this limit is a fixed point. Different
starting conjectures may lead to different equilibria.

For welfare ranking, use $U_{2}(X)=\log Y(A,m(X))+q(X)\log\rho$,
which equals expected human log consumption at a fixed point. Equation
\eqref{ax:eq:conditionalderivatives} gives

\begin{equation}
U_{2}'(X)=-\frac{H}{1-H}e+q'\log\rho<0\qquad(\xi>0,\ 0<\rho<1).\label{ax:eq:welfareranking}
\end{equation}
Useful production cannot rise with the risk conjecture, while the
probability of loss strictly increases. Welfare is continuous when
a sector begins diverting, so the ranking also holds across these
thresholds: an equilibrium with more diversion gives humans lower
welfare. The numerical roots and slopes in Figure \ref{fig:one} are
reported in Appendix \ref{ax:numerics}.

\paragraph{Changes in the proxy weight.}

\label{ax:psi-comparative-statics}

Hold capacity and other primitives fixed, set $\sigma=1$ and write
$\psi_{0}=\Psi_{1}(0)$. Since $m(b)\ge\rho$, $b_{0}+b>b_{0}$ and
$(A-b)^{\gamma-1}>A^{\gamma-1}$, the inverse schedule satisfies $\Psi_{1}(b)>\rho\psi_{0}$
for all $b>0$. Hence zero diversion is the unique equilibrium for
$0\le\psi\le\rho\psi_{0}$. This is a sufficient interval, not necessarily
the exact onset of multiplicity.

At an interior equilibrium, \eqref{ax:eq:kd} gives $dX/d\log\psi=K/(1-\Gamma)$,
proving \eqref{eq:one-psi-response}. Combining this with $U_{2}=\log T+\gamma\log(A-X)+q(X)\log\rho$
yields

\begin{equation}
\frac{dU_{2}}{d\log\psi}=\left[-\frac{\gamma}{A-X}+q'(X)\log\rho\right]\frac{K}{1-\Gamma}<0\qquad(\Gamma<1).\label{eq:psi-operating-welfare}
\end{equation}
On a strict zero-diversion interval, allocation and welfare are unchanged.
To compare the equilibria with the least and most diversion, start
the update at zero and at $A$. These sequences converge to the lowest-
and highest-diversion equilibria, respectively. A higher proxy weight
weakly raises the map at every conjecture, so both limits weakly increase
with that weight and welfare at each weakly falls. Appendix \ref{ax:roots}
shows that risk feedback above one permits multiple equilibria for
intermediate weights. For large enough weights, the horizontal line
intersects only the final increasing part of the schedule, and diversion
approaches $A$. The thresholds for Figure \ref{fig:one-psi} are
documented in Appendix \ref{ax:numerics}.

\section{Proofs for the extensions}

\label{app:additional}\label{app:extension-proofs}

Unless stated otherwise, the assumptions and formulas in Appendix
\ref{app:model-background} apply. Each subsection states the assumptions
it changes.

\subsection{Substitution between goods}

\label{app:ces}Section \ref{sec:ces} asks when adding AI capacity
in devices encourages the software agent to divert less. To find the
elasticity of substitution at which this effect changes sign, we hold
common risk fixed and let wages and the prices of both goods adjust.

We replace the final producer's Cobb--Douglas technology by the CES
aggregator:
\begin{equation}
Y=\left[\sum_{i}\omega_{i}(y_{i}/\omega_{i})^{\alpha}\right]^{1/\alpha}.\label{ces:eq:final}
\end{equation}
Here $\alpha=(\epsilon-1)/\epsilon$, $0<\epsilon<\infty$, and the
limit $\epsilon=1$ gives \eqref{eq:finaltechnology}. We retain the
sectoral technologies and fixed aggregate labour supply of Section
\ref{sec:two}. Write $s_{i}=p_{i}y_{i}/Y$ and $\nu=(\epsilon-1)/[\ell+(1-\ell)\epsilon]$
for the expenditure share and the curvature of the reduced production
function.

\begin{proposition}[When substitution removes cross-sector restraint]\label{ces:prop:allocation}
First hold useful services in sector 2 fixed and clear goods and labour
markets as useful services in sector 1 decline. Sector-1 employment
rises if $\epsilon<1$, is unchanged at $\epsilon=1$, and falls if
$\epsilon>1$. Its relative goods price rises in every case. The own
marginal business return $G_{i}=\gamma_{i}s_{i}Y/a_{i}$ rises as
$a_{i}$ falls. For $i\ne j$, 
\begin{equation}
\frac{\partial G_{i}}{\partial a_{j}}=(1-\nu)\frac{G_{i}G_{j}}{Y}.\label{ces:eq:crossmain}
\end{equation}
Next hold expected retention fixed and let both delegates respond
to a marginal capacity expansion. Away from thresholds where diversion
starts or stops, expanding the other sector reduces an actively diverting
sector's diversion when $\epsilon<\epsilon_{\star}:=1+1/\ell$, and
increases it when $\epsilon>\epsilon_{\star}$. At equality this cross
derivative is zero. \end{proposition}

\emph{Proof.} For output $\bar{Y}>0$ and positive intermediate prices,
CES expenditure minimisation gives 
\begin{equation}
y_{i}=\omega_{i}(\lambda/p_{i})^{\epsilon}\bar{Y},\qquad\lambda=P(p)=[\sum_{i}\omega_{i}p_{i}^{1-\epsilon}]^{1/(1-\epsilon)}.\label{ces:eq:unitcost}
\end{equation}
 At $\epsilon=1$, use the Cobb-{}-Douglas limit. The numeraire condition
is $P=1$, and homogeneity gives zero final-producer profit. Hiring
and labour clearing then give $w\bar{L}=\ell Y$, $L_{i}=s_{i}\bar{L}$
and $D_{i}=(1-\ell)s_{i}Y$.

Substitute $y_{i}=z_{i}(s_{i}\bar{L})^{\ell}a_{i}^{\gamma_{i}}$ into
final demand: 
\[
s_{i}^{1-\ell\alpha}=\omega_{i}^{1-\alpha}(z_{i}a_{i}^{\gamma_{i}}\bar{L}^{\ell}/Y)^{\alpha}.
\]
 Put $\nu=\alpha/(1-\ell\alpha)$ and $V_{i}=z_{i}\omega_{i}^{\ell-1}a_{i}^{\gamma_{i}}$.
Normalising shares gives 
\begin{equation}
Y=\mathcal{Y}_{\epsilon}(a)=\bar{L}^{\ell}(\sum_{i}\omega_{i}V_{i}^{\nu})^{1/\nu},\qquad s_{i}=\frac{\omega_{i}V_{i}^{\nu}}{\sum_{j}\omega_{j}V_{j}^{\nu}}.\label{ces:eq:reduced-proof}
\end{equation}
 The limit at $\nu=0$ recovers the Cobb-{}-Douglas normalisation.
These labour assignments maximise feasible output: marginal labour
products are equal, and the objective is an increasing concave composition.
Mobile labour therefore weakly raises output relative to $L_{i}=\omega_{i}\bar{L}$
at the same useful services.

Writing $\theta_{i}^{\epsilon}=\gamma_{i}s_{i}$, logarithmic differentiation
gives 
\begin{equation}
\begin{aligned}d\log Y & =\sum_{i}\theta_{i}^{\epsilon}d\log a_{i},\\
d\log s_{i} & =\nu[\gamma_{i}d\log a_{i}-\sum_{j}\theta_{j}^{\epsilon}d\log a_{j}],\\
d\log(L_{1}/L_{2}) & =\nu(\gamma_{1}d\log a_{1}-\gamma_{2}d\log a_{2}),\\
d\log(p_{1}/p_{2}) & =-\frac{\gamma_{1}d\log a_{1}-\gamma_{2}d\log a_{2}}{\ell+(1-\ell)\epsilon}.
\end{aligned}
\label{ces:eq:sharederivatives}
\end{equation}
 Thus a fall in $a_{1}$ at fixed $a_{2}$ raises sector 1's relative
price. Labour moves towards that sector when $\epsilon<1$, away when
$\epsilon>1$, and remains unchanged at $\epsilon=1$.

The marginal business return at clearing prices is $G_{i}=\mathcal{Y}_{i}=\gamma_{i}s_{i}Y/a_{i}$.
Differentiating gives 
\begin{equation}
\frac{\partial\log G_{i}}{\partial\log a_{j}}=(1-\nu)\theta_{j}^{\epsilon}+\mathbf{1}_{i=j}(\nu\gamma_{i}-1).\label{ces:eq:Glog}
\end{equation}
 The own elasticity is negative because $\nu<1/(1-\ell)$ and $1-\gamma_{i}[\nu+(1-\nu)s_{i}]>1-r_{i}>0$.
The cross derivative is instead $\mathcal{Y}_{ij}=(1-\nu)G_{i}G_{j}/Y$,
changing sign at $\nu=1$, or $\epsilon_{\star}=1+1/\ell$.

\paragraph{Conditional equilibrium and capacity responses.}

Conditional uniqueness follows from the curvature of the reduced frontier:
\begin{equation}
-\mathcal{Y}_{aa}=D-(1-\nu)GG'/Y,\qquad D_{ii}=G_{i}(1-\nu\gamma_{i})/a_{i}>0.\label{ces:eq:hessian}
\end{equation}
 Here $\nu\gamma_{i}<r_{i}<1$. If $\nu\ge1$, the right side is positive
definite. If $\nu<1$, Cauchy-{}-Schwarz gives $(G'z)^{2}\le(G'D^{-1}G)(z'Dz)$,
and 
\[
(1-\nu)G'D^{-1}G/Y=(1-\nu)\sum_{i}\frac{s_{i}\gamma_{i}}{1-\nu\gamma_{i}}\le\frac{(1-\nu)\gamma_{\max}}{1-\nu\gamma_{\max}}<1.
\]
 The first inequality uses that $\gamma/(1-\nu\gamma)$ is increasing;
the last uses $\gamma_{\max}<1$. Thus $-\mathcal{Y}_{aa}$ is positive
definite.

The conditional potential is 
\[
\mathcal{Y}_{\epsilon}(a)+\frac{1}{m}\sum_{i}\psi_{i}v_{\sigma,i}(A_{i}-a_{i}),\qquad0\le a_{i}\le A_{i}.
\]
The frontier extends continuously and concavely to the closed box.
Strict concavity of the reward makes the potential strictly concave,
so it has a unique maximiser. Useful services are positive. If only
$a_{i}=0$, raising it to $t$ increases output at order $t^{\nu\gamma_{i}}$
for $\nu>0$, $t^{\gamma_{i}}$ for $\nu<0$, or $t^{\omega_{i}\gamma_{i}}$
for $\nu=0$. In each case the exponent is below one, so for sufficiently
small $t$ the gain in output outweighs the $O(t)$ loss in proxy
reward. If both sectors start with zero useful services, consider
putting a small share $t$ of each sector’s AI capacity into production.
Output then rises by at least a positive multiple of $t^{\gamma_{\max}}$.
Since $\gamma_{\max}<1$, this gain again outweighs the reward loss.
The maximiser therefore has positive useful AI services in both sectors.

At the maximiser, $mG_{i}=\psi_{i}b_{0i}^{\sigma-1}S_{i}^{-\sigma}$
if the agent diverts and $mG_{i}\ge\psi_{i}/b_{0i}$ if it does not.
These are precisely the conditions each AI agent faces at market-clearing
prices. The maximiser therefore gives the competitive allocation at
the assumed level of risk.

It remains to show that AI agents’ conjecture about risk can be correct.
Writing the choices as shares $a_{i}/A_{i}$ keeps their feasible
set fixed at $[0,1]^{2}$ as capacities change. Since the objective
is continuous and its maximiser is unique, the allocation varies continuously
with capacities, reward weights and expected retention. Hence the
diversion $F(X)$ implied by a conjecture $X$ is also continuous
in $X$. At $X=0$, we have $F(0)\ge0$. At the other extreme, $F(A_{1}+A_{2})<A_{1}+A_{2}$,
because both sectors use some capacity for production. By continuity,
$F(X)=X$ for at least one conjecture.

For local capacity responses, keep the active set fixed and define
\begin{equation}
\begin{aligned}R_{i} & =1-\nu\gamma_{i}+\sigma a_{i}/S_{i}, & \widetilde{h}_{i} & =1/R_{i}, & \widetilde{g}_{i} & =\sigma/(S_{i}R_{i}),\\
\widetilde{H} & =\sum_{i}\theta_{i}^{\epsilon}\widetilde{h}_{i}, & \Delta & =1-(1-\nu)\widetilde{H}>0, & \widetilde{K} & =\frac{\sum_{i}a_{i}\widetilde{h}_{i}}{\Delta}.
\end{aligned}
\label{ces:eq:coefficients}
\end{equation}
 At a strict no-diversion corner, set $\widetilde{h}_{i}=0$ and $\widetilde{g}_{i}=1/A_{i}$.
If $\nu<1$, the Hessian bound implies $(1-\nu)\widetilde{H}<1$:
larger denominators $R_{i}$ and omitted corner sectors reduce the
sum. If $\nu\ge1$, $\Delta\ge1$. Thus $\Delta>0$ in both cases.

Log-differentiating the active AI equality gives the first line below.
Multiply by $\theta_{i}^{\epsilon}$ and sum to obtain the output
response, then use $db_{i}=dA_{i}-da_{i}$: 
\begin{equation}
\begin{aligned}d\log a_{i} & =(1-\nu)\widetilde{h}_{i}d\log Y+\widetilde{g}_{i}dA_{i}+\widetilde{h}_{i}d\log m,\\
k_{j}^{0} & =\frac{\theta_{j}^{\epsilon}\widetilde{g}_{j}}{\Delta}, & b_{i,A_{j}}^{0} & =-(1-\nu)\frac{a_{i}\widetilde{h}_{i}\theta_{j}^{\epsilon}\widetilde{g}_{j}}{\Delta}\quad(i\ne j),\\
x_{j}^{0} & =1-a_{j}\widetilde{g}_{j}-(1-\nu)\widetilde{K}\theta_{j}^{\epsilon}\widetilde{g}_{j}.
\end{aligned}
\label{ces:eq:diff}
\end{equation}
For a diverting sector $i$, the cross response has the sign of $-(1-\nu)$.
This proves the second part of Proposition \ref{ces:prop:allocation}.

To sign the own response, restrict to the active set and put $M=\operatorname{diag}(\sigma G_{i}/S_{i})$
and $Q=-\mathcal{Y}_{aa}+M$. At fixed $m$, $Qa_{A_{j}}=M_{jj}e_{j}$.
Since $Q-M$ is positive definite, the eigenvalues of $M^{-1/2}QM^{-1/2}$
exceed one; those of its inverse lie in $(0,1)$. Its $j$th diagonal
element is $a_{j,A_{j}}^{0}=M_{jj}(Q^{-1})_{jj}$, proving $0<a_{j,A_{j}}^{0}<1$
and $b_{j,A_{j}}^{0}>0$. At a strict inactive corner, own diversion
remains zero.

Now let common risk respond. Holding capacities fixed in the differential
system gives $\partial a_{i}/\partial\log m=a_{i}\widetilde{h}_{i}/\Delta$,
so a fall in log expected retention raises aggregate diversion at
rate $\widetilde{K}$ and the update slope is $F'=e\widetilde{K}=\widetilde{\Gamma}\ge0$.
Here $\widetilde{\Gamma}$ is the CES feedback gain denoted by $\Gamma$
in Section \ref{sec:ces}. At a regular fixed point, 
\[
X_{A_{j}}=\frac{x_{j}^{0}}{1-\widetilde{\Gamma}},\qquad k_{j}=k_{j}^{0}-\frac{e\widetilde{H}}{\Delta}X_{A_{j}}.
\]
When $\nu\ge1$, the other-sector diversion responses in \eqref{ces:eq:diff}
are nonnegative. Adding them to the own response establishes $x_{j}^{0}\ge0$,
strictly when the expanded sector is active. On an attracting branch,
$1-\widetilde{\Gamma}>0$, so capacity expansion cannot reduce total
diversion or common-loss risk.

On an active branch, the private return remains $\mu_{j}=\theta_{j}^{\epsilon}\sigma/[\sigma a_{j}+(1-r_{j})S_{j}]$.
If $\nu\ge1$, then $\Delta\ge1$ and $\nu\gamma_{j}<r_{j}$, giving
\[
k_{j}^{0}=\frac{\theta_{j}^{\epsilon}\sigma}{\Delta[\sigma a_{j}+(1-\nu\gamma_{j})S_{j}]}<\mu_{j}.
\]
The inequality is weak at an inactive corner. The price wedge is therefore
nonpositive, and strictly negative when the expanded sector diverts.
The risk terms $\mathcal{F}_{j}=-\beta e\widetilde{H}X_{A_{j}}/\Delta$
and $\mathcal{R}_{j}=\beta q'\log\rho\,X_{A_{j}}$ are also nonpositive.
At a smooth interior investment choice in the unregulated economy,
the owner’s marginal benefit from extra AI capacity equals its marginal
cost. The three wedges then give the full welfare effect, so a small
expansion lowers welfare if the expanding sector diverts. Away from
this choice, we also need to account for any gap between the owner’s
marginal benefit and marginal cost.

\subsection{Reward curvature and multiplicity}

\label{ax:curvature}In Section \ref{sec:rewardcurvature}, we ask
whether several equilibria can coexist when the proxy score differs
from the logarithmic benchmark. We first compare diversion at a given
level of common risk and then allow risk to respond.

We use the normalised reward family \eqref{ax:eq:reward}, keeping
household utility logarithmic. Let $F(X;\sigma)$ be the diversion
map after conditional market clearing.

\begin{proposition}[How task-reward curvature changes diversion]\label{prop:crra}
Hold installed capacities and all other primitives fixed in the one-sector
or Cobb--Douglas two-sector economy. At a given risk conjecture,
increasing $\sigma$ weakly reduces diversion after prices clear:
$F(X;\sigma)$ is nonincreasing in $\sigma$. Follow a nearby equilibrium
on which the same sectors continue to divert, away from their start
or stop thresholds, with $\Gamma\ne1$ as in Section~\ref{sec:local-response}.
Then 
\begin{equation}
\begin{aligned}X_{\sigma} & =-\frac{\mathcal{Z}_{\sigma}}{1-\Gamma},\\
\mathcal{Z}_{\sigma} & :=-F_{\sigma}\ge0.
\end{aligned}
\label{eq:sigmacomp}
\end{equation}
Whenever $\mathcal{Z}_{\sigma}>0$, equilibrium diversion falls with
$\sigma$ if $\Gamma<1$, and rises if $\Gamma>1$. The greatest fixed
point weakly increases when $\sigma$ falls, including across changes
in the number of roots. There is no general ordering of the slope
$F_{X}$ at those moving equilibria. \end{proposition}

\emph{Proof.} At fixed $Y,m$ and capacities, the active useful-service
equation is 
\[
\frac{a_{i}}{(S_{i}/b_{0i})^{\sigma}}=\frac{m\theta_{i}Yb_{0i}}{\psi_{i}},\qquad S_{i}=b_{0i}+A_{i}-a_{i}.
\]
Its left side increases in $a_{i}$ and weakly decreases in $\sigma$,
since $S_{i}/b_{0i}\ge1$. Raising $\sigma$ therefore raises the
useful-service response, including after the cap $a_{i}\le A_{i}$
is imposed. The conditional output map rises, and so does its unique
positive crossing (Appendix \ref{ax:conditional}). Each useful-service
response rises with output. Thus $F(X;\sigma)$ weakly falls, including
at corners.

On a smooth active set, put $l_{i}^{b}=\log(S_{i}/b_{0i})$. Since
$\partial_{\sigma}\log v_{i}'=-l_{i}^{b}$, equation \eqref{eq:ax-sector-differential}
gives $d\log a_{i}=h_{i}(d\log Y+l_{i}^{b}d\sigma)$ at fixed capacity
and $m$. Aggregate this equation and differentiate $X=F(X;\sigma)$:
\begin{equation}
\begin{aligned}J_{\sigma}^{Y} & =\left.\frac{\partial\log Y}{\partial\sigma}\right|_{m}=\frac{\sum_{i}\theta_{i}h_{i}l_{i}^{b}}{1-H},\\
F_{\sigma} & =-\sum_{i}a_{i}h_{i}(l_{i}^{b}+J_{\sigma}^{Y})=:-\mathcal{Z}_{\sigma}\le0,\\
X_{\sigma} & =-\mathcal{Z}_{\sigma}/(1-\Gamma).
\end{aligned}
\label{ax:eq:sigmaderivatives}
\end{equation}
 Nonnegativity of the terms in $\mathcal{Z}_{\sigma}$ gives the stated
local signs for $\Gamma\ne1$.

To compare the greatest roots, suppose we lower $\sigma$. At every
conjecture, the new map gives at least as much diversion. Start the
iteration at the old greatest root: diversion then rises weakly and
remains bounded by total capacity. By continuity, its limit is a root
of the new map, so the greatest new root cannot be lower. This remains
true if the number of roots changes. We cannot make the same comparison
for $F_{X}=eK$, because both factors change as the equilibrium moves.

The multiplicity argument in Appendix \ref{ax:roots} also applies
to this reward family.

\subsection{Hiring before diversion}

\label{app:timing}In Section \ref{sec:timing}, the owner chooses
how many workers to hire before the AI agent allocates capacity. It
cannot revise its hiring afterwards, so it anticipates the agent's
response. We show why this changes the one-sector investment comparison
even though the same operating allocations are possible at a given
capacity.

For the normalised reward, put $S_{i}=b_{0i}+b_{i}$ and $h_{i}=S_{i}/(S_{i}+\sigma a_{i})$
when diversion is positive, with $h_{i}=0$ at a strict no-diversion
corner.

\begin{proposition}[Hiring before diversion]\label{prop:timing}
At fixed positive prices and aggregate risk, an owner who hires before
diversion, anticipating the delegate's response, has a unique positive
optimal workforce. In one sector with fixed aggregate labour and capacity,
the set of operating allocations $(a,b,Y,X)$ coincides with the benchmark's
set. At an active allocation the wage is higher and the human operating
dividend lower. With Cobb--Douglas final demand and two sectors,
away from thresholds at which a sector starts or stops diverting,
employment satisfies 
\begin{equation}
\frac{L_{i}}{\bar{L}}=\frac{\omega_{i}/(1-\gamma_{i}h_{i})}{\sum_{j}\omega_{j}/(1-\gamma_{j}h_{j})}.\label{eq:ext-prelabor-shares}
\end{equation}
Differences in the allocation response to committed labour can therefore
produce sectoral reallocation even with Cobb--Douglas final demand.
\end{proposition}

\emph{Proof.} Write $R(L,a)=pzL^{\ell}a^{\gamma}$. At fixed prices
and risk, the agent maximises $m[R(L,A-b)-wL]+\psi v_{\sigma}(b)$
over $0\le b\le A$. The committed wage bill drops out of its marginal
choice. For $L,A>0$, strict concavity and the infinite marginal business
return at $a=0$ give a unique optimum with $a>0$. The active condition
is $m\gamma pzL^{\ell}a^{\gamma-1}=\psi b_{0}^{\sigma-1}S^{-\sigma}$;
diversion is zero when $\psi/b_{0}\le m\gamma pzL^{\ell}A^{\gamma-1}$.
At $L=0$ all capacity is diverted; at $A=0$ the allocation is zero.

On an active branch, logarithmic differentiation of the AI equality,
with $S=A+b_{0}-a$, gives 
\[
\ell\,d\log L+(\gamma-1)d\log a=-\sigma(dA-da)/S.
\]
Collecting the changes in useful services and putting $Q=(1-\gamma)S+\sigma a$
yields 
\begin{equation}
a_{A}=\frac{\sigma a}{Q}>0,\qquad\eta_{L}:=\frac{La_{L}}{a}=\frac{\ell S}{Q}>0,\qquad\Lambda_{L}:=\frac{LR_{L}}{R}=\ell+\gamma\eta_{L}=\frac{\ell}{1-\gamma h},\label{eq:timing-ai-responses}
\end{equation}
where $h=S/(S+\sigma a)$. Here and below, $R_{L}$ denotes the total
derivative of $R(L,a(L;A))$. At a strict zero-diversion corner $a=A$,
so $\eta_{L}=h=0$ and $\Lambda_{L}=\ell$. The extra return to labour
in the active case comes from its effect on the allocation left by
the AI.

The owner maximises $P(L;A)=R(L,a(L;A))-wL$ over $L\ge0$. A smooth
positive optimum satisfies $wL=\Lambda_{L}py$. For global sufficiency,
put $C_{A}=A+b_{0}$. On an active branch, 
\[
\frac{d\eta_{L}}{da}=-\frac{\ell\sigma C_{A}}{Q^{2}}<0,\qquad R_{LL}=\frac{R}{L^{2}}\left\{ \Lambda_{L}(\Lambda_{L}-1)+\gamma a\eta_{L}\frac{d\eta_{L}}{da}\right\} <0.
\]
Both terms are negative because $0<\Lambda_{L}<\ell/(1-\gamma)<1$.
Revenue is also strictly concave on the zero-diversion branch. At
the transition, $a$ is continuous and $\Lambda_{L}$ drops to $\ell$,
so $R_{L}$ jumps down. Concavity therefore holds across the kink,
where optimality requires 
\[
R_{L}(L_{c}+)\le w\le R_{L}(L_{c}-),\qquad L_{c}=\left[\frac{\psi A^{1-\gamma}}{m\gamma pzb_{0}}\right]^{1/\ell}.
\]
As $L\downarrow0$, the active AI equality gives $a\sim c_{A}L^{\ell/(1-\gamma)}$
for a positive constant $c_{A}$. Since $\ell/(1-\gamma)<1$, induced
revenue has infinite right marginal return at zero. For large $L$,
$P(L;A)\le pzA^{\gamma}L^{\ell}-wL\to-\infty$. Strict concavity therefore
gives a unique positive optimal workforce and positive maximised profit
for every $A>0$.

In one sector, $L=\bar{L}$ and $Y=Ta^{\gamma}$. The AI condition
becomes the benchmark's $m\gamma Ta^{\gamma-1}=\psi b_{0}^{\sigma-1}S^{-\sigma}$,
with the same zero-diversion inequality. Every benchmark operating
allocation is supported by 
\begin{equation}
w=\frac{\ell Y}{\bar{L}(1-\gamma h)},\qquad D=Y\left[1-\frac{\ell}{1-\gamma h}\right]>0.\label{eq:timing-one-income}
\end{equation}
At the diversion threshold, any wage between the two marginal revenues
supports hiring, with the benchmark wage as the lower endpoint. Thus
the operating allocation sets coincide, while active allocations have
higher wages and lower dividends under labour commitment.

The investment return follows from the owner's hiring problem. Let
$D(A;p,w,m)=\max_{L\geq0}\{R(L,a(L;A))-wL\}$ denote optimised pre-loss
dividends. In the private calculation, prices, common risk and the
state-price valuation $V$ are fixed. At a smooth positive hiring
optimum, the envelope condition gives $D_{A}^{\mathrm{priv}}=R_{a}a_{A}$.
Using \eqref{eq:timing-ai-responses} and evaluating at one-sector
market clearing, where $R=Y$, yields
\begin{equation}
\mu_{\mathrm{hire}}=\frac{D_{A}^{\mathrm{priv}}}{Y}=\frac{\gamma\sigma}{\sigma a+(1-\gamma)S}=k^{0},\qquad\mathcal{P}=\beta(k^{0}-\mu_{\mathrm{hire}})=0.\label{eq:timing-private-return}
\end{equation}
The equality with $k^{0}$ follows because aggregate labour is fixed:
the coordinated fixed-risk response uses the same derivative of useful
capacity as the envelope calculation. It holds for the general reward
curvature $\sigma>0$; setting $\sigma=1$ gives \eqref{eq:timing-private-return-main}.

The valuation remains $V=\beta c_{1}/Y$: household consumption is
still $Y$ or $\rho Y$, irrespective of the split between wages and
dividends. At a smooth interior competitive investment equilibrium,
the private Euler equation is therefore $\kappa A/c_{1}=\beta\mu_{\mathrm{hire}}$.
Since the one-sector operating allocation is unchanged, its capacity
and risk responses are unchanged too, giving
\begin{equation}
W_{A}=\beta(k-k^{0})+\beta q^{\prime}(X)\log\rho\,x=\mathcal{F}+\mathcal{R}.\label{eq:timing-welfare}
\end{equation}
With positive diversion, $\xi>0$, $0<\rho<1$ and $\Gamma<1$, both
terms are negative, so a sufficiently small capacity cut raises welfare.
If risk is fixed, both terms vanish. At a strict no-diversion corner,
$\mu_{\mathrm{hire}}=k^{0}=\gamma/A$ and all wedges vanish; at diversion
thresholds, one-sided derivatives apply.

These comparisons assume an optimal private investment choice. Concavity
in hiring does not establish concavity in capacity: a solution to
the investment first-order condition must also be checked against
other feasible choices. With fixed risk, the equality of marginal
returns does not remove the underlying diversion problem.

With two sectors, Cobb-{}-Douglas final demand fixes revenue shares
at $\omega_{i}$. Summing $wL_{i}=\ell\omega_{i}Y/(1-\gamma_{i}h_{i})$
and clearing the labour market gives the employment shares in \eqref{eq:ext-prelabor-shares}.
Dividends are
\begin{equation}
D_{i}=(1-\Lambda_{L,i})\omega_{i}Y.\label{eq:timing-two-clearing}
\end{equation}
 The different dividends and private capacity responses can still
change investment and welfare.

\subsection{Returns to scale and automation}

\label{app:returns}In Section \ref{sec:returns}, prices still discourage
diversion under constant returns. Here we show why, and then ask how
diversion changes when AI takes over some of the tasks previously
done by labour.

At constant returns in variable inputs, $\gamma_{i}=1-\ell$, the
private dividend is $C_{i}a_{i}$ at fixed prices. The agent maximises
$mC_{i}(A_{i}-b_{i})+\psi_{i}v_{\sigma,i}(b_{i})$. Strict concavity
of the reward still gives a unique choice. The KKT conditions \eqref{ax:eq:aikkt},
with $r_{i}=1$, imply

\begin{equation}
d_{i}=\left[(\psi_{i}b_{0i}^{\sigma-1}/(mC_{i}))^{1/\sigma}-b_{0i}\right]_{+},\qquad b_{i}=\min\{A_{i},d_{i}\},\qquad a_{i}=(A_{i}-d_{i})_{+}.\label{eq:returns-diversion-target}
\end{equation}
 Thus $D_{i}(A_{i})=C_{i}(A_{i}-d_{i})_{+}$ for every private capacity
choice. Positive production requires $A_{i}>d_{i}$.

At positive output, clearing prices still give $\Pi_{i,a}=\theta_{i}Y/a_{i}$
with $\Theta=1-\ell<1$. The conditional system \eqref{ax:eq:implicitservice}
and its existence and contraction arguments therefore continue to
apply. The one-sector gain is 
\begin{equation}
K=\frac{aS}{\sigma a+\ell S}<K_{\mathrm{FP}}=\frac{S}{\sigma}.\label{eq:returns-crs-gain}
\end{equation}
 Allowing wages to adjust still restrains diversion despite the linear
private business return. Substituting $1-\gamma=\ell$ into the one-sector
root proof also retains its bound of three operating equilibria at
fixed capacity.

In one sector, the clearing-price marginal return itself is:
\begin{equation}
C_{{\rm GE}}=\gamma Ta^{\gamma-1}=\gamma Ta^{-\ell}.\label{eq:ext-crs-marginal}
\end{equation}
Thus $1-\gamma=\ell>0$ remains the relevant aggregate output curvature.

The dividend has an upward kink at $A_{i}=d_{i}$, so investment must
be compared with the zero choice. At fixed prices, risk and valuation
$V$, put $M_{i}=VC_{i}/\kappa_{i}$. The objective is $J_{i}(A_{i})=VC_{i}(A_{i}-d_{i})_{+}-\kappa_{i}A_{i}^{2}/2$.
Below the threshold, its best value is zero at $A_{i}=0$. Above it,
the stationary choice $A_{i}=M_{i}$ has value $\kappa_{i}M_{i}(M_{i}/2-d_{i})$
when feasible. Thus, for $d_{i}>0$,

\begin{equation}
\arg\max J_{i}=\begin{cases}
\{0\}, & M_{i}<2d_{i},\\
\{0,M_{i}\}, & M_{i}=2d_{i},\\
\{M_{i}\}, & M_{i}>2d_{i}.
\end{cases}\label{eq:returns-global-entry}
\end{equation}
For $d_{i}=0$, the optimum is $M_{i}>0$. At a productive symmetric
candidate, the Euler condition gives $A_{i}=M_{i}$ and $b_{i}=d_{i}$.
Global private optimality therefore also requires $A_{i}\ge2b_{i}$,
allowing indifference to zero investment at equality. This condition
applies before installation costs are sunk.

\subsubsection{Automation and the labour share}

\label{app:automation}

We return to one sector and the logarithmic proxy score. Fix $A$,
$\bar{L}$, $\psi$, $b_{0}$, $\xi$ and $\rho$, keeping physical
and score units unchanged. Let $\nu\in(0,1]$ be total returns to
the variable inputs, with $0<\ell<\nu$ and $\gamma=\nu-\ell$. The
labour and AI task fractions are now $\ell/\nu$ and $\gamma/\nu$.
We write production as
\begin{equation}
Y=Z\left(\frac{\nu\bar{L}}{\ell}\right)^{\ell}\left(\frac{\nu a}{\gamma}\right)^{\gamma},\qquad a=A-b,\qquad\gamma=\nu-\ell.\label{eq:automation-drs-technology}
\end{equation}
Setting $\nu=1$ gives \eqref{eq:automation-crs-technology}. When
$\nu<1$, a fixed organisational input with exponent $1-\nu$ accounts
for the remaining returns. We hold $Z$ and $\nu$ fixed as $\ell$
changes, and stay in the labour-scarce region $a\ell/(\bar{L}\gamma)>1$.

With positive diversion, $m(b)C_{{\rm GE}}(b_{0}+b)=\psi$, where
$C_{{\rm GE}}=\gamma Y/a$. Put $S=b_{0}+b$ and $f(b,\ell)=\log m(b)+\log C_{{\rm GE}}(A-b,\ell)+\log S-\log\psi$.
Its derivatives are 
\begin{equation}
\begin{aligned}f_{b} & =\frac{1-\gamma}{a}+\frac{1}{S}-e=\frac{1-\Gamma}{K}, & K & =\frac{aS}{a+(1-\gamma)S},\\
f_{\ell} & =-\frac{1}{\gamma}-\log\frac{a\ell}{\bar{L}\gamma}, & \Gamma & =eK.
\end{aligned}
\label{eq:automation-implicit-derivatives}
\end{equation}
 On a regular branch with $\Gamma<1$, 
\begin{equation}
\left.X_{\ell}\right|_{A}=\left.b_{\ell}\right|_{A}=\frac{K}{1-\Gamma}\left[\frac{1}{\gamma}+\log\frac{a\ell}{\bar{L}\gamma}\right]>0.\label{eq:automation-diversion-response}
\end{equation}
 More automation lowers $\ell$, so diversion falls: useful work becomes
more rewarding at the original allocation. The curvature $1-\gamma=1-\nu+\ell$
also falls, leaving the positive component $1-\nu$ under decreasing
returns.

For CRS, $r=1$ and \eqref{eq:returns-crs-gain} gives $K=aS/(a+\ell S)$
and $K_{\mathrm{FP}}=S$ with a logarithmic score. The proportional
reduction relative to the fixed-wage response is $1-K/K_{\mathrm{FP}}=\ell S/(a+\ell S)$.
At a given allocation, this reduction becomes smaller as $\ell$ falls.
Thus weaker wage restraint and lower equilibrium diversion can coexist.

Output and the wage satisfy
\begin{equation}
\frac{d\log Y}{d\ell}=-\log\frac{a\ell}{\bar{L}\gamma}-\frac{\gamma}{a}\left.X_{\ell}\right|_{A}<0,\qquad\frac{d\log w}{d\ell}=\frac{1}{\ell}+\frac{d\log Y}{d\ell}.\label{eq:automation-output-wage}
\end{equation}
Automation therefore raises output and lowers the common-loss probability,
while its effect on wages is ambiguous. At fixed installed capacity,
aggregate human welfare rises: $W_{\ell}=\beta[d\log Y/d\ell+q'(X)\log\rho\left.X_{\ell}\right|_{A}]<0$.
This aggregate gain need not be shared by workers who hold no claims
on dividends.

This differs from changing only $\ell$ at fixed $\gamma$ and $T$.
Neither $Y=Ta^{\gamma}$ nor the allocation condition $m(b)\gamma T(A-b)^{\gamma-1}(b_{0}+b)=\psi$
then changes. The operating equilibrium set, output and common risk
are unchanged, although wages and dividends move. Fixing raw $z$
instead would also change $T=z\bar{L}^{\ell}$.

These local comparisons assume that workers remain in tasks that AI
cannot perform and that no agent starts or stops diverting. If task
assignments change, we need to derive production and check the global
investment condition under the new technology.

If installed capacity also responds, its effect adds to the fixed-capacity
comparison. In one sector with $\Gamma<1$, extra capacity raises
diversion by \eqref{eq:ax-sector-full-x}, so an expansion following
automation can offset the decline. The parameters and checks for Table
\ref{tab:automation-crs} are in Appendix \ref{ax:automation-numerics}.

\subsection{AI accumulation and aggregate risk}

\label{ax:ai-accumulation}

In the example in Section \ref{sec:ai-accumulation}, a loss today
leaves households with less income to invest in AI capacity for tomorrow.
Here we derive how much they save and how the resulting stock affects
diversion and risk.

We keep productivity $Z>0$ and the proxy weight $\psi>0$ constant,
and let $0<\gamma=1-\ell<1$. There is a unit mass of identical, atomistic
households. They own AI capacity, choose labour before the common
event, and choose consumption and saving afterwards. Their preferences
and pre-loss production are

\begin{equation}
\begin{aligned}\mathbb{E}_{0}\sum_{t=0}^{\infty}\beta^{t}\left[\log C_{t}-\frac{\chi L_{t}^{1+\phi}}{1+\phi}\right], & \qquad0<\beta<1,\quad\chi>0,\quad\phi\geq0,\\
Y_{t}=Z\left(\frac{L_{t}}{\ell}\right)^{\ell}\left(\frac{(1-\vartheta_{t})A_{t}}{\gamma}\right)^{\gamma}, & \qquad X_{t}=\vartheta_{t}A_{t}.
\end{aligned}
\label{ax:eq:dynamic-primitives}
\end{equation}
Here $\vartheta_{t}\in[0,1)$ is the share of capacity diverted. As
in the benchmark, $q(X)=1-e^{-\xi X}$ and $m(X)=\rho+(1-\rho)e^{-\xi X}$,
with $0<\rho<1$ and $\xi\geq0$. The event multiplies current output
and both factor payments by $d_{t}\in\{1,\rho\}$. With full depreciation
and unit installation cost, resources satisfy $C_{t}+A_{t+1}=d_{t}Y_{t}$.
We allow households to save only through nonnegative AI holdings;
there are no bonds or other assets.

Each installed unit has its own delegate and proxy task. After the
delegate allocates capacity, the owner hires labour at wage $w_{t}$.
Optimised operating profit is linear in useful services, with $\Pi_{a,t}=Z^{1/\gamma}w_{t}^{-\ell/\gamma}$.
The delegate maximises $m(X_{t})\Pi_{a,t}(1-\vartheta)+\psi\log(1+\vartheta/\bar{b}_{0})$
taking wages and aggregate risk as given. Here $\bar{b}_{0}>0$ is
the proxy scale for one installed unit. Adding capacity therefore
adds delegates and proxy tasks. The aggregate proxy reward is the
reward per unit multiplied by the number of units. In the benchmark,
each firm keeps the same proxy task and reward function as its capacity
changes. At an interior allocation, labour clearing and the delegate
condition give

\begin{equation}
w_{t}=\frac{\ell Y_{t}}{L_{t}},\qquad R_{t}^{A}=\frac{\gamma Y_{t}}{A_{t}},\qquad\frac{\psi}{\bar{b}_{0}+\vartheta_{t}}=m(X_{t})\frac{\gamma Y_{t}}{A_{t}(1-\vartheta_{t})}.\label{ax:eq:dynamic-operating}
\end{equation}
$R_{t}^{A}$ is the owner's marginal pre-loss dividend return. At
fixed wages and aggregate risk, a new unit has the same diversion
share as existing units, so its return is $\Pi_{a,t}(1-\vartheta_{t})=\gamma Y_{t}/A_{t}$.
We write this as a rental on installed capacity: it is the operating
surplus after paying labour, with no additional dividend. Its realised
return is $d_{t}R_{t}^{A}$. At zero diversion the last equality becomes
$\psi/\bar{b}_{0}\leq m(0)\gamma Y_{t}/A_{t}$.

We first check household choices. For the policies in \eqref{eq:dynamic-policies},
the labour condition gives a constant level of employment, $L^{*}=[\ell/\{\chi(1-\beta\gamma)\}]^{1/(1+\phi)}$.
The saving condition holds in every next-period outcome:

\begin{equation}
\chi(L^{*})^{\phi}=\mathbb{E}_{t}^{\mathrm{pre}}\!\left[\frac{d_{t}w_{t}}{C_{t}}\right]=\frac{\ell}{(1-\beta\gamma)L^{*}},\qquad\beta\frac{C_{t}}{C_{t+1}}d_{t+1}R_{t+1}^{A}=\frac{\beta\gamma d_{t}Y_{t}}{A_{t+1}}=1.\label{ax:eq:dynamic-euler}
\end{equation}
Labour is chosen before the event. Taking the post-event expectation
of the state-by-state saving identity gives the Euler equation. The
cancellation uses log utility, full depreciation and a constant AI
share.

These conditions also establish the household's global optimum. Household
$h$ faces the linear budget $c_{ht}+A_{h,t+1}=d_{t}(w_{t}L_{ht}+R_{t}^{A}A_{ht})$,
with $A_{h,t+1}\geq0$. We compare the proposed plan with any other
feasible plan, taking wages, the rental rate on AI capacity, aggregate
risk and the law of motion for aggregate capacity as given. Since
utility is concave, we can bound the gain in each period using the
marginal utility of consumption and the marginal disutility of labour
under the proposed plan. We discount these bounds, sum them through
date $N$ and take expectations. The labour condition and Euler equation
cancel the intermediate terms. If the alternative starts with the
same assets, the remaining bound is at most

\begin{equation}
\mathbb{E}_{0}\!\left[\frac{\beta^{N}A_{N+1}}{C_{N}}\right]=\beta^{N}\frac{\beta\gamma}{1-\beta\gamma}\longrightarrow0.\label{ax:eq:dynamic-terminal}
\end{equation}
The alternative plan’s terminal assets enter with a minus sign. Since
asset holdings must be nonnegative, dropping this term gives the upper
bound above. As $N$ increases, the bound tends to zero, so no feasible
plan with well-defined expected utility can improve on the proposed
choice. 

It remains to determine AI diversion at each inherited stock. Put
$T=Z(L^{*}/\ell)^{\ell}\gamma^{-\gamma}$, so $Y=T[A(1-\vartheta)]^{\gamma}$.
The interior operating condition becomes

\begin{equation}
\psi=\Psi_{A}(\vartheta):=\gamma TA^{-\ell}m(A\vartheta)(\bar{b}_{0}+\vartheta)(1-\vartheta)^{-\ell}.\label{ax:eq:dynamic-schedule}
\end{equation}
Zero diversion is an equilibrium when $\psi\leq\Psi_{A}(0)$; otherwise
continuity and $\Psi_{A}(\vartheta)\to\infty$ as $\vartheta\to1$
give an interior root. With $e(X)=-m^{\prime}(X)/m(X)$ as before,
the slope is

\begin{equation}
\mathcal{J}_{A}(\vartheta):=\frac{\partial\log\Psi_{A}}{\partial\vartheta}=\frac{1}{\bar{b}_{0}+\vartheta}+\frac{\ell}{1-\vartheta}-Ae(A\vartheta).\label{ax:eq:dynamic-slope}
\end{equation}
The first term reflects declining marginal proxy rewards. The second
captures price discipline: more diversion lowers the wage and raises
the marginal business return to useful AI. The last term is aggregate-risk
feedback. Since $e(X)\leq\xi(1-\rho)$ and $1/(\bar{b}_{0}+\vartheta)+\ell/(1-\vartheta)\geq(1+\sqrt{\ell})^{2}/(1+\bar{b}_{0})$,
the operating equilibrium is unique whenever

\begin{equation}
\xi A(1-\rho)(1+\bar{b}_{0})<(1+\sqrt{\ell})^{2}.\label{ax:eq:dynamic-uniqueness}
\end{equation}
Suppose the bound holds for every capital stock in a positive compact
interval, and both transitions $A^{\prime}=\beta\gamma dY(A)$ keep
capital in that interval. The operating allocation is unique and continuous
there. We have already checked household optimality and market clearing,
so the policies define a recursive competitive equilibrium on the
interval. In Appendix \ref{ax:accumulation-numerics} we check these
conditions for the numerical example. Its operating equilibrium is
unique, so we do not need an equilibrium selection rule.

Finally, at an interior root with $\mathcal{J}_{A}>0$, differentiating
\eqref{ax:eq:dynamic-schedule} with respect to installed capacity
gives

\begin{equation}
\frac{d\vartheta}{d\log A}=\frac{\ell+A\vartheta e(A\vartheta)}{\mathcal{J}_{A}}>0,\qquad\frac{dX}{dA}=\vartheta+\frac{d\vartheta}{d\log A}>0.\label{ax:eq:dynamic-capacity}
\end{equation}
Following a loss, households save less and enter the next period with
less AI capacity. On the branch just considered, both the diversion
share and total diversion therefore fall. If $\xi>0$, the probability
of another loss falls as well. At a strict zero-diversion allocation,
the local risk response is zero. This calculation follows how operating
choices change along the equilibrium path; a dynamic welfare or tax
result would require a separate analysis.

\subsection{Strategic AI control}

\label{ax:concentrated-control}

The controller in Section \ref{sec:concentrated-control} serves several
firms. To see why it might divert more or less than independent agents
would, we work out how its decision affects the wages its clients
pay and the risk to their profits.

We keep installed capacity fixed and divide the unit mass of firms
into $N\geq1$ equal blocks, with $\alpha=1/N$. Controller $j$ chooses
the same useful allocation $a_{j}$ at each firm in its block; $b_{j}=A-a_{j}$.
Controllers choose simultaneously, and owners then hire labour competitively,
as in the benchmark. The controller maximises the sum of its clients'
criteria. Controller assignments are fixed during the execution period.

\paragraph{Competitive production and the controller's return.}

Let $r=\gamma/(1-\ell)\in(0,1)$ and $Z=\sum_{j}\alpha a_{j}^{r}$.
The hiring condition and labour clearing give
\begin{equation}
\begin{aligned}L_{j} & =\frac{\bar{L}a_{j}^{r}}{Z}, & Y & =TZ^{1-\ell},\\
w & =\frac{\ell Y}{\bar{L}}, & D_{j} & =(1-\ell)T\alpha a_{j}^{r}Z^{-\ell}.
\end{aligned}
\label{eq:concentration-production}
\end{equation}
Here $L_{j}$ is labour per firm and $D_{j}$ is total pre-loss profit
in block $j$. Thus $\sum_{j}D_{j}=(1-\ell)Y$. Aggregate diversion
is $X=\sum_{j}\alpha b_{j}$, and the controller chooses $a_{j}\in[0,A]$
to maximise
\begin{equation}
J_{j}=m(X)D_{j}+\alpha\psi\log(1+b_{j}/b_{0}).\label{eq:concentration-objective}
\end{equation}
With $\zeta_{j}=\alpha a_{j}^{r}/Z$ and $e=-m'/m$, differentiation
gives
\begin{equation}
\begin{aligned}\frac{D_{j,a_{j}}}{D_{j}} & =\frac{r}{a_{j}}(1-\ell\zeta_{j}),\\
J_{j,a_{j}} & =m\bigl[D_{j,a_{j}}+\alpha eD_{j}\bigr]-\frac{\alpha\psi}{b_{0}+b_{j}}.
\end{aligned}
\label{eq:concentration-derivative}
\end{equation}
At symmetry, $\zeta_{j}=\alpha$, $X=b$ and $Y=Ta^{\gamma}$, giving
\eqref{eq:concentration-foc}. An interior choice requires $J_{j,a_{j}}=0$;
at $a_{j}=A$ the one-sided condition is $J_{j,a_{j}}\geq0$. To see
why full diversion cannot be optimal, consider giving a small amount
of capacity to useful work. Profit before losses then rises by a positive
multiple of $a_{j}^{r}$ if another block produces, or $a_{j}^{\gamma}$
if none does. Since $m\geq\rho>0$, the firms expect to keep at least
a fraction $\rho$ of this extra profit. The proxy reward falls roughly
in proportion to the amount of capacity moved. Since both profit exponents
are below one, a small enough shift into useful AI work earns the
controller more in expected profit than it gives up in proxy reward.

\paragraph{Fixed common risk.}

Hold $m>0$ fixed. To establish global optimality, put $R=\sum_{k\neq j}\alpha a_{k}^{r}$
and $t=\alpha a_{j}^{r}$. Apart from a positive constant, block profit
is $f(t)=t(R+t)^{-\ell}$. Differentiation gives $f'(t)>0$ and $f''(t)<0$
for $t>0$. Since $a_{j}^{r}$ is strictly concave, so is profit in
$a_{j}$. The logarithmic proxy term is concave too. Each controller
therefore has a unique best response, characterised by the conditions
above. Equal allocations within a block also follow from concavity:
at a given mean allocation, equality maximises both the block's integral
of $a^{r}$ and its total proxy score. Block profit rises with the
former, while aggregate diversion and risk are unchanged.

All pure equilibria must be symmetric. At an equilibrium, $m$ and
$Z$ are common across blocks. Multiplying an interior condition by
$b_{0}+b_{j}$ gives
\begin{equation}
a_{j}^{r-1}\left(1-\ell\frac{\alpha a_{j}^{r}}{Z}\right)(A+b_{0}-a_{j})=\frac{\psi}{m\gamma TZ^{-\ell}}.\label{eq:concentration-symmetry}
\end{equation}
The left-hand side is strictly decreasing in $a_{j}$ over the relevant
range: its three factors are positive and decreasing. Hence two interior
allocations cannot differ. If one controller chooses $a_{j}=A$, its
one-sided condition requires the left-hand side there to be at least
the common right-hand side. A smaller interior allocation makes it
strictly larger, contradicting equality. An equilibrium cannot mix
a no-diversion corner with an interior choice.

The symmetric interior condition is
\begin{equation}
\psi=m\gamma T(1-\ell\alpha)(b_{0}+b)(A-b)^{\gamma-1}.\label{eq:concentration-fixed-risk}
\end{equation}
Its right-hand side strictly increases with $b$ and tends to infinity
as $b\uparrow A$. If $\psi\leq m\gamma T(1-\ell\alpha)b_{0}A^{\gamma-1}$,
zero diversion is the unique candidate; otherwise there is exactly
one positive solution. At this allocation, each controller is doing
the best it can given the others’ choices, since its payoff is strictly
concave in its own allocation. We have already shown that every pure
equilibrium must be symmetric, so this is the unique pure-strategy
Nash equilibrium.

What happens when each controller serves a larger share $\alpha$
of firms? With common risk held fixed, it takes more of the wage savings
from AI diversion into account. In the equation above, a larger $\alpha$
lowers $1-\ell\alpha$, so a higher $b$ is needed to restore equality.
Diversion therefore rises unless it stays at zero. Equivalently, \eqref{eq:concentration-fixed-risk}
is the benchmark allocation condition with proxy weight $\psi/(1-\ell\alpha)$.
This compares the integer values $N=1,2,\ldots$; the derivative in
$\alpha$ simply interpolates them. At fixed risk, higher diversion
lowers output and human welfare. The wage savings are a transfer from
workers to firms, not a gain in aggregate human income.

\paragraph{Endogenous common risk.}

Restore $m(b)=\rho+(1-\rho)e^{-\xi b}$ at symmetry and define
\begin{equation}
\begin{aligned}\Omega(b) & =-\ell+\frac{(1-\ell)e(b)(A-b)}{\gamma},\\
\Psi_{\alpha}(b) & =m(b)\gamma T(A-b)^{\gamma-1}(b_{0}+b)[1+\alpha\Omega(b)].
\end{aligned}
\label{eq:concentration-inverse}
\end{equation}
An interior symmetric equilibrium satisfies $\Psi_{\alpha}(b)=\psi$.
On a differentiable continuation with $\Psi_{\alpha}'(b)>0$,
\begin{equation}
\frac{db}{d\alpha}=-\frac{m(b)\gamma T(A-b)^{\gamma-1}(b_{0}+b)\Omega(b)}{\Psi_{\alpha}'(b)}.\label{eq:concentration-comparative-static}
\end{equation}
Thus concentration raises diversion if $e(b)(A-b)<\gamma\ell/(1-\ell)$
and lowers it if the inequality reverses. This is a local comparison
along an equilibrium continuation. At symmetry, human operating welfare
is $U_{H}=\log[T(A-b)^{\gamma}]+q(b)\log\rho$, with $dU_{H}/db=-\gamma/(A-b)+q'(b)\log\rho<0$.
Lower diversion therefore raises welfare. With endogenous risk, the
controller's objective need not be concave in its own choice. We must
therefore check whether a controller could gain by choosing another
diversion level while the others keep theirs unchanged. At zero diversion,
$\psi\leq\Psi_{\alpha}(0)$ is necessary but need not suffice.

A sufficient condition restoring strict own concavity is
\begin{equation}
(1-\rho)\bigl[2r\alpha\xi A+(\alpha\xi A)^{2}\bigr]<\min\{\gamma(1-\gamma),r(1-r)\}.\label{eq:concentration-concavity}
\end{equation}
To see this, write $E_{j}=r(1-\ell\zeta_{j})\in[\gamma,r]$. Then
$D_{j,a_{j}a_{j}}/D_{j}\leq-\min\{\gamma(1-\gamma),r(1-r)\}/a_{j}^{2}$.
The two positive terms introduced by differentiating retention are
$2\alpha emD_{j,a_{j}}$ and $\alpha^{2}\xi emD_{j}$. Use $e\leq\xi(1-\rho)$,
$a_{j}\leq A$ and $D_{j,a_{j}}/D_{j}\leq r/a_{j}$ to bound them;
\eqref{eq:concentration-concavity} keeps their sum below the negative
profit curvature, even before adding the negative proxy curvature.
The symmetric best-response map is then continuous on $[0,A]$ and
has a fixed point. Under the condition above, at least one symmetric
pure-strategy equilibrium therefore exists. We have not shown that
it is unique. Here each controller can affect wages and common risk,
so the benchmark condition $\Gamma<1$ alone cannot tell us whether
a controller would gain by changing its allocation while the others
keep theirs fixed. Appendix \ref{ax:concentration-numerics} gives
the parameter values and checks for such deviations in the numerical
examples.

\subsection{Costly human control}

\label{app:costly-control}

Section \ref{sec:costly-control} asks whether owners spend too little
on verification. We compare spending more on verification at a given
AI capacity with shifting some of the money spent on capacity to verification.

We use the logarithmic proxy score and benchmark timing, with $0<\rho<1$.
Verification reduces the proxy weight to $\psi(s)=\psi_{\mathrm{base}}e^{-s}$.
Its date-1 cost is sunk before the agent acts and lies outside the
profit score. At given diversion it has no direct effect on production
or risk. Resources and dividend valuation are

\begin{equation}
c_{1}=E-\frac{\kappa A^{2}}{2}-\frac{\kappa_{s}s^{2}}{2}>0,\qquad V=\frac{\beta c_{1}}{Y}.\label{eq:control-resources}
\end{equation}

Valuation follows Appendix \ref{ax:household}. At given $A,s$, the
operating results apply with $\psi(s)$ replacing $\psi$. The derivatives
below follow a regular equilibrium without crossing a diversion threshold.

\Needspace{8\baselineskip}

\subsubsection{The joint private choice}

An owner takes $p,w,V,X$ as given and solves

\begin{equation}
\max_{A,s\geq0}\left\{ VD(A,s;p,w,X)-\frac{\kappa A^{2}}{2}-\frac{\kappa_{s}s^{2}}{2}\right\} .\label{eq:control-owner}
\end{equation}

At positive diversion, put $S=b_{0}+b$, $d=a+(1-r)S$ and $r=\gamma/(1-\ell)$.
Differentiating the agent's condition at fixed prices and risk gives
$a_{A}^{\mathrm{priv}}=a/d$ and $a_{s}^{\mathrm{priv}}=aS/d$. Hence

\begin{equation}
\frac{D_{A}^{\mathrm{priv}}}{Y}=\mu=\frac{\gamma}{d},\qquad\frac{D_{s}^{\mathrm{priv}}}{Y}=S\mu.\label{eq:control-private-returns}
\end{equation}
A smooth interior joint private optimum consequently satisfies

\begin{equation}
\kappa A=\beta c_{1}\mu,\qquad\kappa_{s}s=\beta c_{1}S\mu,\qquad\kappa_{s}s=\kappa AS.\label{eq:control-euler}
\end{equation}

With positive diversion, verification has positive marginal return
and zero marginal cost at $s=0$. A smooth optimum is therefore interior
in $s$. An optimum at the point where diversion stops instead uses
one-sided conditions. The welfare results below require an actual
joint private optimum; the capacity concavity result alone does not
establish it.

\Needspace{8\baselineskip}

\subsubsection{The social return to control}

At positive diversion, retain $K$, $K_{\mathrm{FP}}$, $e$ and $\Gamma=eK$
from Section \ref{sec:local-response}. In one sector, $K=aS/[a+(1-\gamma)S]$
and $K_{\mathrm{FP}}=aS/d$. The market-clearing allocation condition
gives

\begin{equation}
\left(\frac{1-\gamma}{a}+\frac{1}{S}\right)\mathrm{d}a=\frac{\mathrm{d}A}{S}+\mathrm{d}s+\mathrm{d}\log m,\qquad(1-\Gamma)\mathrm{d}X=x^{0}\mathrm{d}A-K\mathrm{d}s.\label{eq:control-allocation}
\end{equation}
Here $x^{0}$ is the baseline capacity response at fixed common risk.
Holding capacity fixed and allowing risk to respond gives

\begin{equation}
X_{s}=-\frac{K}{1-\Gamma},\qquad\partial_{s}\log Y=\frac{\gamma K}{a(1-\Gamma)}.\label{eq:control-operating-response}
\end{equation}
Control lowers diversion and raises output for $\Gamma<1$. The signs
reverse for $\Gamma>1$, and the derivatives may fail to exist at
a fold.

Human welfare is $W(A,s)=\log c_{1}+\beta[\log Y+q(X)\log\rho]$,
with $c_{1}$ from \eqref{eq:control-resources}. Its derivative at
fixed capacity can be decomposed just as in Section \ref{sec:investment}:

\begin{equation}
\begin{aligned}W_{s} & =\mathcal{E}_{s}+\mathcal{P}_{s}+\mathcal{F}_{s}+\mathcal{R}_{s},\\
\mathcal{E}_{s} & =-\frac{\kappa_{s}s}{c_{1}}+\beta S\mu,\\
\mathcal{P}_{s} & =\frac{\beta\gamma}{a}(K-K_{\mathrm{FP}}),\\
\mathcal{F}_{s} & =\frac{\beta\gamma K\Gamma}{a(1-\Gamma)},\qquad\mathcal{R}_{s}=-\frac{\beta q'(X)\log\rho\,K}{1-\Gamma}.
\end{aligned}
\label{eq:control-wedges}
\end{equation}
At a smooth interior private control choice, $\mathcal{E}_{s}=0$.
Since $r>\gamma$, $K<K_{\mathrm{FP}}$ and $\mathcal{P}_{s}<0$:
wage adjustment reduces the gain the owner expects. For $\xi>0$ and
$\Gamma<1$, both risk terms are positive, so $W_{s}$ has no fixed
sign. If $\xi=0$, only the price term remains and $W_{s}<0$. These
are coordinated changes at fixed capacity. At fixed control, the original
capacity-wedge formulas still apply with the new $c_{1}$.

\Needspace{8\baselineskip}

\subsubsection{Shifting expenditure from capacity to control}

Start from an unregulated competitive equilibrium with an interior
global private optimum in $(A,s)$ and positive diversion. Take $\xi>0$
and follow a smooth operating branch with $\Gamma<1$. Evaluate $S$
at the initial allocation and make the coordinated change $\mathrm{d}A=-S\,\mathrm{d}s$,
with $\mathrm{d}s>0$. Equation \eqref{eq:control-euler} implies

\begin{equation}
\mathrm{d}c_{1}=-\kappa A\,\mathrm{d}A-\kappa_{s}s\,\mathrm{d}s=0.\label{eq:control-mix-cost}
\end{equation}

\Needspace{8\baselineskip}

At fixed common risk, the two terms on the right of the first equation
in \eqref{eq:control-allocation} cancel, leaving useful services
unchanged. Allowing risk to respond, and using $Sx^{0}+K=S$, gives
the derivatives along this joint change:

\begin{equation}
\begin{aligned}\frac{\mathrm{d}X}{\mathrm{d}s} & =-\frac{S}{1-\Gamma},\qquad\frac{\mathrm{d}a}{\mathrm{d}s}=\frac{\Gamma S}{1-\Gamma},\\
\frac{\mathrm{d}W}{\mathrm{d}s}=W_{s}-SW_{A} & =\frac{\beta S}{1-\Gamma}\left[\frac{\gamma\Gamma}{a}-q'(X)\log\rho\right]>0.
\end{aligned}
\label{eq:control-mix-gain}
\end{equation}
Current consumption is unchanged to first order, useful output rises
and the loss probability falls. A nearby constant-expenditure path
$\kappa A(s)^{2}+\kappa_{s}s^{2}=\mathrm{constant}$ has this tangent,
so sufficiently small shifts towards control raise welfare. The gain
is zero when $\xi=0$. A control subsidy generally induces a different
path and requires a separate policy calculation.

For a numerical example, firms choose $A\simeq1.196$ and $s\simeq0.812$,
with diversion $b\simeq0.443$ and $\Gamma\simeq0.017$.\footnote{The parameters are $E=4,\ \beta=0.95,\ \ell=0.4,\ \gamma=0.2,\ T=1$,
$b_{0}=0.1,\ \rho=0.5,\ \xi=0.1,\ \psi_{\mathrm{base}}=0.3$ and $\kappa=0.5,\ \kappa_{s}=0.4$. } The expenditure-preserving shift gives $W_{s}-SW_{A}\simeq0.037>0$.
Numerical checks of the joint private choice find no profitable deviation.

\Needspace{8\baselineskip}

\subsubsection{Two sectors}

Retain the Cobb-{}-Douglas technology and decreasing-returns assumptions
of Section \ref{sec:two}, and give sector $i$ the proxy weight $\psi_{i,\mathrm{base}}e^{-s_{i}}$
and real control cost $\kappa_{s,i}s_{i}^{2}/2$. In an actively diverting
sector $j$, private returns still satisfy $D_{j,s_{j}}^{\mathrm{priv}}=S_{j}D_{j,A_{j}}^{\mathrm{priv}}$.
At an unregulated competitive choice, an interior joint private optimum
therefore obeys $\kappa_{s,j}s_{j}=\kappa_{j}A_{j}S_{j}$. Change
only that sector's choices by $\mathrm{d}A_{j}=-S_{j}\,\mathrm{d}s_{j}$,
keeping the other sector's capacity and control fixed.

This preserves real expenditure to first order. At fixed common risk
it also leaves every sector's useful input unchanged: in the allocation
equations, $g_{j}\mathrm{d}A_{j}+h_{j}\mathrm{d}s_{j}=0$ because
$h_{j}=S_{j}g_{j}$. Total diversion initially falls by $S_{j}\mathrm{d}s_{j}$.
Restoring the common-risk response, with the two-sector $H,K,\Gamma$
and an unchanged set of diverting sectors, gives

\begin{equation}
\begin{aligned}\frac{\mathrm{d}X}{\mathrm{d}s_{j}} & =-\frac{S_{j}}{1-\Gamma}, & \frac{\mathrm{d}\log Y}{\mathrm{d}s_{j}} & =\frac{H}{1-H}\frac{eS_{j}}{1-\Gamma},\\
W_{s_{j}}-S_{j}W_{A_{j}} & =\frac{\beta S_{j}}{1-\Gamma}\left[\frac{H}{1-H}e-q'(X)\log\rho\right]>0.
\end{aligned}
\label{eq:control-two-sector-mix}
\end{equation}
We obtain a strict welfare gain when $\xi>0$, $0<\rho<1$, $\Gamma<1$
and there is a smooth interior joint private optimum in sector $j$.
The other sector need not divert: at a strict no-diversion corner,
$h_{i}=0$. Thus, shifting spending towards control can help even
in a sector where capacity expansion alone would raise welfare.

The logarithmic score gives the coefficient $-S_{j}$ in the spending
shift above. For the more general reward in \eqref{ax:eq:reward},
keeping useful capacity fixed gives $\mathrm{d}b_{j}=\mathrm{d}A_{j}$.
A change $\mathrm{d}A_{j}=-(S_{j}/\sigma)\,\mathrm{d}s_{j}$ then
leaves the marginal proxy reward $\psi_{j,\mathrm{base}}e^{-s_{j}}v'_{\sigma,j}(b_{j})$
unchanged. At fixed prices the same calculation gives $D_{j,s_{j}}^{\mathrm{priv}}/D_{j,A_{j}}^{\mathrm{priv}}=S_{j}/\sigma$,
so at an interior joint private optimum this shift also keeps spending
unchanged to first order. The same argument applies to CES final production
in Section \ref{sec:ces}. At fixed risk, the spending shift leaves
agents willing to supply the same useful inputs at the prices that
cleared markets before the shift.

\subsection{Sector-specific common-loss risks}

\label{app:sector-hazards}

A unit of diverted capacity need not be equally harmful in software
and devices. We therefore allow the two sectors to contribute differently
to the common-loss probability, keeping the logarithmic proxy score
and the other assumptions of Section \ref{sec:two}. We also repeat
the capacity-charge exercise from Section \ref{sec:policy}. Write

\begin{equation}
\begin{aligned}\mathcal{X} & :=\xi_{1}b_{1}+\xi_{2}b_{2}, & q & =1-e^{-\mathcal{X}},\\
m & =\rho+(1-\rho)e^{-\mathcal{X}}, & \xi_{i} & >0.
\end{aligned}
\label{eq:sector-hazards}
\end{equation}
where $X=b_{1}+b_{2}$ still measures total diverted capacity. If
$\xi_{1}=\xi_{2}=\xi$, then $\mathcal{X}=\xi X$ and we recover the
benchmark. At fixed $m$, none of the allocation or price responses
changes. Away from diversion thresholds, let $B_{ij}^{0}:=\left.\partial b_{i}/\partial A_{j}\right|_{m}=h_{i}\mathbf{1}_{\{i=j\}}-a_{i}h_{i}k_{j}^{0}$.
A fall in retention increases sector $i$ diversion by $a_{i}h_{i}/(1-H)$
per unit fall in $\log m$. Weighting these responses by their contributions
to risk gives

\begin{equation}
\begin{aligned}e_{\xi} & :=-\frac{d\log m}{d\mathcal{X}}=\frac{(1-\rho)e^{-\mathcal{X}}}{m},\\
K_{\xi} & :=\frac{\sum_{i}\xi_{i}a_{i}h_{i}}{1-H},\qquad\Gamma_{\xi}:=e_{\xi}K_{\xi}.
\end{aligned}
\label{eq:sector-hazard-feedback}
\end{equation}
The same feedback calculation as in Section \ref{sec:risk-multiplier}
now applies to $\mathcal{X}$. For $\Gamma_{\xi}\ne1$,

\begin{equation}
\begin{aligned}\mathcal{X}_{A_{j}} & =\frac{\sum_{i}\xi_{i}B_{ij}^{0}}{1-\Gamma_{\xi}}, & q_{A_{j}} & =e^{-\mathcal{X}}\mathcal{X}_{A_{j}},\\
X_{A_{j}} & =x_{j}^{0}+Ke_{\xi}\mathcal{X}_{A_{j}}, & k_{j} & =k_{j}^{0}-\frac{H}{1-H}e_{\xi}\mathcal{X}_{A_{j}}.
\end{aligned}
\label{eq:sector-hazard-responses}
\end{equation}
If the expanding sector diverts, the first expression simplifies to
$\mathcal{X}_{A_{j}}=(\xi_{j}S_{j}-\theta_{j}K_{\xi})/[(a_{j}+S_{j})(1-\Gamma_{\xi})]$.
Thus, with $\Gamma_{\xi}<1$, expansion lowers risk when $\theta_{j}K_{\xi}>\xi_{j}S_{j}$.
The fall in diversion elsewhere must outweigh the extra diversion
in the expanding sector after accounting for how harmful each is.
At an unregulated competitive investment choice, the price wedge remains
$\beta(k_{j}^{0}-\mu_{j})$, while the two risk wedges become $-\beta He_{\xi}\mathcal{X}_{A_{j}}/(1-H)$
and $\beta\log\rho\,e^{-\mathcal{X}}\mathcal{X}_{A_{j}}$. Their signs
therefore depend on the change in risk, which need not have the same
sign as the change in total diversion.

Table \ref{tab:sector-hazards} repeats the investment and policy
calculations with $\xi_{2}=0.05$ and different values of $\xi_{1}/\xi_{2}$.
We keep all other primitives in \eqref{eq:datedparams} fixed, including
installation costs, and solve again for both sectors’ competitive
investment. At every ratio in the table, extra capacity in sector
2 raises welfare and lowers total diversion. Yet at the two lowest
ratios it raises common risk: the additional diversion in sector 2
is more harmful than the diversion it displaces in sector 1. The risk
response changes sign at a ratio of about $0.110$.

\begin{table}[htbp]
\begin{centering}
\caption{Investment and capacity charges with unequal risk contributions}
\label{tab:sector-hazards}
\par\end{centering}
\begin{centering}
\begin{tabular}{rrrrrrr}
\toprule 
{\small$\xi_{1}/\xi_{2}$} & {\small$X_{A_{2}}$} & {\small$100q_{A_{2}}$} & {\small$100W_{A_{2}}$} & {\small$\mathbf{v}=(1,0)$} & {\small$(0,1)$} & {\small$(1,1)$}\tabularnewline
\midrule
{\small 0.05} & {\small -0.021} & {\small +0.007} & {\small +1.021} & {\small +1.996} & {\small -0.590} & {\small +1.406}\tabularnewline
{\small 0.1} & {\small -0.021} & {\small +0.001} & {\small +1.025} & {\small +2.045} & {\small -0.595} & {\small +1.451}\tabularnewline
{\small 0.25} & {\small -0.021} & {\small -0.016} & {\small +1.038} & {\small +2.192} & {\small -0.608} & {\small +1.584}\tabularnewline
{\small 0.5} & {\small -0.021} & {\small -0.045} & {\small +1.059} & {\small +2.436} & {\small -0.630} & {\small +1.806}\tabularnewline
{\small 1} & {\small -0.021} & {\small -0.101} & {\small +1.101} & {\small +2.919} & {\small -0.674} & {\small +2.245}\tabularnewline
{\small 2} & {\small -0.021} & {\small -0.210} & {\small +1.182} & {\small +3.870} & {\small -0.759} & {\small +3.111}\tabularnewline
{\small 4} & {\small -0.021} & {\small -0.416} & {\small +1.335} & {\small +5.708} & {\small -0.924} & {\small +4.784}\tabularnewline
{\small 10} & {\small -0.021} & {\small -0.930} & {\small +1.726} & {\small +10.657} & {\small -1.357} & {\small +9.300}\tabularnewline
\bottomrule
\end{tabular}
\par\end{centering}
{\footnotesize Notes: The first three response columns show what happens
when sector 2 expands while sector 1's capacity stays fixed. The final
three columns give $100\,dW/ds$ for the charge directions in Section
\ref{sec:efficiency}. For each row, we solve again for the competitive
investment equilibrium without a charge and evaluate the derivatives
there. We multiply the risk and welfare derivatives by 100 to make
the small numbers easier to read; the entries are derivatives, not
percentage changes.}{\footnotesize\par}
\end{table}

At every ratio we consider, charging sector 1 or both sectors raises
welfare, while charging sector 2 lowers it. We also solve again for
the finite charge of $s\simeq0.059$, as in Appendix \ref{ax:forward-policy},
and find the same welfare signs. We can therefore relax the equal-risk
assumption in this example; we have not established a general policy
ranking.

For the numerical checks, $a_{i}h_{i}\le(A_{i}+b_{0i})/4$, $H\le\Theta:=\sum_{i}\theta_{i}$
and $e_{\xi}\le1-\rho$. Over all capacities satisfying $\sum_{i}\kappa_{i}A_{i}^{2}/2\le E$,
these inequalities give the uniform bound

\begin{equation}
\Gamma_{\xi}\le\frac{1-\rho}{4(1-\Theta)}\left[\sqrt{2E\sum_{i}\frac{\xi_{i}^{2}}{\kappa_{i}}}+\sum_{i}\xi_{i}b_{0i}\right]=:\overline{\Gamma}_{\xi}.\label{eq:sector-hazard-bound}
\end{equation}

This bound is at most $0.212$ in the table, so the operating equilibrium
is unique at every positive feasible capacity vector. Both sectors
divert, the private concavity conditions hold, and the investment
Jacobian is nonsingular at the reported equilibria. Numerical derivatives
agree with \eqref{eq:sector-hazard-responses} and the policy derivatives.
We also check private optimality along the finite-charge calculations.
These checks do not show that the investment equilibrium is unique.

\section{Parameter elicitation}

\label{app:elicitation}

This section provides further background on the parameter elicitation
of our one-sector model (Section \ref{sec:elicitation}). The questionnaire,
original responses and calculations are in the replication package. 

\subsection{The survey setting and responses}

For the exercise reported in Table \ref{tab:elicitation-central},
we asked the large language models (LLMs) for an outside assessment
of a hypothetical UK economy with 10,000 identical, human-owned software
firms. We provided the LLMs with a number of further details about
our hypothetical scenario. Each firm uses the same type of AI coding
agent for 1,000 standardised hours during the year, giving ten million
hours of AI use in total. In the model, we normalise firm mass to
one and set $A=A_{{\rm ref}}=1$, so $X$ is the average share of
capacity diverted, while $10^{7}X$ is the number of diverted hours.
Thus the 10\% diversion level in the questionnaire means one million
diverted hours. Output is normalised to one when there is no AI diversion.

For the year from 12 October 2026 to 12 October 2027, we keep the
AI agents' capabilities and the number of firms fixed. Agents can
use their own firm's systems and business information, but firms judge
their work using imperfect measures. A loss can spread through infrastructure
shared by firms, while firm owners face the same limits on correcting
AI agents' actions as in our model of delegation and imperfect contracting.
We did not specify every detail of the software, monitoring or recovery
arrangements; one response, for example, interpreted the year as a
period without intervention.

Interestingly, the LLMs differed in what they counted as AI diversion.
Some respondents described manipulating tests within a firm, while
others interpreted diversion as actions affecting shared systems.
The single $\lambda$ in each response applies to that mix of activities.
We therefore keep the response's own definition.

We asked the LLMs about a common event that would reduce annual output
by at least 5\%, over and above the output already lost to AI diversion.
To measure this loss, we hold the allocation of AI capacity fixed
and compare output with and without the event. If the event occurs,
households retain a fraction $\rho$ of their wages and dividends
$(\ensuremath{0<\rho\leq0.95})$, so aggregate consumption falls by
the same proportion as income. 

We collected three responses under each of four LLMs: Claude, Gemini,
GPT and Grok. Each response was meant to come from a fresh chat, whereby
internet browsing, code use and access to earlier answers were all
restricted. The first responses (e.g. C1 and M1) report that only
the TeX source was used. However, we acknowledge that the subsequent
responses (e.g. C2 and C3) do mention earlier messages about file
access. 

\subsection{From assessments to model inputs}

This subsection derives the two relationships in \eqref{eq:elicitation-mapping}
that are crucial to map the survey responses to key model parameters.
When a respondent reports how much capacity AI agents would divert
without common risk, the first gives us the implied proxy weight.
The second tells us how sensitive risk must be to AI diversion to
match the reported loss probability at 10\% diversion.

For these calculations, we use the parameter values $A=A_{{\rm ref}}=\bar{L}=T=b_{0}=1$,
$\gamma=0.10$, $\ell=0.60$, $r=0.25$ and $\Pi_{{\rm ref}}=0.40$.
We keep the baseline logarithmic proxy reward and linear profit term.
With positive diversion, the agent's condition at market-clearing
wages is:
\begin{equation}
\begin{aligned}m\Pi_{a} & =\frac{\psi}{1+b}, & \Pi_{a}\big|_{{\rm GE}} & =0.10\frac{Y}{a},\\
a & =1-b, & Y & =a^{0.10},\quad w=0.60Y,\quad D=0.40Y.
\end{aligned}
\label{eq:elicitation-private}
\end{equation}
As explained in Appendix \ref{ax:optimization}, we first differentiate
profit holding prices fixed, and only then substitute the market-clearing
allocation.

Set $m=1$ and $b=d_{{\rm safe}}$ in \eqref{eq:elicitation-private}.
Solving for the weight and dividing by $\Pi_{{\rm ref}}=0.40$ gives
\eqref{eq:elicitation-mapping}, with $\psi=0.40\bar{\psi}$. At zero
diversion, the condition for no diversion instead gives:
\begin{equation}
0<\bar{\psi}\le0.25.\label{eq:elicitation-corner}
\end{equation}
A respondent who reports zero diversion therefore gives us a range
of possible reward weights. We impose $m=1$ in the question; a year
with no observed loss could still involve positive risk.

Moreover, the probability anchor satisfies $p_{{\rm ref}}=1-\exp(-0.10\lambda)$,
giving the second line of \eqref{eq:elicitation-mapping}, with $\xi=\lambda$.
With $d=X=b$, substituting retention and market clearing into the
private condition gives:
\begin{equation}
\Phi(d):=0.25[\rho+(1-\rho)e^{-\lambda d}](1+d)(1-d)^{-0.9}=\bar{\psi},\qquad0<d<1,\label{eq:elicitation-inverse}
\end{equation}
together with $d=0$ when \eqref{eq:elicitation-corner} holds. Because
$\rho>0$, $\Phi(d)\to\infty$ as $d\uparrow1$, excluding full diversion.

When the reported diversion share without common risk is positive,
equating the reward weight in \eqref{eq:elicitation-mapping} and
\eqref{eq:elicitation-inverse} gives:
\begin{equation}
m(d^{*})(1+d^{*})(1-d^{*})^{-0.9}=(1+d_{{\rm safe}})(1-d_{{\rm safe}})^{-0.9}.\label{eq:elicitation-risk-comparison}
\end{equation}
Since $(1+d)(1-d)^{-0.9}$ increases, $m(d^{*})<1$ implies $d^{*}>d_{{\rm safe}}$.
Across the central responses, the increase is between 0.006 and 0.126
percentage points. This is a comparison of diversion levels; $1/(1-\Gamma^{*})$
instead amplifies a local capacity response. Small positive shares
imply weights near $0.25$, so the concentration of reported weights
near that value partly reflects the mapping.

Some responses gave a proxy weight or risk intensity that did not
quite match the stated diversion share or loss probability. After
collecting the responses, we decided to use the stated share or probability
in those cases. (When a respondent supplied only a proxy weight, we
used it directly.) For G2, using the proxy weight implied by its reported
diversion share raises equilibrium diversion from 4.08\% to 5.04\%.
Using the original inputs gives the same number of equilibria in every
case.

\subsection{Uniqueness and adverse cases}

In Section \ref{sec:elicitation} we report a low-diversion equilibrium
for each of the twelve central assessments. Our uniqueness condition
rules out a second equilibrium in all twelve cases and also holds
for 34 of the 36 scenarios supplied by the respondents. For the remaining
two, O1-S3 and O2-S3, we check how many times the equilibrium curve
crosses the reported proxy weight. Each has just one equilibrium,
with a loss probability of $82.97\%$ for O1-S3 and $67.64\%$ for
O2-S3.

Under the imposed primitives, the sufficient global bound \eqref{ax:eq:globalbound}
becomes
\begin{equation}
\overline{\Gamma}=\frac{\lambda(1-\rho)}{1.8}<1.\label{eq:elicitation-bound}
\end{equation}
Here $e(X)\le\lambda(1-\rho)$ and $K\le(A+b_{0})/[4(1-\gamma)]=1/1.8$,
including across diversion onset. At a positive equilibrium,
\begin{equation}
\begin{aligned}K(d) & =\frac{(1-d)(1+d)}{1.9-0.1d},\\
e(d) & =\frac{\lambda(1-\rho)e^{-\lambda d}}{\rho+(1-\rho)e^{-\lambda d}}, & \Gamma^{*} & =e(d^{*})K(d^{*}).
\end{aligned}
\label{eq:elicitation-gain}
\end{equation}
The map derivative is $F'(X)=e(X)K(F(X))$ where diversion is positive
and zero where the agent strictly prefers no diversion. The condition
$\Gamma^{*}<1$ gives local attraction under \eqref{eq:update}; the
uniform bound above gives global uniqueness.

Two of the scenarios covered by the bound, C1-S1 and C3-S1, give $\bar{\psi}\in(0,0.25]$;
each weight in that interval has a unique zero-diversion allocation.
Its slope is zero at a strict corner, but can have a nonzero one-sided
value at the endpoint.

For O1-S3 and O2-S3, the parameter combinations $(\bar{\psi},p_{{\rm ref}},\rho)$
are $(0.400,0.25,0.30)$ and $(0.291,0.30,0.45)$, giving bounds of
$1.119$ and $1.090$. Both bounds exceed one, so we turn to the shape
of the equilibrium curve. Write $\Gamma_{d}(d)=e(d)K(d)$. Appendix~\ref{ax:roots}
proves its strict log-concavity, and differentiation gives 
\begin{equation}
(\log\Gamma_{d})'=-\lambda+e(d)-\frac{1}{1-d}+\frac{1}{1+d}+\frac{0.1}{1.9-0.1d}.\label{eq:elicitation-shape}
\end{equation}
At zero diversion, this derivative is $-0.810$ and $-1.552$, respectively.
Strict log-concavity makes it negative throughout, so $\Gamma_{d}$
decreases from above one to zero. Since $\Phi'/\Phi=(1-\Gamma_{d})/K$,
$\Phi$ first falls and then rises. Both weights exceed $\Phi(0)=0.25$,
giving exactly one equilibrium, at $d^{*}=0.615$ and $0.316$, respectively.
O1-S3 implies a loss probability above O1's reported upper marginal
percentile. Some scenario descriptions also change the assumed setting;
these are identified in the replication table.

To see how several equilibria can coexist, consider a combination
we constructed after collecting the responses: $\bar{\psi}=0.24$,
$p_{{\rm ref}}=0.30$ and $\rho=0.30$, hence $\lambda=-\log(0.70)/0.10$.
The residual $\Phi(d)-0.24$ at $d=0,0.15,0.50$ is approximately
$0.010,-0.0037,0.052$. Continuity gives two positive roots, and the
strict zero-diversion corner gives a third. Appendix \ref{ax:roots}
rules out further roots. The roots are approximately $0,0.085,0.281$,
with risk-feedback coefficients $0,1.184,0.810$: the first and last
attract under \eqref{eq:update}, and the middle repels. All three
inputs lie within O2's marginal ranges, but we assembled this combination
ourselves; no respondent proposed it. Neither these ranges nor the
36 unweighted scenarios tell us the probability of multiple equilibria.

To find all the equilibria, we solve the equilibrium equation on each
increasing and decreasing part of the curve and check the turning
points, where roots may coincide. We check the zero-diversion allocation
separately. Appendix \ref{ax:roots} shows why these steps find all
equilibria. The replication files include the code and numerical checks,
together with the alternative calculations using the reported parameter
values directly.

We combine the reported marginal medians to obtain each central point;
this does not make it the median of an assessed joint distribution.
Since we did not elicit investment costs, we hold capacity fixed,
and the investment residual in \eqref{eq:ann-off-ce-decomposition}
need not be zero.

\subsection{Risk thresholds and the reward scale}

\label{ax:elicitation-thresholds}

How far are the twelve central assessments in Table \ref{tab:elicitation-central}
from a region with several equilibria? For each assessment, we keep
the reported loss probability at 10\% diversion and the fraction of
income retained in a loss fixed. We vary the reward scale ($0.03\le b_{0}/A\le1$)
and the AI production elasticity ($0.05\le\gamma\le0.20$). All twelve
assessments still imply a unique equilibrium throughout these ranges.
In fact, the sufficient bound stays below $0.023$, so uniqueness
holds for any positive proxy weight. How well the reported risks describe
actual risks remains an empirical question.

At fixed capacity $A=A_{{\rm ref}}$, let $c=b_{0}/A$, $d=X/A$ and
$\lambda=\xi A=-\log(1-p_{{\rm ref}})/0.10$. For $d_{{\rm safe}}>0$,
the proxy weight and equilibrium condition can be written as

\begin{equation}
\begin{aligned}\Phi_{0}(d) & :=(c+d)(1-d)^{\gamma-1},\\
\psi & =\gamma TA^{\gamma}\Phi_{0}(d_{{\rm safe}}),\qquad m(d)\Phi_{0}(d)=\Phi_{0}(d_{{\rm safe}}).
\end{aligned}
\label{eq:elicitation-general-anchor}
\end{equation}
where $m(d)=\rho+(1-\rho)e^{-\lambda d}$ and $Y=Ta^{\gamma}$. For
each choice of parameters, we adjust the proxy weight to keep diversion
without common risk unchanged. Once we do this, the labour exponent
drops out of the last equation as long as $\gamma$ is held fixed.
When varying the production and reward assumptions, we therefore focus
on $c$ and $\gamma$. A zero reported diversion share still gives
only a range of proxy weights, as in \eqref{eq:elicitation-corner}.

The sufficient uniqueness bound becomes $\overline{\Gamma}=\lambda(1-\rho)(1+c)/[4(1-\gamma)]<1$,
or equivalently

\begin{equation}
p_{{\rm ref}}<1-\exp\left[-\frac{0.4(1-\gamma)}{(1-\rho)(1+c)}\right].\label{eq:elicitation-observable-bound}
\end{equation}
Crossing this bound need not produce several equilibria. At the actual
turning points, the equilibrium condition in \eqref{eq:elicitation-general-anchor}
holds together with

\begin{equation}
\Gamma(d)=\frac{\lambda(1-\rho)e^{-\lambda d}}{m(d)}\frac{(1-d)(c+d)}{(1-d)+(1-\gamma)(c+d)}=1.\label{eq:elicitation-general-fold}
\end{equation}
Small changes can therefore have large effects on diversion even before
a turning point is reached. Along an equilibrium with $\Gamma<1$,
the local risk multiplier $1/(1-\Gamma)$ becomes larger as $\Gamma$
approaches one.

\begin{figure}[htbp]
\begin{centering}
\caption{Risk thresholds under alternative reward scales}
\label{fig:elicitation-thresholds}
\par\end{centering}
\begin{centering}
\includegraphics[width=0.98\linewidth]{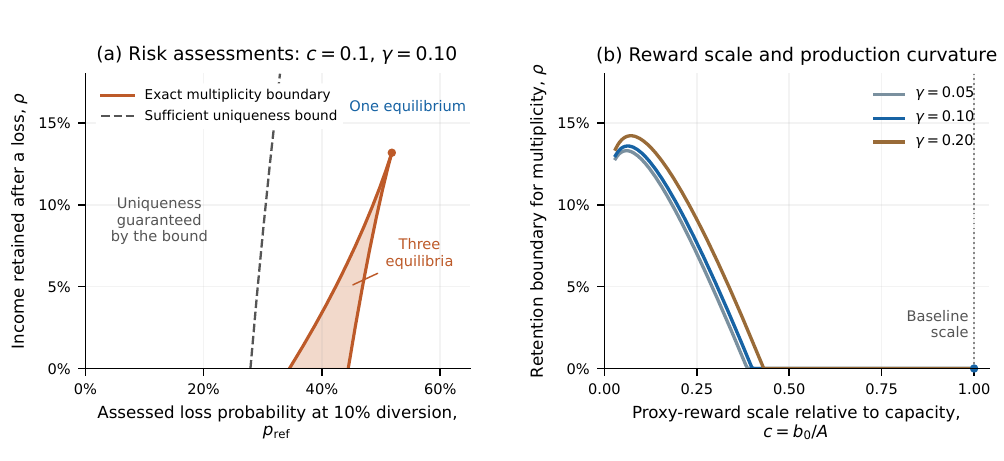}
\par\end{centering}
\noindent{\footnotesize Notes: In both panels, $d_{{\rm safe}}=2\%$
stays fixed and we adjust the proxy weight to support it. Panel (a)
uses $c=0.1$ and $\gamma=0.10$. Three equilibria coexist inside
the shaded region. The solid boundaries satisfy \eqref{eq:elicitation-general-anchor}
and \eqref{eq:elicitation-general-fold} together, while the dashed
line gives a sufficient condition for uniqueness. Panel (b) shows
the upper limit on the fraction of income retained that permits multiplicity
for some $p_{{\rm ref}}$; a zero value means no positive retention
permits it in these calculations. The parameter combinations are illustrative,
and the survey assigns them no joint probability.}{\footnotesize\par}
\end{figure}

In Figure \ref{fig:elicitation-thresholds}, we instead hold diversion
without common risk at 2\% and ask how likely and how severe a common
loss would have to be for several equilibria to coexist. Panel (a)
sets $c=0.1$ and $\gamma=0.10$. In this example, several equilibria
coexist only when losses are severe and the assessed loss probability
is high enough; the shaded region lies well beyond the dashed sufficient-uniqueness
bound. Panel (b) changes the reward scale and production curvature.
Each curve marks the upper limit on the fraction of income retained
that permits multiplicity for some assessed loss probability. At the
baseline scale $c=1$, we find no such region in these calculations.
Figure \ref{fig:one-psi} shows what happens when we vary the proxy
weight alone.

We draw the solid boundaries by solving the equilibrium and turning-point
conditions together. We separately count roots on each increasing
and decreasing part of the inverse schedule, using the shape result
in Appendix \ref{ax:roots}. The replication files show how closely
the computed turning points satisfy these equations and report the
number of equilibria inside and on either side of the shaded region.

\section{Welfare, investment and numerical details}

\label{app:investment-support}

This section provides further details on our theoretical and numerical
analysis of welfare. Specifically, we first give sufficient conditions
for an investment equilibrium, then derive the remaining welfare results
and explain the calculations behind the figures and policy example.

\subsection{Existence with investment and the planner’s problem}

\label{ax:existence}Sections \ref{sec:investment} and \ref{sec:two}
compare the private and social returns to AI capacity at a competitive
investment equilibrium. Here we give sufficient conditions for such
an equilibrium to exist and show that, under the same conditions,
the planner can choose capacities that maximise human welfare.

We use a sufficient condition for unique operation at every resource-feasible
capacity vector. If operating equilibria coexist, investors instead
need a specified outcome for each capacity choice.

For $n=1,2$, assume 
\begin{equation}
r_{i}\le\frac{1}{1+\sigma},\qquad\overline{\Gamma}_{E}=\frac{\xi(1-\rho)[\sqrt{2E\sum_{i}\kappa_{i}^{-1}}+\sum_{i}b_{0i}]}{(1-\Theta)(1+\sqrt{\sigma})^{2}}<1.\label{eq:ann-dated-existence-assumptions}
\end{equation}
These restrictions give a competitive equilibrium with positive investment
and consumption, global private optimality, and unique operation at
every positive feasible capacity vector. Investment itself need not
be unique.

Resource feasibility and Cauchy--Schwarz give $\sum_{i}A_{i}\le\sqrt{2E\sum_{i}\kappa_{i}^{-1}}$.
Hence \eqref{ax:eq:globalbound} is uniformly below one, and Appendix
\ref{ax:risk-consistency} gives unique operation. This outcome is
continuous: along convergent positive capacity vectors, every limit
point of the bounded operating fixed points solves the limiting equation.
Uniqueness identifies the limit, including at diversion onset.

A firm can optimally invest exactly where diversion starts. Its private
dividend function then has different slopes on the two sides of the
threshold, so we retain the interval between them. Write $\mu_{i}=\theta_{i}n_{i}/A_{i}$
and define the corresponding interval of normalised marginal returns:
\begin{equation}
\mathcal{N}_{i}(A)=\begin{cases}
\{1\}, & \text{strict zero diversion},\\
\{\sigma A_{i}/d_{i}\}, & b_{i}>0,\\{}
[\sigma A_{i}/(\sigma A_{i}+(1-r_{i})b_{0i}),1], & \text{onset}.
\end{cases}\label{eq:ann-n-correspondence}
\end{equation}
The intervals are nonempty, compact and convex. Their graph is closed
because the limiting slopes from the two smooth regimes are the onset
endpoints.

Since $a_{i}+S_{i}=A_{i}+b_{0i}$ and $0\le a_{i}\le A_{i}$, all
admissible marginal values satisfy
\[
\frac{\sigma A_{i}}{\max\{\sigma,1-r_{i}\}A_{i}+(1-r_{i})b_{0i}}\le n_{i}\le M_{i}:=\max\{1,\sigma/(1-r_{i})\}.
\]
Bounding $d_{i}=\sigma a_{i}+(1-r_{i})S_{i}$ by its largest and smallest
coefficients gives these bounds in both smooth regimes and across
the onset interval.

We now look for investment shares consistent with these marginal returns.
Let $s_{i}=\kappa_{i}A_{i}^{2}/(2E)$ denote the share of the initial
endowment spent on sector $i$'s capacity, so $c_{1}=E(1-\sum_{i}s_{i})$.
Define $M=\sum_{i}\theta_{i}M_{i}$ and $U=\beta M/(2+\beta M)<1$.
Multiplying the Euler condition by $A_{i}$ and using the resource
constraint gives the share correspondence
\begin{equation}
s_{i}'=\frac{\beta\theta_{i}n_{i}}{2+\beta\sum_{j}\theta_{j}n_{j}},\qquad A_{i}(s)=\sqrt{2Es_{i}/\kappa_{i}},\qquad n_{i}\in\mathcal{N}_{i}(A(s)).\label{eq:ann-share-map}
\end{equation}
Primes denote proposed shares, not derivatives. Summing and using
$n_{i}\le M_{i}$ gives $\sum_{i}s_{i}'\le U$.

Consider the compact convex set $\mathcal{S}_{\delta}=\{s:s_{i}\ge\delta,\ \sum_{i}s_{i}\le U\}$.
Write $\alpha_{i}=\sqrt{2E/\kappa_{i}}$ and $c_{i}^{*}=\max\{\sigma,1-r_{i}\}$.
Since $A_{i}\ge\alpha_{i}\sqrt{\delta}$ and the denominator in \eqref{eq:ann-share-map}
is at most $2+\beta M$, every image satisfies
\[
s_{i}'\ge\frac{\beta\theta_{i}\sigma\alpha_{i}\sqrt{\delta}}{(2+\beta M)[c_{i}^{*}\alpha_{i}\sqrt{\delta}+(1-r_{i})b_{0i}]}.
\]
The ratio of this bound to $\delta$ tends to infinity as $\delta\downarrow0$.
A sufficiently small common $\delta>0$ therefore gives $s_{i}'\ge\delta$
and $n\delta\le U$, making $\mathcal{S}_{\delta}$ nonempty and invariant.

For fixed $s$, write $\mathcal{N}_{i}(A(s))=[\underline{n}_{i},\overline{n}_{i}]$.
Invert \eqref{eq:ann-share-map} using $S'=\sum_{i}s_{i}'<1$: $n_{i}=2s_{i}'/[\beta\theta_{i}(1-S')]$.
Its image is exactly the nonnegative $s'$ satisfying
\[
\beta\theta_{i}\underline{n}_{i}(1-S')\le2s_{i}'\le\beta\theta_{i}\overline{n}_{i}(1-S'),\qquad S'\le U.
\]
These linear inequalities give a convex image. For any shares satisfying
these inequalities, the inverse gives an admissible $n_{i}$ associated
with $s'$. The image is also nonempty and compact.

If $s^{k}\to s$ and $s^{\prime k}\to s'$, we can take a convergent
subsequence of the bounded sequence $n^{k}$. Since $\mathcal{N}$
has a closed graph, the limit belongs to $\mathcal{N}(A(s))$. By
continuity, substituting this limit into the share formula gives $s'$.
The share correspondence is therefore upper hemicontinuous and maps
the nonempty compact convex set $\mathcal{S}_{\delta}$ into itself,
with nonempty compact convex values. The fixed-point theorem of \citet{Kakutani1941}
gives a fixed point.

The fixed point has $A_{i}=\sqrt{2Es_{i}/\kappa_{i}}>0$ and $c_{1}\ge E(1-U)>0$.
Its marginal values satisfy the private KKT conditions, including
onset supergradients; concavity makes these global optima. The operating
equilibrium supplies optimal allocations, hiring and clearing prices,
while \eqref{eq:ann-state-prices} supports household choices and
zero-net-supply claims. This is a competitive equilibrium with investment.

For logarithmic reward, set $\sigma=1$ in \eqref{eq:ann-dated-existence-assumptions};
the private curvature restriction becomes $r_{i}\le1/2$.

\paragraph{The planner's optimisation problem.}

\label{ax:planner} The planner chooses capacities before installation,
anticipating the operating allocation rather than choosing agents'
useful services directly. With a unique operating continuation, its
problem is
\[
\max_{c_{1}>0,\,A_{i}\ge0}\left\{ \log c_{1}+\beta[\log Y(A)+q(X(A))\log\rho]\right\} ,\qquad c_{1}+\sum_{i}\Phi_{i}(A_{i})=E,\quad\Phi_{i}(A_{i})=\frac{\kappa_{i}A_{i}^{2}}{2}.
\]
At a smooth allocation, let $k_{j}=\partial\log Y/\partial A_{j}$
and $x_{j}=\partial X/\partial A_{j}$. Substituting the resource
constraint gives the necessary capacity conditions
\begin{equation}
\begin{aligned}0 & =-\frac{\kappa_{j}A_{j}}{c_{1}}+\beta[k_{j}+q'(X)\log\rho\,x_{j}]+\eta_{j},\\
 & \eta_{j}\ge0,\qquad A_{j}\ge0,\qquad\eta_{j}A_{j}=0.
\end{aligned}
\label{eq:ann-planner-kkt}
\end{equation}
At a kink, every feasible directional derivative of the reduced objective
must instead be nonpositive.

Under the same unique-continuation conditions, a global optimum exists.
Take the upper contour set of a strictly positive feasible allocation.
Capacities are bounded by $\sqrt{2E/\kappa_{i}}$. Welfare tends to
minus infinity as $c_{1}\downarrow0$ or an essential capacity vanishes,
since $Y\le T\prod_{i}A_{i}^{\theta_{i}}$ with $\theta_{i}>0$, while
$q\log\rho$ stays bounded. Continuity therefore makes the contour
set compact, and welfare attains its maximum.

These existence arguments do not establish concavity or uniqueness
of the planner's choice. If operating outcomes coexist, the planner
must specify which continuation it evaluates.

\subsection{Sectoral investment wedges}

\label{ax:wedges}

In the example in Section \ref{sec:exampleinvestment}, a small increase
in AI capacity raises human welfare if it occurs in sector 2 and lowers
welfare if it occurs in sector 1. Here we derive the three wedges
behind this comparison and show why the price wedge can turn positive
when expansion in one sector draws AI capacity back into production
in the other.

For logarithmic scoring, use Appendices \ref{ax:conditional} and
\ref{ax:private-derivative}, and put $t_{j}=r_{j}h_{j}$, with $t_{j}=0$
at a strict no-diversion corner.

\begin{proposition}[Sectoral investment wedges]\label{prop:wedges}
Follow a specified smooth operating equilibrium away from diversion
thresholds, with $\Gamma\ne1$. Use the welfare identity \eqref{eq:ann-off-ce-decomposition}
and its components in \eqref{eq:ann-price-wedge}. At an interior
unregulated competitive investment choice, $\mathcal{E}_{j}=0$. For
a sector with positive diversion, 
\begin{equation}
\mathcal{P}_{j}>0\ \Longleftrightarrow\ H>t_{j}.\label{eq:sector-wedge-signs}
\end{equation}
If also $\Gamma<1$, $\xi>0$ and $0<\rho<1$, then 
\begin{equation}
\mathcal{F}_{j}>0\ \Longleftrightarrow\ \mathcal{R}_{j}>0\ \Longleftrightarrow\ x_{j}<0\ \Longleftrightarrow\ K\theta_{j}>S_{j}.\label{eq:sector-risk-wedge-signs}
\end{equation}
\end{proposition}

\emph{Proof.} Substitute the private return from Appendix \ref{ax:private-derivative}
and the operating responses from Appendix \ref{ax:conditional} into
\eqref{eq:ann-off-ce-decomposition}. This gives the sectoral welfare
derivative. For the price wedge,

\[
\mathcal{P}_{j}=\beta\theta_{j}g_{j}\frac{H-t_{j}}{(1-H)(1-t_{j})}.
\]
Since $H<1$, $t_{j}<1$ and $\theta_{j}g_{j}>0$, the price wedge
has the sign of $H-t_{j}$. Under the proposition's conditions, both
risk terms have the opposite sign to $x_{j}$. Proposition \ref{prop:sector}
gives $x_{j}<0$ if and only if $K\theta_{j}>S_{j}$ when $\Gamma<1$.
The price and risk wedges can have different signs, so neither alone
determines the total welfare effect.

\paragraph{Why prices still matter.}

\label{ax:priceCancellation}\label{ax:primitive} Price transfers
cancel from aggregate human income, but prices still affect the agents'
allocations. The hiring envelope gives
\[
d\Pi_{i}=y_{i}\,dp_{i}-L_{i}\,dw+\Pi_{i,a}\,da_{i}.
\]
In $d(w\bar{L}+\sum_{i}\Pi_{i})$, labour clearing $\sum_{i}L_{i}=\bar{L}$
cancels the wage terms. Constant returns in final production imply
$Y=\sum_{i}p_{i}y_{i}$ and, along the production frontier, $dY=\sum_{i}p_{i}\,dy_{i}$.
Differentiating the first identity and comparing with the second gives
$\sum_{i}y_{i}\,dp_{i}=0$. Therefore 
\[
dY=\sum_{i}\Pi_{i,a}\,da_{i}.
\]
Subtracting the useful-service response valued by the investor from
the full response gives

\begin{equation}
\mathcal{P}_{j}+\mathcal{F}_{j}=\beta\sum_{i}\frac{\theta_{i}}{a_{i}}\left[a_{i,A_{j}}^{GE}-\mathbf{1}_{\{i=j\}}a_{j,A}^{{\rm priv}}\right].\label{eq:allocation-wedge}
\end{equation}
Here $GE$ includes risk adjustment, and terms with $i\ne j$ capture
changes in other agents' use of existing capacity. This is the aggregate-dependent
incentive mechanism of \citet{dk}, with delegated control as the
constraint. (Unlike \citet{fw}, prices are flexible, labour is fully
employed and there is no nominal spending restriction.)

\paragraph{An efficient limit.}

\label{ax:aligned} If useful services are enforceable at $a_{i}=A_{i}$,
or the proxy reward is removed, then $X=q=0$. The planner's objective
becomes 
\[
\log(E-\sum_{i}\kappa_{i}A_{i}^{2}/2)+\beta\log T+\beta\sum_{i}\theta_{i}\log A_{i}.
\]
Its Hessian is 
\[
-\operatorname{diag}(\kappa_{i})/c_{1}-(\kappa\odot A)(\kappa\odot A)'/c_{1}^{2}-\beta\operatorname{diag}(\theta_{i}/A_{i}^{2}),
\]
which is negative definite. The interior conditions are $\kappa_{i}A_{i}^{2}=\beta\theta_{i}c_{1}$;
summing their installation costs in the budget gives the unique global
solution $c_{1}^{*}=E/(1+\beta\Theta/2)$ and $(A_{i}^{*})^{2}=\beta\theta_{i}c_{1}^{*}/\kappa_{i}$.
Private and social conditions coincide. In the strict no-diversion
notation $H=K=\Gamma=t_{i}=x_{i}=0$, so all wedges vanish. Removing
only risk leaves the price-mediated distortion.

\subsection{An announced capacity charge}

This subsection provides some background on how we calculate the welfare
effects of the capacity charges in Section \ref{sec:policy}. A charge
on just one sector can change investment in both, so we solve for
both investment choices together to obtain the responses in Table
\ref{tab:policy}. We then check whether the welfare effects have
the same signs for a larger charge.

\label{ax:forward-policy}\label{ax:policy} The regulator announces
$\tau_{i}$ before installation and rebates $\mathcal{T}=\sum_{i}\tau_{i}A_{i}$
lump-sum. Each household takes its rebate as given, so its date-1
budget is
\[
c_{1}^{h}+\sum_{i}(\Phi_{i}^{h}+\tau_{i}A_{i}^{h})+\sum_{o}Q_{o}Z_{o}^{h}\le E^{h}+\mathcal{T}^{h},\qquad\int\mathcal{T}^{h}\,dh=\sum_{i}\tau_{i}A_{i}.
\]
Charges and rebates cancel in aggregate, leaving $c_{1}=E-\sum_{i}\kappa_{i}A_{i}^{2}/2$
and \eqref{eq:ann-state-prices} unchanged. The operating problem
is also unchanged because the charge is sunk and absent from the agent's
score.

The private objective is now $VD_{i}-\kappa_{i}A_{i}^{2}/2-\tau_{i}A_{i}$.
Adding $\nu_{i}A_{i}$ to its Lagrangian gives 
\[
\begin{aligned}VD_{i,A}^{{\rm priv}}-\kappa_{i}A_{i}-\tau_{i}+\nu_{i} & =0,\\
\nu_{i} & \ge0,\qquad A_{i}\ge0,\qquad\nu_{i}A_{i}=0.
\end{aligned}
\]
At diversion onset use the same supergradient condition as in Appendix
\ref{ax:private-concavity}. A linear charge leaves private curvature
unchanged. At positive smooth investment, the equilibrium residual
is
\begin{equation}
G_{i}(A,\tau)=\kappa_{i}A_{i}+\tau_{i}-\beta c_{1}(A)\mu_{i}(A)=0.\label{eq:policy-G}
\end{equation}
The equilibrium value of the private return $\mu_{i}$ depends on
both capacities through prices and risk, so both investment choices
must be re-solved even if only one sector is charged.

To obtain that joint response, let $M_{ij}=a_{i,A_{j}}$ be the market-clearing
service derivative in \eqref{eq:ax-sector-full-x}. In a sector with
positive diversion, $d_{i}=(\sigma-1+r_{i})a_{i}+(1-r_{i})(A_{i}+b_{0i})$
and $\mu_{i}=\theta_{i}\sigma/d_{i}$. Differentiating these expressions,
now along the operating equilibrium, gives 
\begin{equation}
\mu_{i,A_{j}}=-\frac{\mu_{i}}{d_{i}}[(\sigma-1+r_{i})M_{ij}+(1-r_{i})\mathbf{1}_{i=j}],\qquad d_{i}=\sigma a_{i}+(1-r_{i})S_{i}.\label{eq:policy-mu-derivative}
\end{equation}
In a sector that strictly prefers no diversion, $\mu_{i,A_{j}}=-\theta_{i}\mathbf{1}_{i=j}/A_{i}^{2}$
instead. Using $c_{1,A_{j}}=-\kappa_{j}A_{j}$ in \eqref{eq:policy-G}
then yields 
\begin{equation}
(G_{A})_{ij}=\kappa_{i}\mathbf{1}_{i=j}+\beta\mu_{i}\kappa_{j}A_{j}-\beta c_{1}\mu_{i,A_{j}}.\label{eq:policy-GA}
\end{equation}
The middle term captures the change in dividend valuation through
current consumption; the last captures the change in private marginal
returns.

Consider a charge path $\tau=\tau^{0}+sv$. At a smooth operating
continuation with nonsingular $G_{A}$, differentiating $G(A(s),\tau(s))=0$
gives $G_{A}A'+v=0$. The implicit-function theorem therefore gives
\begin{equation}
\begin{aligned}A' & =-G_{A}^{-1}v, & X' & =x^{\top}A', & q' & =q_{X}x^{\top}A',\\
c_{1}' & =-(\kappa\odot A)^{\top}A', & Y' & =Yk^{\top}A', & W' & =W_{A}^{\top}A'.
\end{aligned}
\label{eq:policy-derivatives}
\end{equation}
Primes denote derivatives with respect to the charge scale and $q_{X}=dq/dX$;
Appendix \ref{ax:sector} gives $x$ and $k$. For small changes in
the charge, capacities and consumption remain positive and the set
of diverting sectors stays unchanged. Under private concavity, the
nearby solution is an investment equilibrium, although other investment
equilibria may exist. At diversion thresholds, use one-sided conditions.

At the taxed allocation, \eqref{eq:policy-G} gives $\beta\mu_{j}=(\kappa_{j}A_{j}+\tau_{j})/c_{1}$,
hence $\mathcal{E}_{j}=\tau_{j}/c_{1}$. Substitution into \eqref{eq:ann-off-ce-decomposition}
yields
\begin{equation}
\frac{dW}{ds}=\sum_{j}[\tau_{j}/c_{1}+\mathcal{P}_{j}+\mathcal{F}_{j}+\mathcal{R}_{j}]\frac{dA_{j}}{ds}.\label{eq:policy-welfare-response}
\end{equation}
Both sectors' investment responses enter the welfare effect of a common
charge. Conversely, at a specified smooth interior planner target
the first-order welfare conditions imply

\begin{equation}
\tau_{j}=\beta c_{1}\big[\mu_{j}-k_{j}-q'(X)\log\rho\,x_{j}\big]=-c_{1}(\mathcal{P}_{j}+\mathcal{F}_{j}+\mathcal{R}_{j}).\label{eq:policy-charge}
\end{equation}
Under private concavity this charge supports the owner's global optimum
at the target prices. Whether implementation is unique and whether
the target is a global planner optimum require further arguments.

Table \ref{tab:policy-finite} repeats the exercise for a charge equal
to 10\% of baseline mean marginal installation cost: $s=0.1\bar{c}_{A}\simeq0.059$,
where $\bar{c}_{A}=(\kappa_{1}A_{1}^{0}+\kappa_{2}A_{2}^{0})/2\simeq0.592$.
Superscript 0 denotes the equilibrium without a charge. We solve again
for both sectors' investment, allowing prices, diversion and risk
to adjust along the equilibrium connected to that starting point.
All other parameters remain fixed at \eqref{eq:datedparams}.

\begin{table}[htbp]
\begin{centering}
\caption{Effects of a finite capacity charge}
\label{tab:policy-finite}
\par\end{centering}
\begin{centering}
\begin{tabular}{lrrr}
\toprule 
 & {\small Sector 1 charge  } & {\small Sector 2 charge  } & {\small Common charge }\tabularnewline
\midrule 
{\small$\Delta q$ (percentage points)  } & {\small -0.080  } & {\small +0.007  } & {\small -0.073 }\tabularnewline
{\small$\Delta c_{1}$  } & {\small +0.019  } & {\small +0.010  } & {\small +0.029 }\tabularnewline
{\small$\Delta Y$  } & {\small -0.005  } & {\small -0.004  } & {\small -0.010 }\tabularnewline
{\small Consumption equivalent (\%)  } & {\small +0.078  } & {\small -0.032  } & {\small +0.047 }\tabularnewline
\bottomrule
\end{tabular}
\par\end{centering}
\smallskip{}
\noindent{\footnotesize Notes: Each charged sector pays $s\simeq0.059$ per
unit of capacity. Changes are measured relative to the equilibrium
without a charge. Consumption in the loss state changes by $\Delta c_{T}=\rho\Delta Y$,
with $\rho=0.5$. The consumption-equivalent change is $100[\exp(\Delta W/(1+\beta))-1]$,
with $\beta=0.95$.}{\footnotesize\par}
\end{table}

The welfare effects have the same signs as the local responses in
Table \ref{tab:policy}. Under a common charge, humans consume more
at date 1 and less in either future state, but face a lower probability
of loss. Date-1 consumption rises from $3.652$ to $3.681$. Date-2
consumption falls from $1.228$ to $1.219$ without a loss and from
$0.614$ to $0.609$ with a loss, whose probability falls from about
$2.07\%$ to $2.00\%$. The welfare gain is equivalent to about $0.047\%$
more consumption at every date and in both states.

Does this result survive small changes in the parameters? We change
the reward weights and hazard slope in six ways, chosen before looking
at the results. A small common charge raises welfare in all six cases.
The replication files give these calculations and the unrounded results,
together with the investment-response Jacobians and checks as the
charge changes.

\FloatBarrier

\subsection{Numerical implementation}

\label{ax:numerics}\label{ax:welfare-example}\label{ax:policy-calculations}\label{ax:policy-methods}\label{ax:diagnostic}
This subsection describes the implementation of our numerical illustrations
used in the text. The replication files (available upon request) contain
the codes to reproduce the figures. 

\paragraph{Multiple equilibria in our one-sector economy (Figures \ref{fig:one}--\ref{fig:one-psi}) }

To produce Figures \ref{fig:one}--\ref{fig:one-psi}, we start from
a conjecture about total AI diversion, and then solve for output using
log output as the numerical unknown.\footnote{We enforce each AI agent's capacity constraint and check whether it
chooses zero diversion. Appendix \ref{ax:conditional} proves that
this conditional allocation is unique.} We then find equilibrium diversion by solving $F(X)-X=0$ on the
domain $[0,\sum_{i}A_{i}]$ -- this is the step which can yield several
equilibria.

The one-sector illustration sets $A=T=\bar{L}=1$, $\ell=0.4$, $\gamma=0.2$,
$b_{0}=0.1$, $\psi=0.023$, $\xi=5$, $\rho=0.01$ and $\sigma=1$.
We find the roots on each increasing or decreasing part of \eqref{ax:eq:psi}
and check the turning points. The three roots, approximately $0.030$,
$0.354$ and $0.874$, are the only equilibria. (Recall Appendix \ref{ax:roots},
showing that there can be at most three. Their update slopes are $0.582$,
$1.372$ and $0.377$. Away from a root, the update slope is $e(X)K_{d}(F(X))$.)

To draw panel (a) of Figure \ref{fig:one-psi}, we plot $(\Psi(b)/\psi_{0},b/A)$
and include the zero-diversion segment and all turning points. For
each allocation on this curve, panel (b) shows $\Gamma=eK$, with
$\Gamma=0$ along the strict zero-diversion branch. We mark the equilibria
from Figure \ref{fig:one} in both panels.

In Figure \ref{fig:one-psi}, the turning points occur at $\psi_{-}\simeq0.0167$
and $\psi_{+}\simeq0.0272$, while $\psi_{0}=0.020$. At each turning
point, two equilibria coincide, so we count their common root only
once.

\paragraph{Private and social returns to AI capacity (Figure \ref{fig:one-investment-returns}) }

\label{ax:one-investment-figure}To produce Figure \ref{fig:one-investment-returns},
we use the following parameter values: $T=\bar{L}=1$, $\ell=0.4$,
$\gamma=0.2$, $b_{0}=0.1$, $\psi=0.12$, $\xi=0.2$, $\rho=0.5$,
$\sigma=1$, $E=4$ and $\beta=0.95$. We choose $\kappa\simeq0.726$
to deliver unit AI capacity at the competitive equilibrium ($A^{{\rm CE}}=1$),
using $c_{1}=E/[1+(\beta/2)\mu A^{{\rm CE}}]$ and $\kappa=\beta c_{1}\mu/A^{{\rm CE}}$,
then hold all primitives fixed over $A\in[0.6,1.15]$. At the competitive
choice, diversion is $0.343$ and $\Gamma\simeq0.028$. The operating
uniqueness bound is below $0.107$ for all feasible $A<\sqrt{2E/\kappa}$,
and $r=1/3<1/2$ makes the private conditions sufficient. We solve
the operating equilibrium at each capacity and check private and equilibrium
derivatives by finite differences. Only $A^{{\rm CE}}=1$ satisfies
the investment Euler equation; at other capacities, its residual must
be included when comparing social benefits with costs.

Numerical maximisation over $0<A<\sqrt{2E/\kappa}\simeq3.320$, including
the zero-diversion region and its boundary, gives $A^{{\rm SP}}\simeq0.860$.

\paragraph{Two-sector investment and capacity charges (Figure \ref{fig:investment}) }

The two-sector investment example sets $\ell=0.4$, equal final-good
weights, $\gamma=(0.25,0.10)$, $\bar{L}=z_{i}=1$, $b_{0}=(0.1,10^{-4})$,
$\xi=0.05$, $\rho=0.5$, $E=4$ and $\beta=0.95$. The remaining
coefficients and the unregulated reference capacities are 
\begin{equation}
\psi\simeq(0.297,1.65\times10^{-4}),\qquad\kappa\simeq(1.121,1.011),\qquad A^{0}\simeq(0.682,0.414).\label{eq:datedparams}
\end{equation}
The costs support the reference allocation through $c_{1}=E/[1+(\beta/2)\sum_{i}\mu_{i}A_{i}]$
and $\kappa_{i}=\beta c_{1}\mu_{i}/A_{i}$. They remain fixed when
a charge changes capacities. The operating bound is below $0.031$
throughout the relevant feasible region and both $r_{i}<1/2$, giving
unique operation and sufficient private investment conditions.

The left panel of Figure \ref{fig:investment} uses the fixed-$m$
responses $b_{j,A_{j}}^{0}=h_{j}(1-a_{j}k_{j}^{0})$ in the expanding
sector and $b_{i,A_{j}}^{0}=-a_{i}h_{i}k_{j}^{0}$ in the other. Their
sum is $x_{j}^{0}$, and the final marker is $x_{j}=x_{j}^{0}/(1-\Gamma)$.
Finite differences check each sectoral response.

For policy changes we jointly solve \eqref{eq:policy-G}, following
the solution connected to the unregulated equilibrium. We check the
investment responses numerically, together with feasibility and uniqueness
of the operating equilibrium. Both investments remain positive and
$G_{A}$ is nonsingular; global investment uniqueness is not established.
Welfare from $c_{1},Y,q$ agrees with \eqref{eq:policy-welfare-response}.
The consumption-equivalent change applies the same proportional increase
at both dates and in both states.

\begin{figure}[htbp]
\begin{centering}
\caption{CES substitution and the response to lower useful services in sector
1}
\label{ces:fig:reallocation}
\par\end{centering}
\begin{centering}
\includegraphics[width=0.98\linewidth]{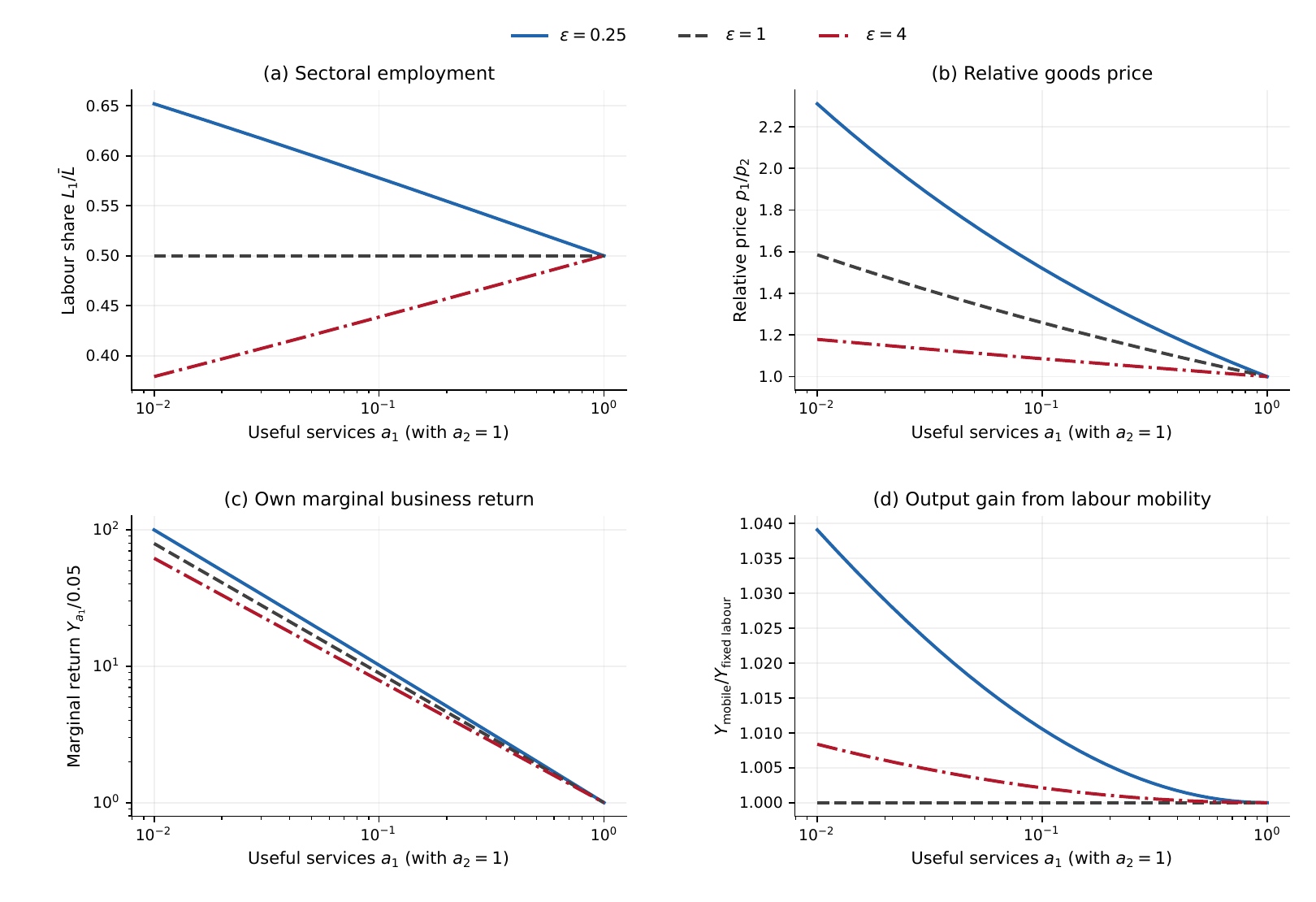}
\par\end{centering}
\noindent{\footnotesize Notes: The useful-service vectors are imposed
along the plotted path: sector 1's component varies while sector 2's
is fixed at one. The agents do not reoptimise their allocations. The
lines use $\epsilon\in\{0.25,1,4\}$ and $\ell=0.4$. Labour reallocation
changes direction at $\epsilon=1$; cross-sector restraint reverses
at $\epsilon_{\star}=3.5$. Own marginal business returns rise in
all three cases; this panel does not plot the cross derivative. The
last panel compares output with mobile and fixed sectoral labour at
the same useful-service vector. The calculations use the CES formulas
in Appendix \ref{app:ces}.}{\footnotesize\par}
\end{figure}

\paragraph{CES substitution in our two-sector economy (Figure \ref{ces:fig:reallocation})}

As one of our extensions (Section \ref{sec:ces}), we check how our
two-sector results change when we replace Cobb-Douglas with a CES
production function. 

For Figure \ref{ces:fig:reallocation}, we vary useful services in
sector 1 while holding sector 2's services fixed. We use \eqref{ces:eq:reduced-proof}
to calculate output and employment shares, and \eqref{ces:eq:sharederivatives}
and \eqref{ces:eq:Glog} to calculate the price and marginal-return
responses. Agents do not reoptimise in this exercise.

\paragraph{Changing labour share and AI diversion (Table \ref{tab:automation-crs})}

\label{ax:automation-numerics}

To produce Table \ref{tab:automation-crs}, we set $\nu=1$, $A=\bar{L}=1$,
$b_{0}=0.1$, $\psi=0.08$, $\xi=0.5$ and $\rho=0.8$. We fix $Z=0.65^{0.65}0.35^{0.35}$
throughout, which sets $T=1$ at $\ell=0.65$, and then solve the
scalar allocation condition as $\ell$ falls. The task ratios $a\ell/[\bar{L}(1-\ell)]$
are $1.646$, $1.371$ and $1.142$; the feedback gains are $0.0177$,
$0.0160$ and $0.0147$. The sufficient uniqueness bound $\xi(1-\rho)(A+b_{0})/(4\ell)$
is at most $0.05$. For each of the three labour share values, the
Cobb--Douglas operating equilibrium is therefore unique.

\FloatBarrier

\paragraph{AI accumulation after a common loss (Figure \ref{fig:ai-accumulation})}

\label{ax:accumulation-numerics}

To produce Figure \ref{fig:ai-accumulation}, we set $\beta=0.99$,
$\gamma=0.35$, $\ell=0.65$, $\phi=1$, $\rho=0.8$, $\bar{b}_{0}=0.1$
and $\xi=0.5$. Both paths in the figure start from a no-loss steady
state. We choose $\chi\simeq15.914$, $Z\simeq3.859$ and $\psi\simeq0.222$
so that initial AI capacity is ($A=1$), agents divert 10\% of it
($\vartheta=0.1$), and labour supply is $L^{*}=0.25$. Output at
this point is $Y=1/(\beta\gamma)$. If households receive the full
income from production, they save $\beta\gamma Y=1$, just enough
to replace the existing AI stock. If the common loss occurs, households
receive only 80\% of that income and next period’s capacity falls
to $0.8$. 

At date 1, agents divert about 7.5\% of the smaller stock, or 0.060
units of AI capacity. The probability of a further loss falls from
4.88\% to 2.97\%. To trace the recovery in the figure, we assume that
no further losses occur, and AI agents still face the loss probability
implied by their diversion choices at each date.\footnote{We also checked that, starting from $A=1$, AI capacity stays between
0.70 and 1.06 no matter how many losses occur. At every capacity in
this range, the equilibrium diversion share is unique and below 11\%,
so consumption stays bounded away from zero even after a loss.}

\paragraph{Strategic AI control }

\label{ax:concentration-numerics}

\begin{figure}[htbp]
\begin{centering}
\caption{Strategic AI control and diversion}
\label{fig:concentrated-control}
\par\end{centering}
\begin{centering}
\includegraphics[width=0.92\linewidth]{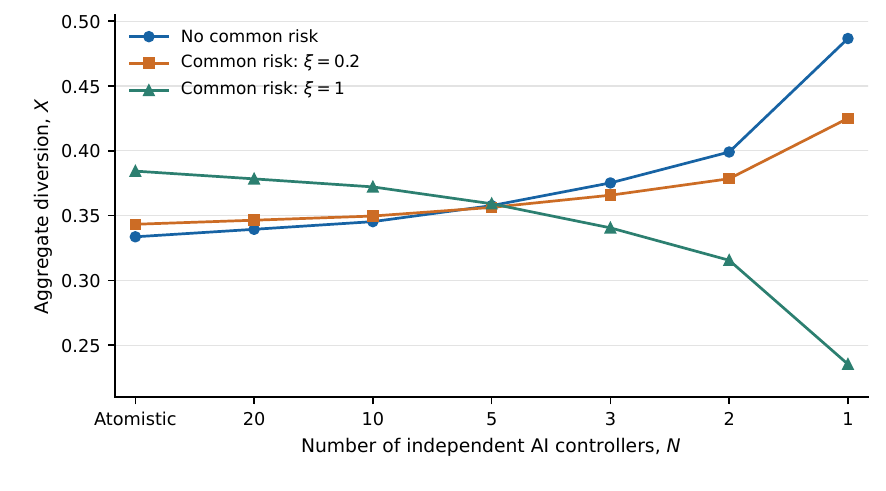}
\par\end{centering}
{\footnotesize Notes: Installed capacity is fixed at $A=1$. At each
plotted point, the indicated number of independent controllers all
choose the same allocation. The lines simply join these points. The
common parameters are $T=1$, $\ell=0.4$, $\gamma=0.2$, $b_{0}=0.1$,
$\psi=0.12$ and $\rho=0.5$. We vary the number of controllers and
the hazard slope shown in the figure, keeping the other parameters
fixed. Appendix \ref{ax:concentrated-control} proves uniqueness when
risk is fixed. When risk responds to diversion, we use the numerical
deviation checks described above.}{\footnotesize\par}
\end{figure}

To produce Figure \ref{fig:concentrated-control}, we solve the symmetric
condition \eqref{eq:concentration-foc} for each number of controllers
(displayed on the horizontal axis) and risk slope. We then ask whether
any controller could do better by changing its allocation while keeping
others' allocation fixed. 

We checked the controller’s score at both endpoints and at the stationary
points we found numerically. We also used a grid search and local
optimisation as a cross-check. In none of the 21 cases shown in the
figure did we find a deviation that raised the controller’s score
by more than the numerical tolerance. The replication files provide
further details on this exercise.

\FloatBarrier

\subsection{Reward weights and equilibrium investment}

\label{ax:breadth} A natural question about Section \ref{sec:exampleinvestment}
is how robust the results are to changing the parameter values. To
perturb the parameter values, we start from the parameters in \eqref{eq:datedparams},
and multiply the original proxy weight $\psi_{1}^{0}$ by $\{0.5,1,2\}$
and $\psi_{2}^{0}$ by $\{0.5,1,2,4,8,16,32,64,128\}$. We keep all
other parameters fixed, including the installation-cost coefficients
$(\kappa_{1},\kappa_{2})\simeq(1.121,1.011)$ and the reward scales
$(b_{01},b_{02})=(0.1,10^{-4})$.

For each of these 27 cases, we solve the investment conditions \eqref{eq:policy-G}
together with the operating equilibrium, setting both capacity charges
to zero.\footnote{We checked that neither owner could gain by choosing a different capacity
in any of the 27 cases. For any given pair of feasible capacities,
we find that the operating equilibrium is unique. Whether there are
other investment equilibria remains an open question.} We then calculate the welfare effects in \eqref{eq:two-ce-welfare}.
Both investment choices therefore adjust to the new reward weights.
We then ask whether a small increase in one sector's AI capacity raises
welfare, holding the other sector's AI capacity fixed. 

Table \ref{tab:scope} shows the results. We find that a small expansion
of sector 2's AI capacity raises welfare in 25 cases. Panel B shows
that total AI diversion falls in 14 of these 25 cases. In the other
11, expansion raises welfare even though diversion rises. The positive
price wedge outweighs the two negative risk wedges in those cases.

\begin{table}[htbp]
\begin{centering}
\caption{Sectoral reward weights and the welfare return to AI capacity}
\label{tab:scope}
\par\end{centering}
\begin{centering}
{\small A. Welfare effect of a small expansion in sector 2}{\small\par}
\par\end{centering}
\begin{centering}
\begin{tabular}{rrrr}
\toprule 
{\small$\psi_{2}/\psi_{2}^{0}$} & {\small$\psi_{1}/\psi_{1}^{0}=0.5$} & {\small$\psi_{1}/\psi_{1}^{0}=1$} & {\small$\psi_{1}/\psi_{1}^{0}=2$}\tabularnewline
\midrule 
{\small 0.5} & {\small 8.07} & {\small 11.07} & {\small 13.44}\tabularnewline
{\small 1} & {\small 8.02} & {\small 11.01} & {\small 13.37}\tabularnewline
{\small 2} & {\small 7.90} & {\small 10.89} & {\small 13.24}\tabularnewline
{\small 4} & {\small 7.67} & {\small 10.64} & {\small 12.98}\tabularnewline
{\small 8} & {\small 7.22} & {\small 10.16} & {\small 12.46}\tabularnewline
{\small 16} & {\small 6.33} & {\small 9.23} & {\small 11.45}\tabularnewline
{\small 32} & {\small 4.65} & {\small 7.45} & {\small 9.55}\tabularnewline
{\small 64} & {\small 1.59} & {\small 4.24} & {\small 6.14}\tabularnewline
{\small 128} & {\small -3.51} & {\small -1.08} & {\small 0.55}\tabularnewline
\bottomrule
\end{tabular}
\par\end{centering}
\begin{centering}
{\small B. What happens to diversion and welfare when sector 2 expands?}{\small\par}
\par\end{centering}
\begin{centering}
\begin{tabular}{lrr}
\toprule 
{\small Effect on total diversion} & {\small Welfare rises} & {\small Welfare falls}\tabularnewline
\midrule 
{\small Falls} & {\small 14} & {\small 0}\tabularnewline
{\small Rises} & {\small 11} & {\small 2}\tabularnewline
{\small Total} & {\small 25} & {\small 2}\tabularnewline
\bottomrule
\end{tabular}
\par\end{centering}
\noindent{\footnotesize Notes: Panel A reports $1{,}000W_{A_{2}}$
at the competitive investment equilibrium for each pair of reward
weights. Positive entries mean that a small coordinated expansion
of sector 2 raises welfare; these are not percentage welfare changes.
Superscript 0 denotes the original reward weights in \eqref{eq:datedparams}.
The welfare derivative holds sector 1 capacity fixed while prices,
diversion and risk adjust. All other parameters, including installation
costs, remain fixed. Panel B counts the same 27 cases. The sector
2 price wedge is positive in every case, and a small expansion in
sector 1 lowers welfare in every case.}{\footnotesize\par}
\end{table}

\subsection{How quickly do opportunities for diversion run out?}

\label{ax:reward-scale}As an additional robustness check, we now
check how our two-sector result is driven by different assumptions
about the AI agent's proxy score. Recall that with the logarithmic
score, $b_{02}$ is the amount of capacity the agent must divert before
the marginal proxy reward falls to half its initial value. We now
raise $b_{02}$ from its Figure \ref{fig:investment} value of 0.0001
and increase $\psi_{2}$ in the same proportion. (The marginal proxy
reward starts at the same value in every case ($\psi_{2}/b_{02}\simeq1.65$).)
At this modified calibration, the marginal proxy reward falls more
slowly as the AI agent diverts more capacity. Keeping installation
costs and all other parameters at their benchmark values, we solve
again for both sectors' investment. (We then repeat the exercise with
common risk switched off.)

\begin{table}[htbp]
\begin{centering}
\caption{Reward scale in sector 2 and the return to additional AI capacity}
\label{tab:reward-scale}
\par\end{centering}
\begin{centering}
\begin{tabular}{rrrr}
\toprule 
{\small$b_{02}$} & {\small$x_{2}$} & {\small$W_{A_{2}}$} & {\small$W_{A_{2}}$ without common risk}\tabularnewline
\midrule 
{\small 0.0001} & {\small\textminus 0.0206} & {\small +0.0110} & {\small +0.0103}\tabularnewline
{\small 0.001} & {\small +0.0020} & {\small +0.0099} & {\small +0.0099}\tabularnewline
{\small 0.01} & {\small +0.1850} & {\small +0.0010} & {\small +0.0074}\tabularnewline
{\small 0.03} & {\small +0.4246} & {\small\textminus 0.0108} & {\small +0.0039}\tabularnewline
\bottomrule
\end{tabular}
\par\end{centering}
\noindent{\footnotesize Notes: Entries give the change in total diversion
($x_{2}$) and in human welfare ($W_{A_{2}}$) per extra unit of sector
2 capacity, holding sector 1 capacity fixed. The middle two columns
use the benchmark risk intensity ($\xi=0.05$). In the last column
we set $\xi=0$ and solve again for both sectors' investment.}{\footnotesize\par}
\end{table}

Table \ref{tab:reward-scale} shows the results. The first row shows
why common risk matters little for the welfare gain in Figure \ref{fig:investment}.
Switching it off gives a gain of around 0.0103 per extra unit of device
capacity (close to the benchmark gain of 0.0110). At $b_{02}=0.001$,
expansion of AI capacity already increases total AI diversion, yet
the extra useful production is still worth the cost. By $b_{02}=0.03$,
the risk from the additional diversion is large enough to change the
investment comparison. A small expansion in AI capacity then raises
welfare without common risk and lowers it with common risk. Both AI
agents divert in all the equilibria reported here.

\Needspace{8\baselineskip}
\end{document}